\documentclass[traditabstract]{aa}
\pdfoutput=1
\usepackage{epsfig}
\usepackage{aalongtable}
\usepackage{natbib}
\bibpunct{(}{)}{;}{a}{}{,}    
\usepackage{graphics}
\usepackage{color} 
\usepackage{ulem}
\usepackage{times}
\usepackage{color}
\newfont{\tss}{cmssdc10 scaled 950}

\begin{document}

\title{Multi-wavelength study of 14000 star-forming galaxies 
from the Sloan Digital Sky Survey\thanks{Tables \ref{tab2} --  
\ref{tab3} and Figures \ref{fig10} -- \ref{fig15}
are available only in the electronic edition.}}


\author{Y. I.\ Izotov \inst{1,2,3}
\and N. G.\ Guseva \inst{1,2}
\and K. J.\ Fricke \inst{1,4}
\and C.\ Henkel \inst{1,5}
}
\offprints{Y.I. Izotov, izotov@mao.kiev.ua}
\institute{          Max-Planck-Institut f\"ur Radioastronomie, 
                     Auf dem H\"ugel 
                     69, 53121 Bonn, Germany
\and
                     Main Astronomical Observatory,
                     Ukrainian National Academy of Sciences,
                     Zabolotnoho 27, Kyiv 03680,  Ukraine
\and
                     LUTH, Observatoire de Paris, CNRS, 
                     Universite Paris Diderot,
                     Place Jules Janssen 92190 Meudon, France
\and
                     Institut f\"ur Astrophysik, 
                     G\"ottingen Universit\"at, Friedrich-Hund-Platz 1, 
                     37077 G\"ottingen, Germany
\and
                     Astronomy Department, King Abdulaziz University, 
                     P.O. Box 80203, Jeddah, Saudi Arabia
}

\date{Received \hskip 2cm; Accepted}
\abstract{
We studied a large sample 
of $\sim$ 14000 dwarf star-forming galaxies with strong emission lines.
These low-metallicity galaxies with oxygen abundances of
12+logO/H $\sim$ 7.4 -- 8.5 are selected from the Sloan Digital Sky Survey 
(SDSS) and distributed in the redshift range of $z$ $\sim$ 0 -- 0.6. 
We modelled spectral energy distributions (SED) of all galaxies
which were based
on the SDSS spectra in the visible range of 0.38$\mu$m -- 0.92$\mu$m
and included both the stellar and ionised gas emission. These SEDs were 
extrapolated to the UV and mid-infrared ranges to cover the wavelength
range of 0.1$\mu$m -- 22$\mu$m. The SDSS spectroscopic data were
supplemented by photometric data from the {\sl GALEX}, SDSS, 2MASS, 
{\sl WISE}, {\sl IRAS}, and NVSS all-sky surveys. Using these data, we derived
global characteristics of the galaxies, such as their element abundances,
luminosities, and stellar masses.
The luminosities and stellar masses range within the sample over 
$\sim$5 orders of magnitude, thereby linking low-mass and low-luminosity blue 
compact dwarf (BCD) galaxies to luminous galaxies, which are
similar to high-redshift Lyman-break galaxies (LBGs).
It was found that the luminosity $L$(H$\beta$) of the H$\beta$ emission line,
a characteristic of the youngest stellar population with an age of a few Myr,
is correlated with luminosities in other wavelength ranges. This implies
that the most recent burst of star formation makes a significant contribution
to the emission in the visible range and dominates in other wavelength ranges.
It was also found that the contribution of the young population to the galaxy
luminosity is higher for galaxies with higher $L$(H$\beta$) and higher
equivalent widths EW(H$\beta$). We found 20 galaxies with very red {\sl WISE} 
mid-infrared $m$(3.4$\mu$m) -- $m$(4.6$\mu$m)
colour ($\geq$ 2 mag), which suggests the important contribution
of the hot (with a temperature of several hundred degree) dust emission
in these galaxies.
Our analysis of the balance between the luminosity in the {\sl WISE} bands
that covered a wavelength range of 3.4$\mu$m -- 22$\mu$m and the luminosity of the
emission absorbed at shorter wavelengths showed that the luminosity
of the hot dust emission is increased with increasing $L$(H$\beta$)
and EW(H$\beta$). 
We demonstrated that the emission emerging from 
young star-forming regions is the dominant dust-heating source for temperatures
to several hundred degrees in the sample star-forming galaxies.}
\titlerunning{Multi-wavelength study of 14000 star-forming galaxies}
\authorrunning{Y.I.Izotov et al.}
\maketitle


\section{Introduction}

The problems related to the formation of first galaxies at high redshifts
from nearly pristine gas and the determination of their physical properties 
were extensively studied in recent years \citep[e.g., ][]{P01,S09,E10,SB10}. 
However, these studies are very difficult because of their
large distances and their faintness, which leads to coarse linear resolutions 
and low signal-to-noise ratios. 
Most of the extensive high-redshift studies \citep[e.g., ][]{Cr09,FS09,E12}
were focussed on the bright massive disc-like galaxies at $z$ $\ga$ 1. 
Recent merger events were frequently detected in these galaxies. On the other
hand, high-redshift dwarf galaxies are too faint to be detected.

The visible wavelength range of 
low-redshift star-forming galaxies is the most informative one for 
diagnostics of the ionised gas and the determination of its chemical 
composition. However, this part of the spectrum is shifted to the infrared 
range in high-redshift objects, which is strongly contaminated by the
emission and absorption lines of the Earth's atmosphere, making studies
with ground-based telescopes difficult or impossible. 

The situation could possibly be improved after the launch of the 
{\sl James Webb} space telescope. However, there is an alternative 
approach to study
the nearby galaxies, which are expected to have properties similar to those of 
distant galaxies. 
These galaxies may play an important role for our
understanding of star-formation processes in low-metallicity environments, and 
they can be considered as local counterparts or analogues of high-redshift 
LBGs. Due to their proximity, they can be
studied in greater detail not resolvable in distant galaxies.

In recent years, \citet{H05} have
identified nearby ($z$ $<$ 0.3) ultraviolet-luminous galaxies (UVLGs) selected
from the {\sl Galaxy Evolution Explorer} ({\sl GALEX}). These compact UVLGs 
were eventually called Lyman-break analogues (LBAs). They resemble LBGs in 
several respects.  In particular, their metallicities are 
subsolar, and their star-formation rates ($SFR$s) of 
$\sim$ 4 -- 25 $M_\odot$ yr$^{-1}$ are overlapping with those for LBGs.
\citet{G10} studied kinematical properties of LBAs and found their 
striking similarities with LBGs.

\citet{C09} selected a sample of 251 compact strongly 
star-forming galaxies
at $z$ $\sim$ 0.112 -- 0.36 on the basis of their green colour on the 
Sloan Digital Sky Survey (SDSS) composite $g,r,i$ 
images (``green pea'' galaxies), which again 
are similar to LBGs because of their low metallicity and high $SFR$s. 

\citet{I11a} extracted a sample of 803 star-forming luminous compact
galaxies (LCGs) with hydrogen H$\beta$ 
luminosities 
$L$(H$\beta$) $\geq$ 3$\times$10$^{40}$ erg s$^{-1}$ and H$\beta$ equivalent
widths EW(H$\beta$) $\geq$ 50\AA\ from SDSS spectroscopic data.
These galaxies have properties 
similar to ``green pea'' galaxies but are distributed over a wider
range of redshifts $z$ $\sim$ 0.02 - 0.63. The $SFR$s of LCGs are high at
$\sim$ 0.7 -- 60 $M_\odot$ yr$^{-1}$ and overlap with those of LBGs.

\citet[see also \citet{G09}]{I11a} showed that LBGs, LCGs, luminous metal-poor
star-forming galaxies 
\citep{Hoyos05}, extremely metal-poor 
emission-line galaxies at $z$ $<$ 1 
\citep{Kakazu07}, 
and low-redshift blue compact
dwarf (BCD) galaxies with strong star-formation activity 
obey a common luminosity-metallicity relation. 
Therefore, it is promising to study nearby 
strongly star-forming galaxies over a wide 
range of luminosities and metallicities to shed light on physical conditions
and star-formation histories in low and high-redshift galaxies. 

\setcounter{table}{0}

\begin{table*}
\centering{
\caption{The number of SDSS emission-line galaxies detected in different bands. \label{tab1}}
\begin{tabular}{lccccccccccccccc} \hline
Property &Spectra&&\multicolumn{13}{c}{Photometry} \\ \cline{2-2} \cline{4-16}
 &SDSS   &&\multicolumn{2}{c}{{\sl GALEX}}&SDSS&\multicolumn{3}{c}{2MASS}&&\multicolumn{4}{c}{{\sl WISE}}&\multicolumn{1}{c}{{\sl IRAS}}&\multicolumn{1}{c}{NVSS} \\
 \cline{4-5} \cline{7-9} \cline{11-14}
 &       &&FUV&NUV& $g$& $J$  & $H$ & $K$ && 3.4$\mu$m& 4.6$\mu$m& 12$\mu$m& 22$\mu$m& 60$\mu$m& 20cm \\ \hline
Total    &14610   &&12610  &12518  &14607  &6256&5405&3952&&13873   &13673   &11468   &9197&337&655 \\ 
Compact  &~\,6958&&~\,6098&~\,6130&~\,6958&3380&3072&2271&&~\,6958&~\,6431&~\,5645&4671&~\,92&134 \\ \hline
\end{tabular}
}
\end{table*}

The completion of the space {\sl GALEX} survey in the UV range \citep{M07},
the Sloan Digital Sky Survey (SDSS) in the visible range \citep{A09}, 
the 2MASS survey in the near-infrared range \citep{S06}, 
the space {\sl Wide-field Infrared Survey Explorer} survey 
\citep[{\sl WISE}, ][]{W10}, and the NRAO VLA Sky Survey (NVSS) in the
radio continuum at 20 cm \citep{C98}
opened the opportunity to study properties of large samples of
star-forming galaxies over the extremely large 
wavelength range of $\sim$ 0.15$\mu$m -- 20 cm.

We used all these data bases with the aim to fit spectral energy distributions
(SEDs) in $\sim$ 14000 star-forming galaxies with strong emission lines
to derive global characteristics of the galaxies and the relations
between them.

\section{The sample \label{sel}}

   We used the whole spectroscopic data base of the SDSS DR7 \citep{A09},
which is comprised of $\sim$ 900000 spectra of galaxies to select 
a sample of 14610 spectra with strong emission lines, 
which do not show evidence of AGN activity. 
Most notable features used for this are broad lines in spectra of
Sy1 galaxies and strong high-ionisation [Ne~{\sc v}] $\lambda$3426, 
He~{\sc ii} $\lambda$4686 emission lines in spectra of Sy2 galaxies.
We select only spectra of star-forming galaxies with equivalent widths 
EW(H$\beta$) $\ga$ 10\AA.
This selection criterion is chosen to make diagnostics of the ionised gas
more reliable. In particular, dust extinction in these spectra is
obtained from decrement of several hydrogen lines. Furthermore,
[O {\sc iii}] $\lambda$4959, 5007 emission lines were present in all
spectra. They are strong in a vast majority of spectra making the 
determination of the chemical composition more reliable.
The [O {\sc iii}]$\lambda$5007/H$\beta$ vs. 
[N {\sc ii}]$\lambda$6583/H$\alpha$ diagram for star-forming galaxies 
with EW(H$\beta$) $\ga$ 10\AA\ is shown in Fig. \ref{fig1} by blue symbols.
For comparison, spectra of all SDSS DR7 galaxies are shown by grey dots.
The red solid line from \citet{K03} separates star-forming galaxies from AGNs.

There is a very small fraction of the galaxies from our sample
which are located to the right from the solid line (Fig. \ref{fig1})
and implies some AGN activity. However,
we decided to keep these galaxies in the sample. \citet{K03} noted that 
the exact separation between star-forming galaxies and AGN is subject to 
considerable uncertainty. Their selection
of the demarcation between star-forming galaxies and AGN
was subjective. It was based on the suggestion 
that star-forming galaxies exhibit little scatter around a single relation in 
the BPT diagram. However, this is not exactly the case, according to 
Fig. \ref{fig1}. There is an appreciable number of star-forming galaxies 
with large scatter.

A small number of selected spectra (224 spectra) represents individual 
H {\sc ii} regions in nearby spirals but
the overwhelming majority is composed of integrated spectra of galaxies from 
farther distances. In particular, approximately half of the galaxies have
a compact structure with angular diameter not exceeding 6\arcsec\ 
(Table \ref{tab1}). We also noted that there are 452 galaxies ($\sim$ 3\%) 
with two or more SDSS spectra of the same H {\sc ii} region that are
obtained in different epochs. 
We excluded spectra of H {\sc ii} regions in nearby spirals and duplicate
spectra.
Finally, our sample included 13934 spectra.
We supplemented this data with photometric data
in five SDSS bands $u$, $g$, $r$, $i$, and $z$.

The selected SDSS objects were cross-identified with
sources from photometric sky surveys in other
wavelength ranges, fully covering the range of equatorial coordinates of
the SDSS. In spite of this we did not cross-identify 
sources from the {\sl Spitzer} and {\sl Herschel} space infrared missions
since they do not fully cover the SDSS region.

In the UV range, we used {\sl Galaxy Evolution Explorer} ({\sl GALEX}) 
FUV ($\lambda_{\rm eff}$ = 1528\AA) and NUV ($\lambda_{\rm eff}$ = 2271\AA)
observations. We identified 86\% of galaxies with {\sl GALEX} sources in both the
FUV and NUV bands, matching the coordinates of the 
objects with selected SDSS spectra within a round aperture of 6\arcsec\
in radius (Table \ref{tab1}).

In the near-infrared range, the SDSS sample was cross-identified with
the 2MASS sources. We identified 43\%, 37\%, and 27\% of galaxies with sources 
in the 
$J$ ($\lambda_{\rm eff}$ = 1.22$\mu$m), $H$ ($\lambda_{\rm eff}$ = 1.63$\mu$m),
and $K_s$ ($\lambda_{\rm eff}$ = 2.19$\mu$m) bands, respectively,
which indicate a lower detectability of 2MASS in comparison to {\sl GALEX}
(Table \ref{tab1}).

The coordinates of SDSS selected galaxies were used to identify sources in 
the {\sl WISE} All-Sky Source Catalog (ASSC) within a circular aperture of 
6\arcsec\ in radius.
In total, 95\%, 94\%, 78\%, and 63\% of SDSS galaxies were identified with
{\sl WISE} sources respectively
in $W1$ (3.4$\mu$m), $W2$ (4.6$\mu$m), $W3$ (12$\mu$m), and $W4$ (22$\mu$m)
bands (Table \ref{tab1}). 

Furthermore, we cross-identified the SDSS sample with sources from the 
{\sl IRAS} far-infrared survey at 60 $\mu$m and from the NVSS survey in the 
continuum at 20 cm within matching radii of 6\arcsec\ and 15\arcsec, 
respectively. Only a small fraction of the SDSS objects were detected 
in these surveys (Table \ref{tab1}).

Finally, the SDSS sample was cross-identified with sources from the {\sl ROSAT}
X-ray survey and the {\sl Planck} sub-mm survey. Only a few objects within 
the matching radius of 20\arcsec\ were found in these surveys. Therefore, we 
did not use the data from the {\sl ROSAT} and {\sl Planck} surveys in this 
paper.

\section{The determination of galaxy parameters}

\subsection{Spectroscopic data from the SDSS}

The SDSS spectra were used to derive emission-line fluxes, equivalent
widths, the extinction
coefficient $C$(H$\beta$), the H {\sc ii} region luminosity in the H$\beta$ 
emission line, and chemical element abundances.
First, the spectra were retrieved from the SDSS database. The observed line
fluxes were obtained using the IRAF\footnote {IRAF is the Image 
Reduction and Analysis Facility distributed by the National Optical Astronomy 
Observatory, which is operated by the Association of Universities for Research 
in Astronomy (AURA) under cooperative agreement with the National Science 
Foundation (NSF).} SPLOT routine. The line-flux errors 
included statistical errors in addition to errors introduced by
the standard-star absolute flux calibration, which we set to 1\% of the
line fluxes. These errors will be later propagated into the calculation
of abundance errors.
The line-flux densities were corrected for two effects: (1) reddening using 
the extinction curve of \citet{C89} and 
(2) underlying hydrogen stellar absorption that is derived simultaneously by an
iterative procedure, as described in \citet{ITL94}.

  The correction for
extinction was done in two steps. First, emission-line intensities with
observed wavelengths were corrected for the Milky Way extinction, 
using values of the extinction  $A$($V$) in the $V$ band from the 
NASA/IPAC Extragalactic Database (NED).
Then, the internal extinction was derived from the Balmer hydrogen emission 
lines after correction for the Milky Way extinction. 
The internal extinction was
applied to correct line intensities at non-redshifted wavelengths.
The extinction coefficients in both cases of the Milky Way and the internal 
extinction are defined as $C$(H$\beta$) = 1.47$E(B-V)$,
where $E(B-V)$ = $A(V)$/3.2 \citep{A84}. 

   To determine element abundances, we generally followed  
the procedures of \citet{ITL94,ITL97} and \citet{TIL95}.
We adopted a two-zone photoionised H {\sc ii}
region model: a high-ionisation zone with temperature $T_{\rm e}$(O~{\sc iii}),
where the [O~{\sc iii}] lines
originate, and a 
low-ionisation zone with temperature $T_{\rm e}$(O~{\sc ii}), where the [O~{\sc ii}]
lines originate. 
In the H {\sc ii} regions with a detected [O~{\sc iii}] $\lambda$4363
emission line, the temperature $T_{\rm e}$(O~{\sc iii}) was calculated using 
the direct method based on the 
[O~{\sc iii}] $\lambda$4363/($\lambda$4959+$\lambda$5007) line ratio.
In total, the [O~{\sc iii}] $\lambda$4363 emission line was detected
in $\sim$5700 spectra. Its flux with accuracy better than 50\% was measured
in $\sim$2800 spectra.

In H {\sc ii} regions where the [O~{\sc iii}] $\lambda$4363 emission line 
was not detected, we used a semi-empirical method described by \citet{IT07} to 
derive $T_{\rm e}$(O~{\sc iii})
from the sum of line fluxes $R_{23}$ =
$I$([O {\sc ii}] $\lambda$3727 + [O {\sc iii}] $\lambda$4959
+ [O {\sc iii}] $\lambda$5007)/$I$(H$\beta$). 

For $T_{\rm e}$(O~{\sc ii}), we used
the relation between the electron temperatures $T_{\rm e}$(O~{\sc iii}) and
$T_{\rm e}$(O~{\sc ii}) obtained by \citet{I06} from
the H {\sc ii} photoionisation models of \citet{SI03}.   
Ionic and total oxygen abundances were derived
using expressions for ionic abundances and ionisation correction factors (ICFs)
obtained by \citet{I06}. 

We checked the validity of the semi-empirical
method for the abundance determination comparing oxygen abundances 
which were derived
by direct and semi-empirical methods in galaxies where both methods can be
applied. We found that oxygen abundances derived by both methods agree within 
0.2 dex (Fig. \ref{fig2}). The largest difference 
between direct and semi-empirical oxygen abundances
was found for galaxies with 12+logO/H $\ga$ 8.0. This is because
\citet{IT07} obtained
the relation between $T_{\rm e}$(O~{\sc iii}) and $R_{23}$ 
for galaxies with 12+log O/H $<$ 8.0
exhibiting ``lower branch'' objects on the $R_{23}$ -- 12+logO/H diagram.
At higher 12+logO/H $\ga$ 8.0 the method described by \citet{IT07} 
is less accurate.

The extinction-corrected luminosity $L$(H$\beta$) was obtained from the
observed H$\beta$ emission-line flux adopting the total extinction, which
included both the Milky Way and internal galaxy extinctions, and the
distance derived from the redshift. For distance determination 
we used the relation $D$ = $f$($z$,$H_0$,$\Omega_{\rm M}$,$\Omega_\Lambda$)
from \citet{R67}, where
the Hubble constant $H_0$ = 67.3 km s$^{-1}$ Mpc$^{-1}$ and cosmological
parameters, $\Omega_{\rm M}$ = 0.273 and $\Omega_\Lambda$ = 0.682, 
were obtained from
the {\sl Planck} mission data \citep{A13}. The equivalent widths EW(H$\beta$)
were reduced to the rest frame.

The SDSS spectra were obtained with a small aperture of 3\arcsec\ in 
diameter. To derive integrated characteristics of the galaxies from their
spectra and to make a comparison with photometric data more accurate, we 
corrected spectroscopic data for the aperture using the relation 
2.5$^{r(3\arcsec)-r}$, where $r$ and $r$(3\arcsec) are the 
SDSS $r$-band total magnitude
and the magnitude within the 3\arcsec\ spectroscopic aperture, respectively.

Throughout the paper,
we used $L$(H$\beta$) in units erg s$^{-1}$.

\setcounter{figure}{0}

\begin{figure}
\includegraphics[angle=-90,width=0.99\linewidth]{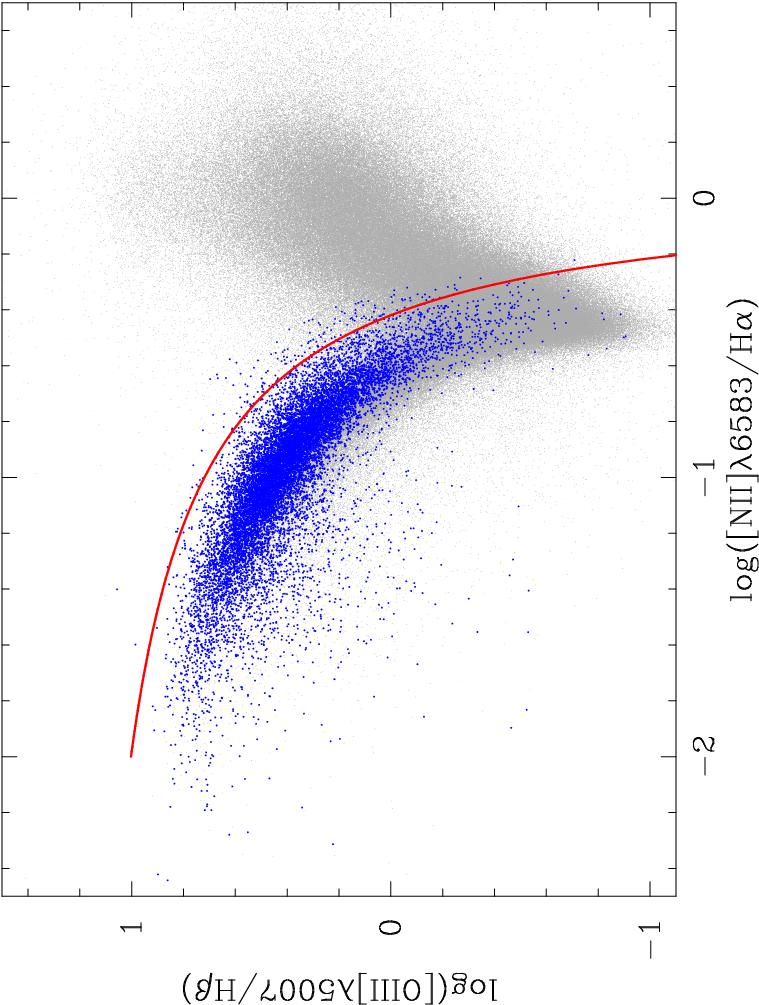}
\caption{The Baldwin-Phillips-Terlevich (BPT) diagram \citep{B81}. 
Selected SDSS star-forming galaxies are shown by blue filled circles.
Also plotted are all emission-line galaxies from SDSS DR7 (cloud
of grey dots).
The red solid line from \cite{K03} separates star-forming 
galaxies from active galactic nuclei.}
\label{fig1}
\end{figure}

\setcounter{figure}{1}

\begin{figure}
\includegraphics[angle=-90,width=0.99\linewidth]{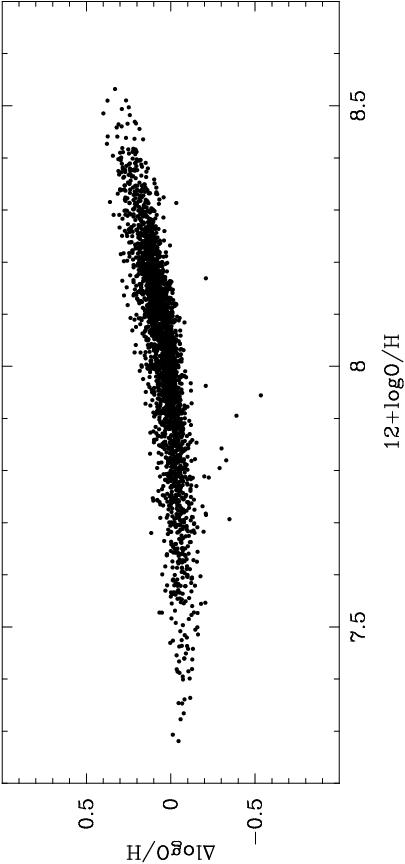}
\caption{Relation between the difference of oxygen abundances
$\Delta$log(O/H) derived by the direct and semi-empirical methods \citep{IT07}
and the oxygen abundance obtained by the direct method. Only 
galaxies where [O {\sc iii}] $\lambda$4363 is measured with accuracy better
than 50\% are shown.
}
\label{fig2}
\end{figure}

\setcounter{figure}{2}

\begin{figure}
\centering{
\hbox{
\includegraphics[angle=-90,width=0.47\linewidth]{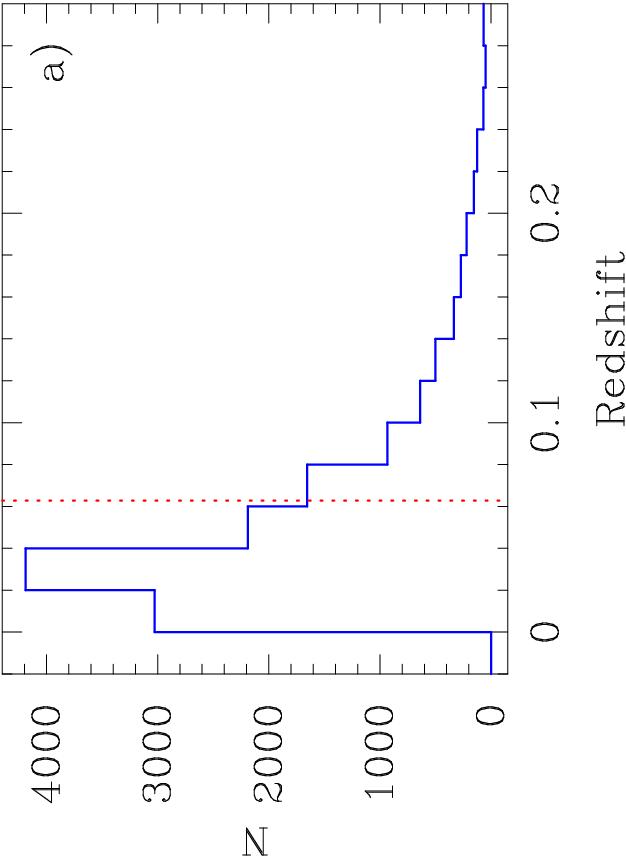}
\hspace{0.2cm}\includegraphics[angle=-90,width=0.47\linewidth]{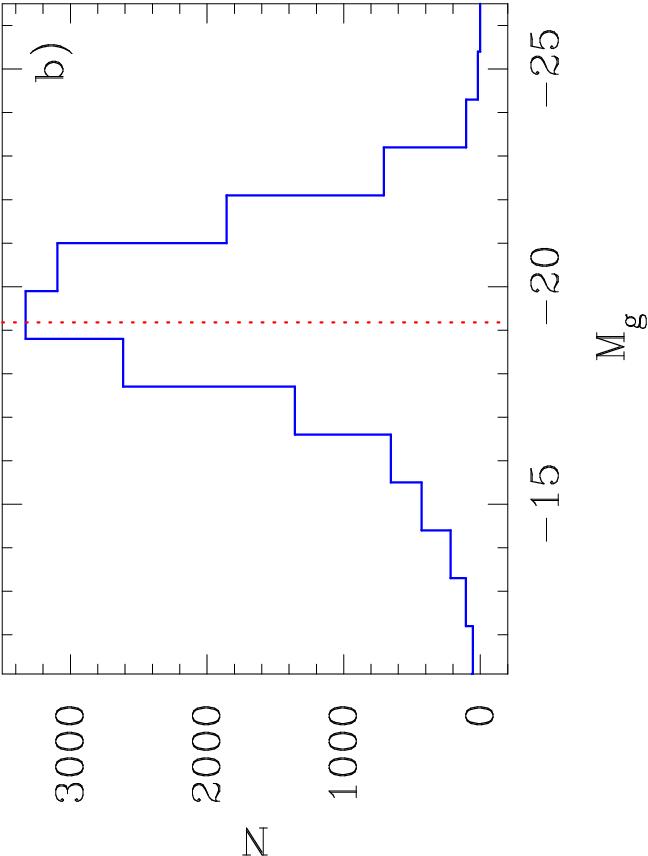}}
\vspace{0.2cm}
\hbox{
\includegraphics[angle=-90,width=0.47\linewidth]{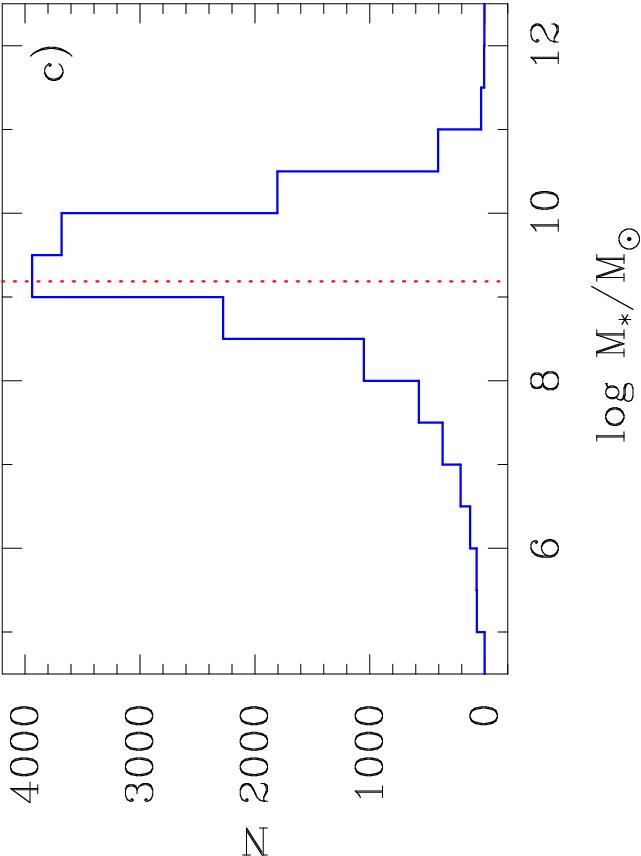}
\hspace{0.2cm}\includegraphics[angle=-90,width=0.47\linewidth]{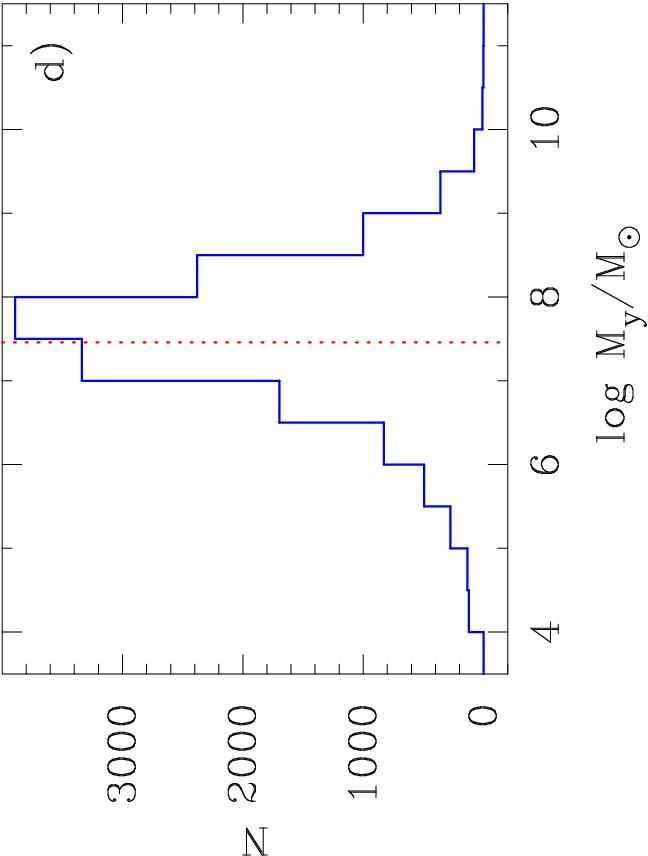}}
\vspace{0.2cm}
\hbox{
\includegraphics[angle=-90,width=0.47\linewidth]{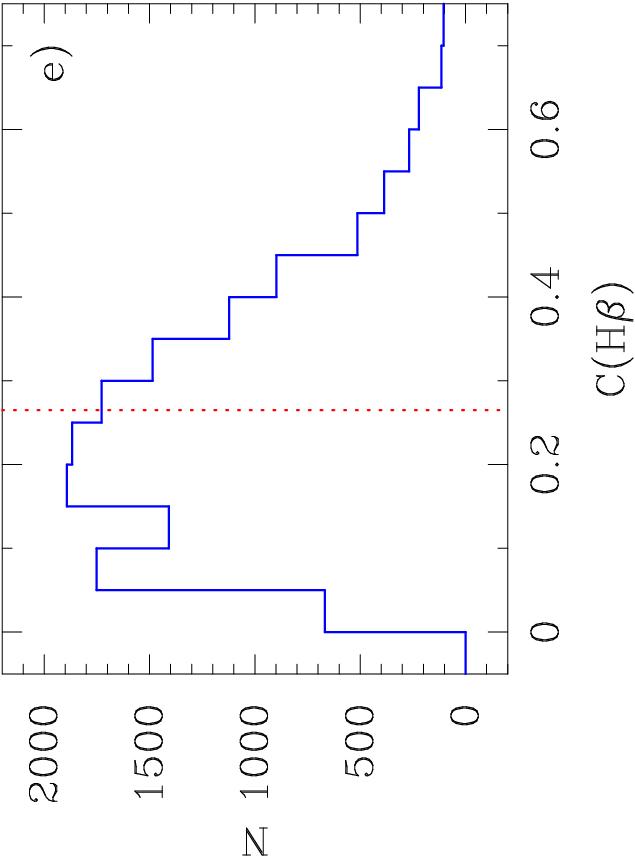}
\hspace{0.2cm}\includegraphics[angle=-90,width=0.47\linewidth]{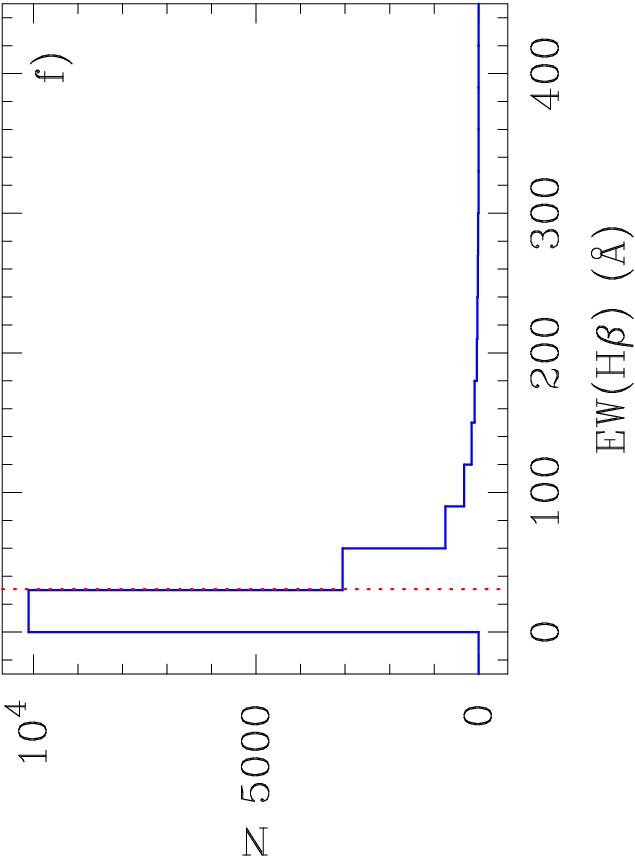}}
\vspace{0.2cm}
\hbox{
\includegraphics[angle=-90,width=0.47\linewidth]{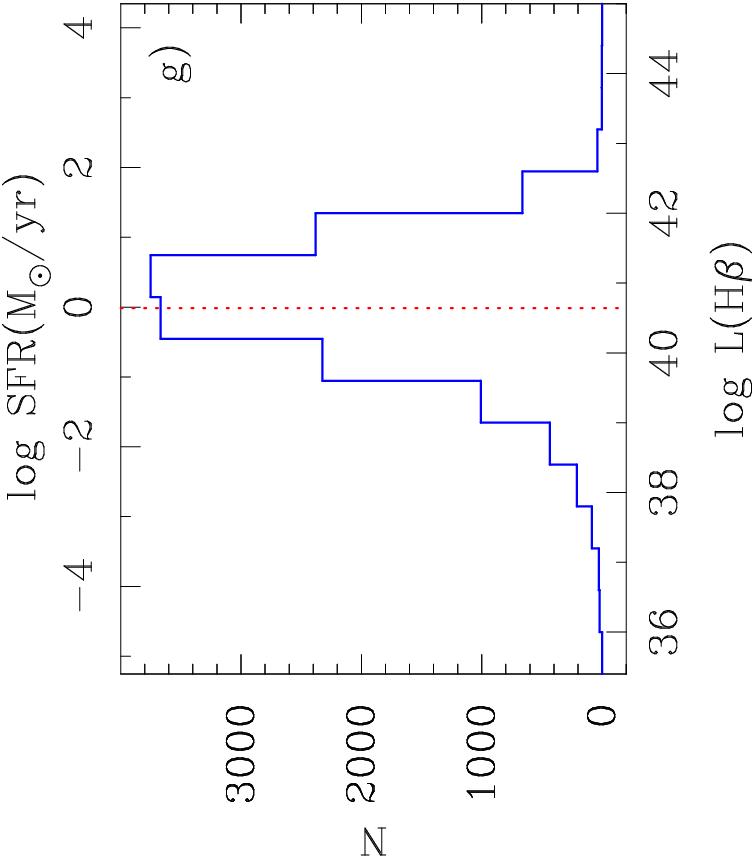}
\hspace{0.2cm}\includegraphics[angle=-90,width=0.47\linewidth]{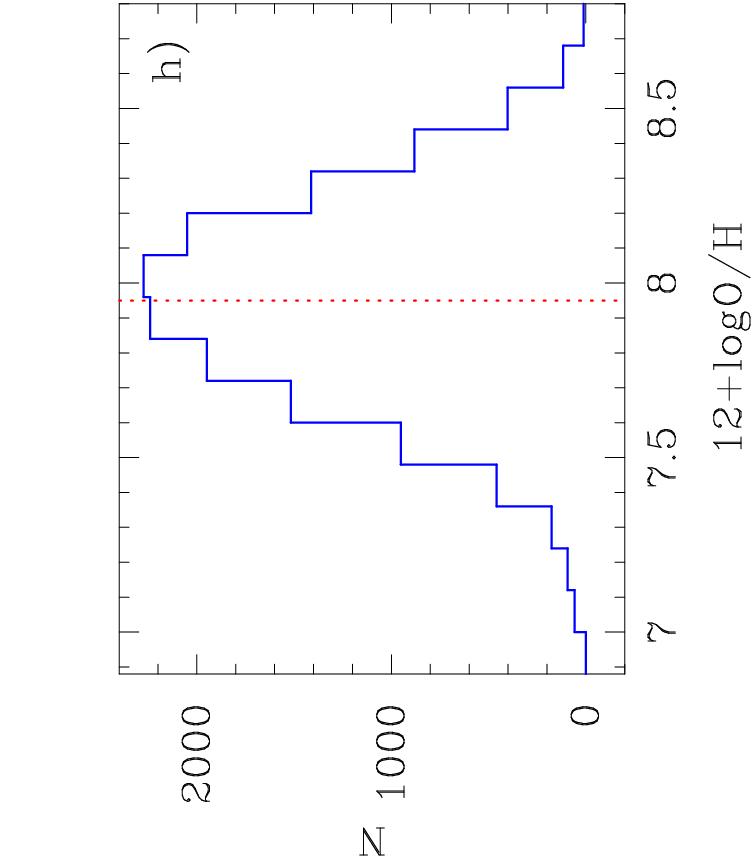}}
} 
\caption{Distribution of the SDSS emission-line galaxies (a) over  
redshift $z$; 
(b) over extinction-corrected absolute- SDSS $g$-magnitude $M_g$;
(c) over total stellar mass $M_*$; (d) over mass $M_{\rm y}$ of the young
stellar population; (e) over the extinction coefficient $C$(H$\beta$);
(f) over the equivalent width EW(H$\beta$) of the H$\beta$ emission line; 
(g) over the extinction- and aperture-corrected H$\beta$ luminosity 
$L$(H$\beta$) (lower axis) and the star-formation rate $SFR$ (upper axis); and
(h) over the oxygen abundance 12+logO/H (h). Dotted vertical lines in all 
panels indicate mean values of the distributions.}
\label{fig3}
\end{figure}

\subsection{Monochromatic luminosities from the photometric data \label{phot}}

Using photometric data from different surveys we transformed them to
observed fluxes in the units of erg s$^{-1}$cm$^{-2}$\AA$^{-1}$ in
all wavelength bands for their direct comparison and the comparison
to the SDSS spectroscopic data.

The fluxes $\widetilde{F}$(FUV) and $\widetilde{F}$(NUV) in the {\sl GALEX} 
catalogue were expressed in $\mu$Jy. We used the effective wavelengths of
1528\AA\ for FUV band and 2271\AA\ for NUV band and the equations
\begin{equation}
\log F({\rm FUV}) = -16.89 + \log \widetilde{F}({\rm FUV})
\end{equation}
and
\begin{equation}
\log F({\rm NUV}) = -17.24 + \log \widetilde{F}({\rm NUV}),
\end{equation}
respectively, to convert fluxes to the adopted units 
erg s$^{-1}$ cm$^{-2}$ \AA$^{-1}$.

The magnitudes $u$, $g$, $r$, $i$, and $z$ in the SDSS bands were used to 
convert them to observed fluxes:

\begin{equation}
\log F(u) = -8.056 - 0.4u,
\end{equation}
\begin{equation}
\log F(g) = -8.326 - 0.4g,
\end{equation}
\begin{equation}
\log F(r) = -8.555 - 0.4r,
\end{equation}
\begin{equation}
\log F(i) = -8.732 - 0.4i,
\end{equation}
\begin{equation}
\log F(z) = -8.882 - 0.4z,
\end{equation}
where the SDSS zeropoints were used from \citet{F96}.

Similarly, the magnitudes in the near-infrared $J$, $H$, and $K$ bands
were converted to the observed fluxes using the following equations:
\begin{equation}
\log F(J) = -9.505 - 0.4J,
\end{equation}
\begin{equation}
\log F(H) = -9.946 - 0.4H,
\end{equation}
\begin{equation}
\log F(K) = -10.368 - 0.4K,
\end{equation}
where zeropoints by \citet{C03} were used.

To convert the {\sl WISE} magnitudes to fluxes, we used zeropoints by 
\citet{W10} and the equations,
\begin{equation}
\log F(W1) = -11.087 - 0.4W1,
\end{equation}
\begin{equation}
\log F(W2) = -11.617 - 0.4W2,
\end{equation}
\begin{equation}
\log F(W3) = -13.186 - 0.4W3,
\end{equation}
\begin{equation}
\log F(W4) = -14.293 - 0.4W4.
\end{equation}

Finally, the {\sl IRAS} fluxes $\widetilde{F}$(60$\mu$m) at 60$\mu$m in Jy, 
and the NVSS fluxes $\widetilde{F}$(20cm)
at 20cm in mJy were converted to the fluxes in the adopted units by
using equations,
\begin{equation}
\log F(60\mu{\rm m}) = -16.079 + \log \widetilde{F}(60\mu{\rm m})
\end{equation}
and
\begin{equation}
\log F(20{\rm cm}) = -26.125 + \log \widetilde{F}(20{\rm cm}).
\end{equation}

The respective extinction-corrected luminosities 
$\lambda$$F_\lambda$$\equiv$$\nu$$F_\nu$ in erg s$^{-1}$ 
were derived from the monochromatic
fluxes using the galaxy redshift, the total extinction coefficient 
$C$(H$\beta$), which included both the Milky Way and intrinsic extinctions, 
and the reddening law by \citet{C89}. Since the extinction at long 
wavelengths is negligible, no extinction correction was applied to
{\sl WISE}, {\sl IRAS}, and NVSS bands. 



\setcounter{figure}{3}

\begin{figure*}
\begin{center}
\hbox{
\includegraphics[angle=-90,width=0.45\linewidth]{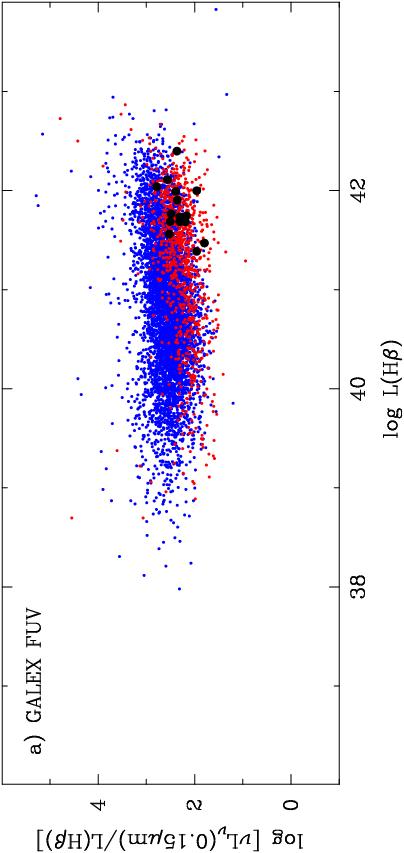}
\hspace{0.2cm}\includegraphics[angle=-90,width=0.45\linewidth]{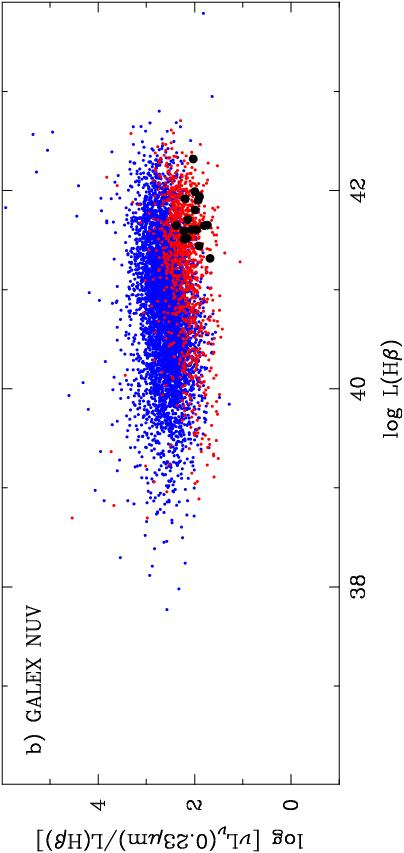}}
\hbox{
\includegraphics[angle=-90,width=0.45\linewidth]{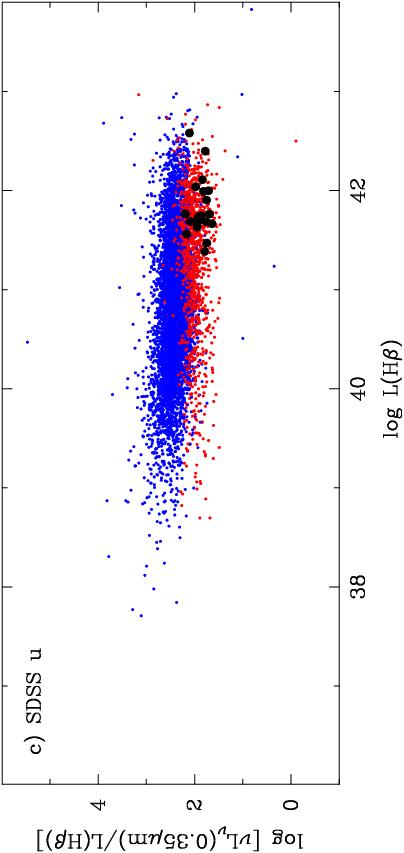}
\hspace{0.2cm}\includegraphics[angle=-90,width=0.45\linewidth]{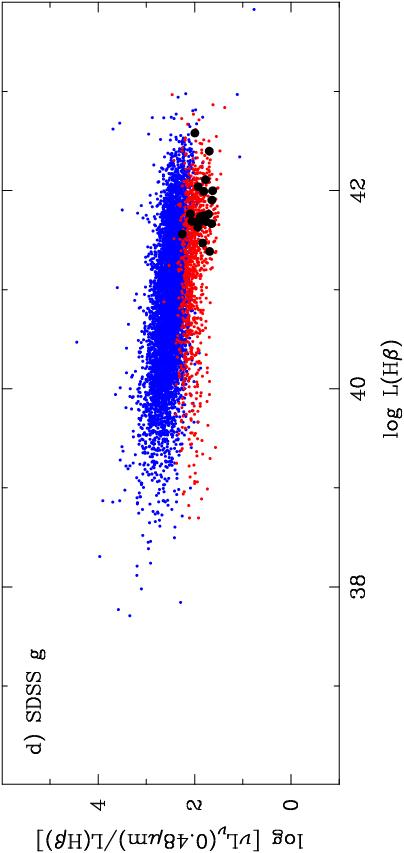}}
\hbox{
\includegraphics[angle=-90,width=0.45\linewidth]{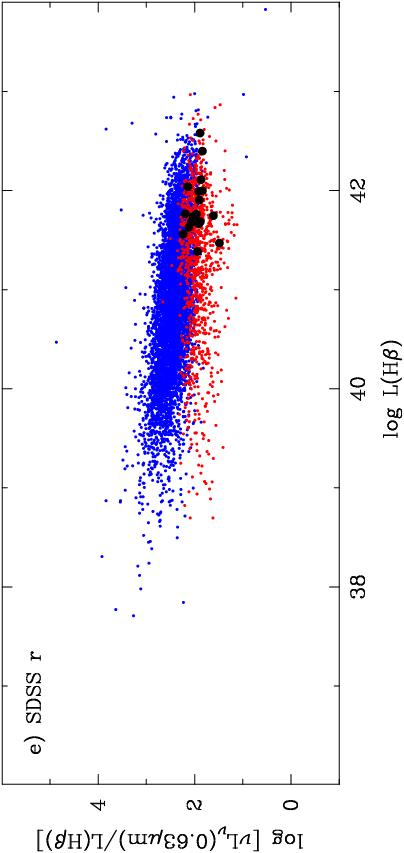}
\hspace{0.2cm}\includegraphics[angle=-90,width=0.45\linewidth]{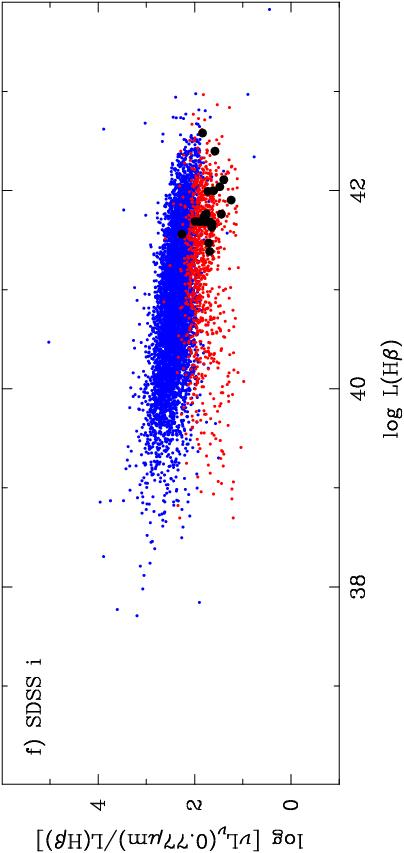}}
\hbox{
\includegraphics[angle=-90,width=0.45\linewidth]{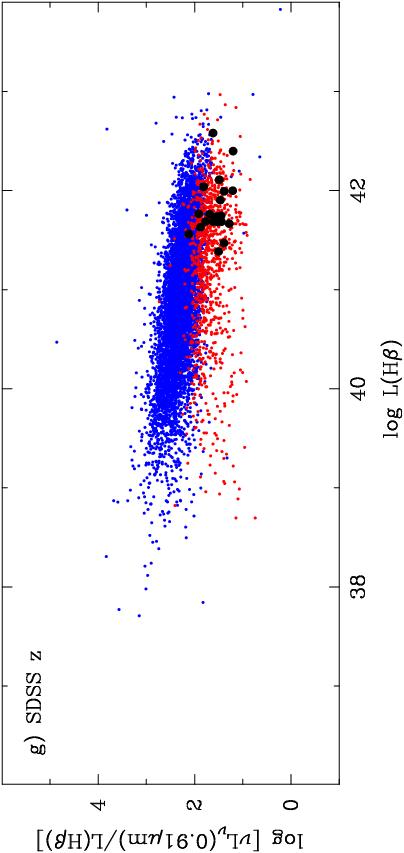}
\hspace{0.2cm}\includegraphics[angle=-90,width=0.45\linewidth]{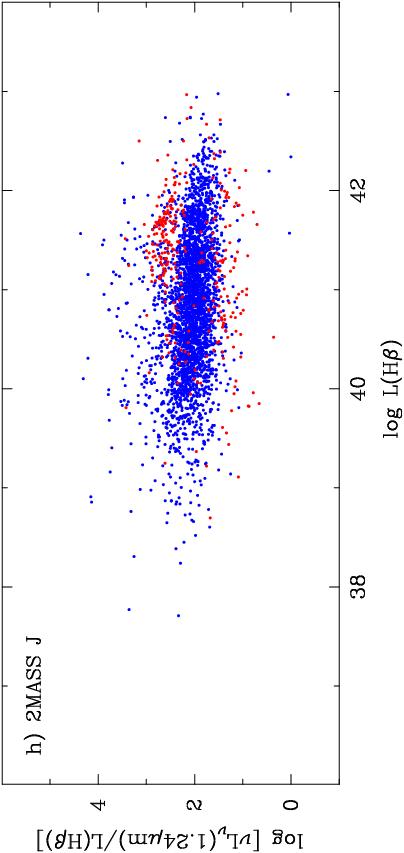}}
\end{center} 
\caption{Relations between the extinction- and aperture-corrected
H$\beta$ luminosity 
$L$(H$\beta$) (abscissa) and different extinction-corrected luminosity
ratios (ordinates). {\bf (a)} The {\sl GALEX}
FUV-to-H$\beta$ luminosity ratio $\nu$$L_\nu$(0.15$\mu$m)/$L$(H$\beta$)
is shown. Galaxies with 
EW(H$\beta$) $\geq$ 50\AA\ and EW(H$\beta$) $<$ 50\AA\ are shown by red and 
blue symbols, respectively. The galaxies with red {\sl WISE} 
$W1-W2$ $\geq$ 2 mag colours are shown by large filled circles.
{\bf (b)} Same as in (a) but the {\sl GALEX}
NUV-to-H$\beta$ luminosity ratio $\nu$$L_\nu$(0.23$\mu$m)/$L$(H$\beta$) 
is shown.
{\bf (c)} Same as in (a) but the SDSS
$u$-to-H$\beta$ luminosity ratio $\nu$$L_\nu$(0.35$\mu$m)/$L$(H$\beta$) 
is shown.
{\bf (d)} Same as in (a) but the SDSS
$g$-to-H$\beta$ luminosity ratio $\nu$$L_\nu$(0.48$\mu$m)/$L$(H$\beta$)
is shown.
{\bf (e)} Same as in (a) but the SDSS
$r$-to-H$\beta$ luminosity ratio $\nu$$L_\nu$(0.63$\mu$m)/$L$(H$\beta$)
is shown.
{\bf (f)} Same as in (a) but the SDSS
$i$-to-H$\beta$ luminosity ratio $\nu$$L_\nu$(0.77$\mu$m)/$L$(H$\beta$) 
is shown.
{\bf (g)} Same as in (a) but the SDSS
$z$-to-H$\beta$ luminosity ratio $\nu$$L_\nu$(0.91$\mu$m)/$L$(H$\beta$)
is shown.
{\bf (h)} Same as in (a) but the 2MASS
$J$-to-H$\beta$ luminosity ratio $\nu$$L_\nu$(1.24$\mu$m)/$L$(H$\beta$) 
is shown.
{\bf (i)} Same as in (a) but the 2MASS
$H$-to-H$\beta$ luminosity ratio $\nu$$L_\nu$(1.66$\mu$m)/$L$(H$\beta$) 
is shown.
{\bf (j)} Same as in (a) but the 2MASS
$K$-to-H$\beta$ luminosity ratio $\nu$$L_\nu$(2.16$\mu$m)/$L$(H$\beta$)
is shown.
{\bf (k)} Same as in (a) but the {\sl WISE}
$W1$-to-H$\beta$ luminosity ratio $\nu$$L_\nu$(3.4$\mu$m)/$L$(H$\beta$) 
is shown.
{\bf (l)} Same as in (a) but the {\sl WISE}
$W2$-to-H$\beta$ luminosity ratio $\nu$$L_\nu$(4.6$\mu$m)/$L$(H$\beta$) 
is shown.
{\bf (m)} Same as in (a) but the {\sl WISE}
$W3$-to-H$\beta$ luminosity ratio $\nu$$L_\nu$(12$\mu$m)/$L$(H$\beta$) 
is shown.
{\bf (n)} Same as in (a) but the {\sl WISE}
$W4$-to-H$\beta$ luminosity ratio $\nu$$L_\nu$(22$\mu$m)/$L$(H$\beta$) 
is shown.
{\bf (o)} Same as in (a) but the {\sl IRAS}
$60\mu$m-to-H$\beta$ luminosity ratio $\nu$$L_\nu$(60$\mu$m)/$L$(H$\beta$) 
is shown.
{\bf (p)} Same as in (a) but the NVSS
20cm-to-H$\beta$ luminosity ratio $\nu$$L_\nu$(20cm)/$L$(H$\beta$) 
is shown.
}
\label{fig4}
\end{figure*}

\setcounter{figure}{3}

\begin{figure*}
\begin{center}
\hbox{
\includegraphics[angle=-90,width=0.45\linewidth]{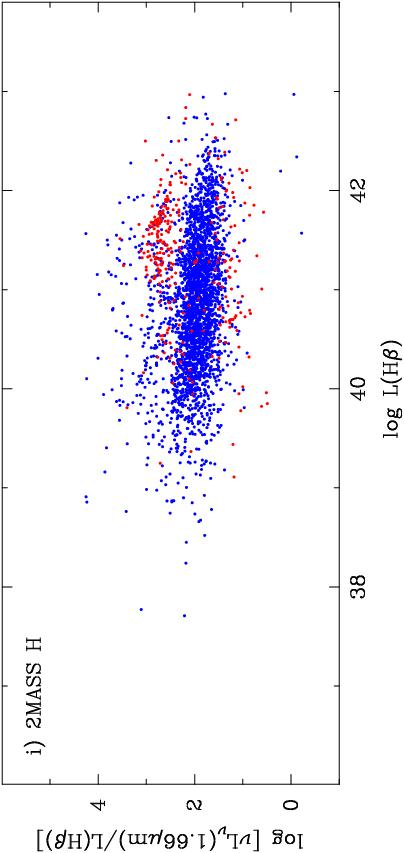}
\hspace{0.2cm}\includegraphics[angle=-90,width=0.45\linewidth]{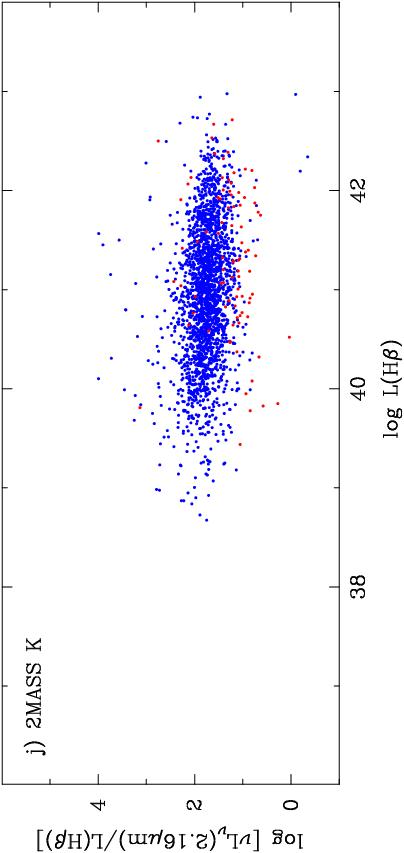}}
\hbox{
\includegraphics[angle=-90,width=0.45\linewidth]{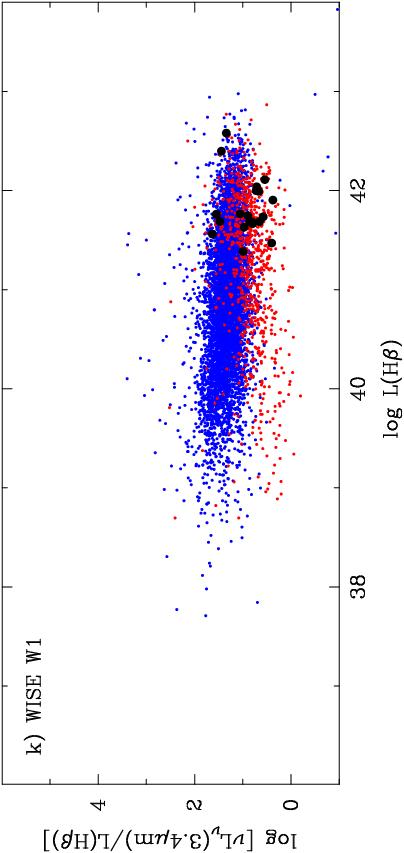}
\hspace{0.2cm}\includegraphics[angle=-90,width=0.45\linewidth]{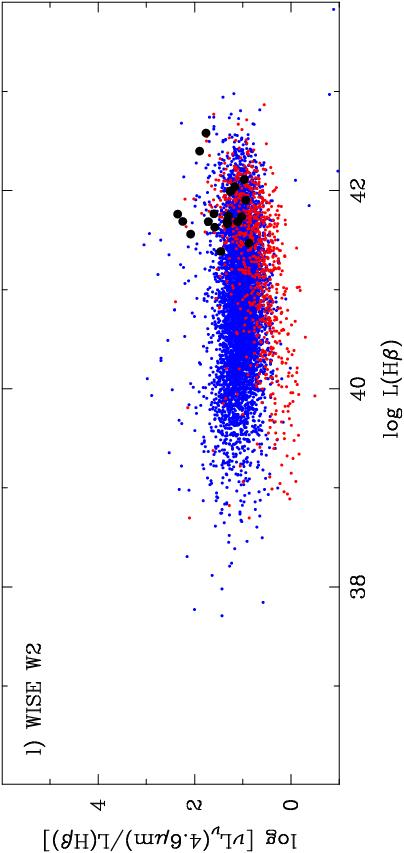}}
\hbox{
\includegraphics[angle=-90,width=0.45\linewidth]{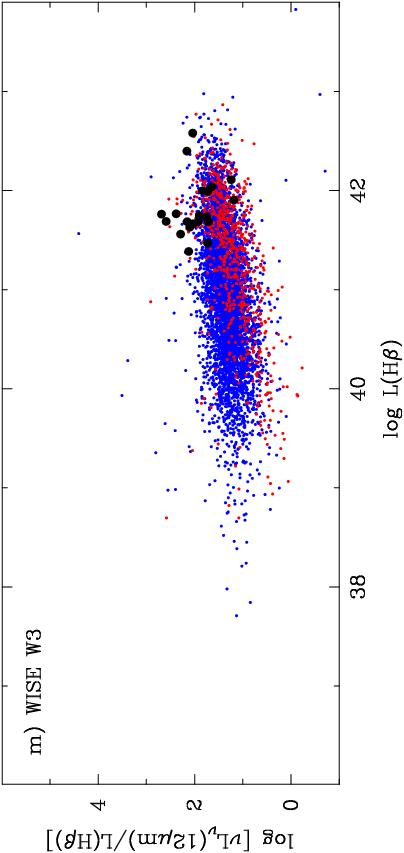}
\hspace{0.2cm}\includegraphics[angle=-90,width=0.45\linewidth]{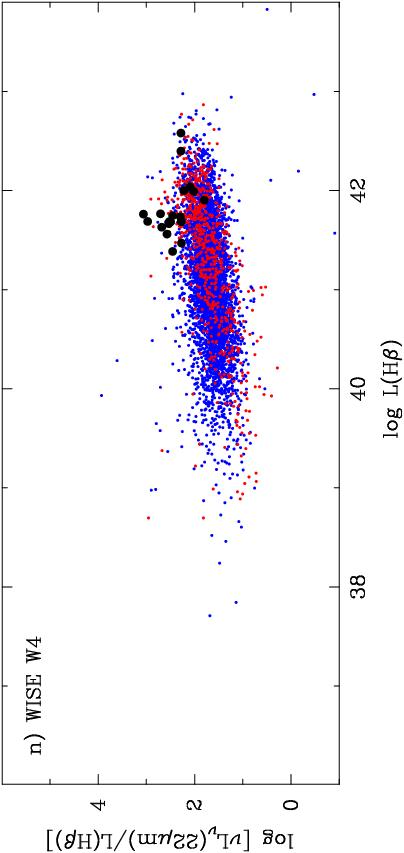}}
\hbox{
\includegraphics[angle=-90,width=0.45\linewidth]{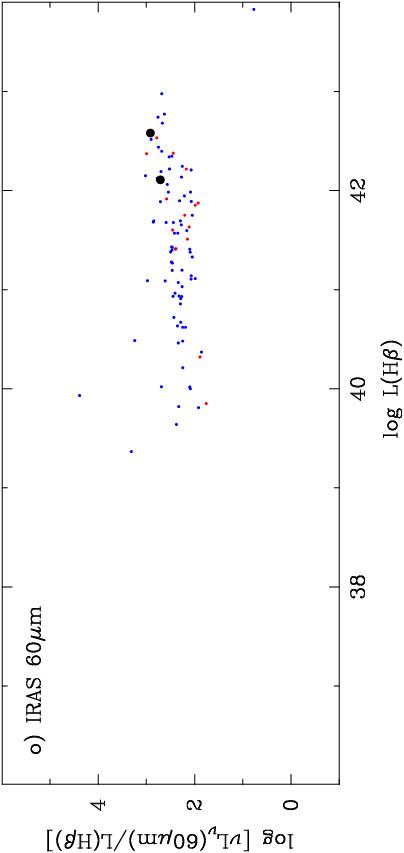}
\hspace{0.2cm}\includegraphics[angle=-90,width=0.45\linewidth]{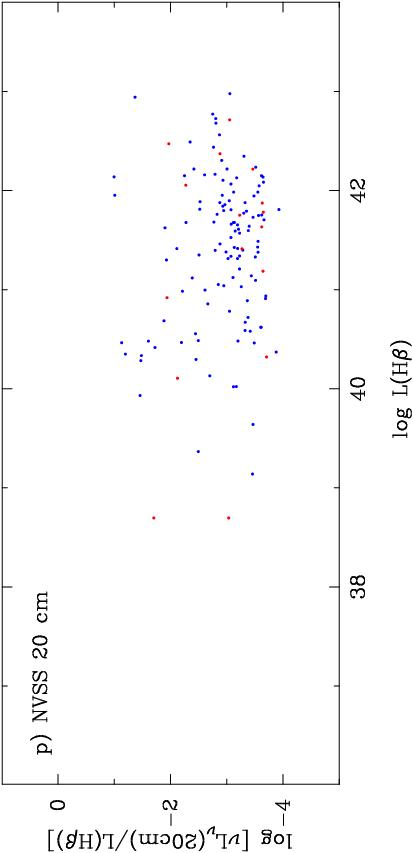}}
\end{center} 
\caption{{\sl ---Continued.}}
\end{figure*}

\subsection{Fitting the spectral energy distribution (SED) from the SDSS
spectra and determining the galaxy's stellar masses}

The stellar mass of a galaxy is one of its most important global
characteristics. It can be derived by modelling the galaxy spectral energy
distribution (SED) which depends on the adopted
star formation history. In the case of strongly star-forming galaxies,
the situation
is however complicated by the presence of strong ionised gas emission, 
which includes the strong emission lines and gaseous continuum. 
Ionised gas emission must be subtracted before determining the stellar masses.
In particular, neglecting the correction
for gaseous continuum emission in the visible range used for the stellar mass
determination would result in an overestimate of
the galaxy stellar mass by $\sim$ 0.4 dex \citep{I11a} for two reasons: 
1) gaseous continuum emission increases the luminosity of the galaxy; 
and 2) the SED of gaseous continuum emission is flatter than that of young 
stars, making the SED redder than expected
for pure stellar emission. Consequently, the fraction of light
from the red old stellar population artificially increases.
To derive the correct stellar mass of the galaxy, we therefore used
the method described below that considers the contribution of 
gaseous continuum emission, which is higher in galaxies with high equivalent
widths EW(H$\beta$).

The method is based on fitting a series of model SEDs to the observed one 
and finding the best
fit. This was described in \citet{G06,G07} and \citet{I11a} and consists of the
following. The fit was performed for each SDSS spectrum over the whole
observed spectral range of $\lambda$$\lambda$3900--9200\AA, which 
included the Balmer jump region ($\lambda$3646\AA) for more distant galaxies 
with $z$ $>$ 0.1 and the Paschen jump region ($\lambda$8207\AA) for
galaxies with $z$ $<$ 0.12. 
As each SED is the sum of both stellar and ionised gas emission,
its shape depends on the relative contribution of these two
components. In galaxies with high EW(H$\beta$) $>$ 100\AA, the 
contribution of the ionised gas emission can be large. However, the 
equivalent widths of the hydrogen emission lines never reach the theoretical 
values for pure gaseous emission, which is $\sim$ 900 -- 1000\AA\ for the 
H$\beta$ emission line and depends slightly on the electron temperature 
of the ionised gas. 

The objects in our SDSS sample span a lower range of 
EW(H$\beta$)s between $\sim$10 and $\sim$500\AA\ due to the
contribution of stellar continuum emission. 
The contribution of gaseous 
emission relative to stellar emission can be parameterized by the equivalent
width, EW(H$\beta$), of the H$\beta$ emission line. Given a temperature
$T_{\rm e}$(H$^+$), the ratio of the gaseous emission to the total 
gaseous and stellar emission is equal 
to the ratio of the observed EW(H$\beta$) to the equivalent width of H$\beta$
expected for pure gaseous emission. The shape of the spectrum
depends also on reddening. 

The extinction coefficient for the ionised
gas $C$(H$\beta$) can be obtained from the observed hydrogen
Balmer decrement. However, there is no direct way to derive the
extinction coefficient for the stellar emission, which can be 
different from the ionised gas extinction coefficient in principle
\citep{C00}. For clarity,
we adopted equal extinction coefficients for ionised gas and stars.
Finally, the SED depends on the star-formation history of the galaxy.

We carried out a series of Monte Carlo simulations to
reproduce the SED of each galaxy in our sample. To calculate
the contribution of the stellar emission to the SEDs, we
adopted the grid of the Padua stellar evolution models by \citet{Gi00}
with heavy element mass fractions $Z$ = 0.001, 0.004, and 0.008. To reproduce 
the SED of the stellar component with any star-formation history, we
used the package PEGASE.2 \citep{FR97} to calculate a grid
of instantaneous burst SEDs for a stellar mass of 1 $M_\odot$ in a wide range 
of ages from 0.5 Myr to 15 Gyr. We adopted a stellar initial mass 
function with a Salpeter slope, an upper mass limit of 100 $M_\odot$, and a 
lower mass limit of 0.1 $M_\odot$. Then the SED with any star-formation 
history can be obtained by integrating the instantaneous burst SEDs over
time with a specified time-varying star-formation rate. 

We approximated the star-formation history in each galaxy by a recent 
short burst with age $t_{\rm y}$ $<$ 10 Myr, which accounts 
for the young stellar population, and a prior continuous star formation 
responsible for the older stars with age ranges
starting at $t_2$ $\equiv$ $t_{\rm o}$, where $t_{\rm o}$ is the age of 
the oldest stars in the galaxy, and ending at $t_1$, where $t_2$ $>$ $t_1$ and 
varies between 10 Myr and 15 Gyr. Note that zero age is now.
The contribution of each stellar population to the SED was parameterized by the
ratio of the masses of the old to young stellar populations, 
$b$ = $M_{\rm y}$/$M_{\rm o}$, which we varied between 0.01 and 1000. 

The total modelled monochromatic
(gaseous and stellar) continuum flux near the H$\beta$ emission line for a 
mass of 1 $M_\odot$ was scaled to fit the monochromatic 
extinction- and aperture-corrected luminosity of 
the galaxy at the same wavelength. The scaling factor is equal to the total
stellar mass $M_*$ in solar units. In our fitting model 
$M_*$= $M_{\rm y}$ + $M_{\rm o}$, $M_{\rm y}$ and $M_{\rm o}$ 
were respectively the masses
of the young and old stellar populations in solar units. 
These masses were derived using $M_*$ and $b$.

The SED of the gaseous continuum was
taken from \citet{A84} for $\lambda$ $\leq$ 1$\mu$m and from \citet{F80}
for $\lambda$ $>$ 1$\mu$m. It included hydrogen and helium free-bound,
free-free, and two-photon emission. In our models, this was
always calculated with the electron temperature $T_{\rm e}$(H$^+$) of the
H$^+$ zone and with the chemical composition derived from the
H {\sc ii} region spectrum. The observed emission lines 
that were corrected for
reddening and scaled using the flux of the H$\beta$ emission
line were added to the calculated gaseous continuum. 
The flux ratio of the gaseous continuum to the total continuum depends
on the adopted electron temperature $T_{\rm e}$(H$^+$) in the H$^+$ zone,
since EW(H$\beta$) for pure gaseous emission decreases with increasing
$T_{\rm e}$(H$^+$). Given that $T_{\rm e}$(H$^+$) is not necessarily equal to
$T_{\rm e}$(O {\sc iii}), we varied it in the range of
(0.7-1.3)$\times$$T_{\rm e}$(O {\sc iii}).
Strong emission lines in SDSS emission-line galaxies were measured
with good accuracy, so the equivalent width of the H$\beta$ emission
line and the extinction coefficient for the ionised gas were accurate
to 5\% and 20\%, respectively. Thus, we varied EW(H$\beta$)
between 0.95 and 1.05 times its nominal value. As for the extinction
coefficient $C$(H$\beta$)$_{\rm SED}$, we varied
it in the range of (0.8-1.2)$\times$$C$(H$\beta$), where $C$(H$\beta$) is the 
extinction coefficient derived from the observed hydrogen Balmer 
decrement. For each galaxy, we calculated 
10$^4$ Monte Carlo models by varying $t_{\rm y}$, $t_1$,
$t_{\rm o}$, $b$, and $T_{\rm e}$(H$^+$) randomly in a large range and EW(H$\beta$) and
$C$(H$\beta$)$_{\rm SED}$ in a relatively smaller range because
the latter quantities are more directly constrained by observations. The best
modelled SED was found from $\chi ^2$ minimization of the deviation between the
modelled and the observed continuum in five wavelength ranges, which are free 
of the emission lines and residuals of the night sky lines. 
Depending on the galaxy redshift, the following ranges were used when
possible: shortward of
the Balmer jump, between the He {\sc i}+H8 $\lambda$3889 and
[Ne {\sc iii}]+H7 $\lambda$3969 emission lines,
between the H$\delta$ and H$\gamma$ emission lines, 
between the He {\sc i} $\lambda$4471 and [Fe {\sc iii}]$\lambda$4658 emission
lines, between the H$\beta$ and He {\sc i} $\lambda$5876 emission lines,
between the He {\sc i} $\lambda$5876 and [O {\sc i}] $\lambda$6300 emission
lines, and longward of the [Ar {\sc iii}] $\lambda$7135 emission line but
shortward of the Paschen jump.

\section{Results \label{res}}

\subsection{Global characteristics of the SDSS sample}

In Fig. \ref{fig3}, we showed the distributions of some global parameters
of all galaxies from the SDSS sample. The galaxies are distributed in a wide
range of redshifts from 0 to 0.65 with an average value of 0.063 
(Fig. \ref{fig3}a). Their average SDSS extinction-corrected absolute $g$ 
magnitude
of $-$19.1 mag is brighter than the brightest magnitude of $-$18 mag 
that is often
used for the galaxy definition as a dwarf galaxy (Fig. \ref{fig3}b).
The brightest galaxies in the sample have $M_g$ as bright as $\sim$ $-$23 mag,
which is comparable to the brightness of the high-redshift LBGs 
and Lyman-$\alpha$ emitting galaxies.

However, the aperture-corrected 
total stellar mass of these galaxies is low with an average
value $<\log M_*/M_\odot>$ of 9.2, which is typical of dwarf galaxies
(Fig. \ref{fig3}c). On the other hand, the mass of the young stellar 
populations with an age of a few
Myr is high. The average $<\log M_{\rm y}/M_\odot>$ is 7.5 implying 
that $\sim$ 2\% of the stellar mass resides in the youngest stellar complexes 
(Fig. \ref{fig3}d).

The total extinction in the direction on the SDSS sample galaxies varies in
a wide range, but it is generally low with the average extinction coefficient
$C$(H$\beta$) of 0.265, corresponding to an extinction $A_V$ of $\sim$ 0.6 mag
in the $V$ band (Fig. \ref{fig3}e). On average, less than 20\% of this 
extinction is caused by the Milky Way and the remaining extinction is 
the galaxy's internal extinction.

The average rest-frame
equivalent width EW(H$\beta$) of the H$\beta$ emission line 
of $\sim$31\AA\ is modest, corresponding to the late stage of a starburst or
implying an important contribution of the non-ionising stellar population
(Fig. \ref{fig3}f). However, there are several thousand galaxies with high 
EW(H$\beta$) $\ga$ 50\AA\ and very strong emission lines. The average
extinction- and aperture-corrected luminosity $L$(H$\beta$) 
of the H$\beta$ emission line is high 
($<\log L({\rm H}\beta>$ $\sim$ 40.6), corresponding to 
the ionising radiation of $\sim$ 10$^4$ O7V stars \citep{L90}, while the entire
range of $L$(H$\beta$) corresponds to the number of O7V stars 
that can range from a few 
to up to 10$^5$ (Fig. \ref{fig3}g). Star-formation rates $SFR$s are
in the range 10$^{-4}$ -- 10$^2$ $M_\odot$ yr$^{-1}$ with an average
value of $\sim$ 1 $M_\odot$ yr$^{-1}$ (upper axis in Fig. \ref{fig3}g).

Finally, the oxygen abundance 12+logO/H with an averaged value of 7.95 
for the entire sample is 
low in the SDSS star-forming galaxies (Fig. \ref{fig3}h). 
Considering that only $\sim$2800 galaxies with the detected
[O {\sc iii}] $\lambda$4363 emission line had a flux was measured with 
an accuracy better than 50\% and
used a direct method, we however obtained slightly higher average oxygen 
abundance 12+logO/H $\sim$ 8.04.

Summarising the comparison of global characteristics, we concluded that the 
SDSS star-forming galaxies are dwarf
galaxies experiencing strong and very strong bursts of star formation.

\subsection{Relations between luminosities of star-forming galaxies}

The relations between the star-forming galaxy luminosities in different 
passbands give an important information on the origin of their radiation. 
In particular, these relations can estimate the contribution of the 
young stellar population at different wavelengths. Furthermore, these
relations can be used to adjust star-formation rates derived
from the luminosities in different bands. As a reference luminosity, 
we used the extinction- and aperture-corrected
luminosity $L$(H$\beta$) of the H$\beta$ emission line, which
characterises the youngest most massive stellar population.
Since we selected only objects with strong emission-lines, the H$\beta$
luminosity was derived for all spectra from the SDSS sample (Table \ref{tab1}).

The high detectability of the SDSS sample objects in 
{\sl GALEX} and {\sl WISE} all-sky surveys allowed us to construct reliable
relations between luminosities based on large samples of star-forming
galaxies. 

To reduce the 
uncertainties due to the aperture corrections of $L$(H$\beta$) 
we considered relations only for compact
objects with typical diameters $\la$ 6\arcsec. Relations between 
the $\nu$$L_\nu$/$L$(H$\beta$) ratios and the H$\beta$ emission-line 
luminosity $L$(H$\beta$) are shown
in Fig. \ref{fig4}. To study the differences between relations
in younger and older bursts we divided the sample of compact star-forming 
galaxies into two parts: those with the rest-frame
H$\beta$ equivalent width 
EW(H$\beta$) $\geq$ 50\AA\ (red dots) and those with EW(H$\beta$) $<$ 50\AA\
(blue dots). 

We noted that the slope of 0 in Fig. \ref{fig4} corresponds to linear 
dependences between luminosities, which implies that radiation in these two
particular wavelength ranges is produced mainly by the same 
young stellar population.

The common feature of the relations in Fig. \ref{fig4} is that the galaxies
with high EW(H$\beta$) $\geq$ 50\AA\ (red dots) are located below the 
galaxies with low  EW(H$\beta$) $<$ 50\AA\ (blue dots) because of higher
H$\beta$ luminosity. Furthermore, the slopes of relations for the galaxies
with high EW(H$\beta$) are shallower, indicating a larger contributions of the
youngest stellar population to the galaxy luminosity. Finally, negative
slopes of relations in some passbands (mainly in the optical range)
again indicate that the contribution of the youngest stellar population
to the galaxy emission is enhanced with increasing H$\beta$ luminosity,
or equivalently, with increasing ongoing star-formation rate.

It is seen in Figs. \ref{fig4}a and \ref{fig4}b that 
the distributions of the $\nu$$L_\nu$(0.15$\mu$m)/$L$(H$\beta$) and 
$\nu$$L_\nu$(0.23$\mu$m)/$L$(H$\beta$) ratios are flat. This indicates
that the emission in the H$\beta$ line and {\sl GALEX} passbands is produced
by the same youngest stellar population. This is contrary to the point
of view that FUV and NUV emission are indicative of an older stellar population
as compared to that responsible for the ionised gas emission. Therefore,
the {\sl GALEX} data in strong-line emission-line galaxies can be used
for the estimation of the star-formation rate of the youngest star formation
episode, in addition to the H$\beta$ emission. However, the dispersion 
of points in Figs. \ref{fig4}a and \ref{fig4}b is high and the $SFR$s estimated
from the H$\beta$ and {\sl GALEX} luminosities may be different by a factor of
up to $\sim$ 10.

The luminosities of the SDSS galaxies with low EW(H$\beta$) $<$ 50\AA\ 
in the visible (SDSS bands) range, near-infrared (2MASS bands) ranges and 
at 3.4$\mu$m ({\sl WISE} $W1$ band) are poor 
indicators of the most recent star
formation because of the presence of negative slopes in relations as shown
in Fig. \ref{fig4}c -- \ref{fig4}l. Contrary to that, the relations for 
galaxies with high  EW(H$\beta$) $\geq$ 50\AA\ are much flatter, indicating
that most of the emission in the 0.35$\mu$m -- 4.6$\mu$m range is also produced
by the youngest stellar population, 
which is similar to that in the {\sl GALEX} bands.

The luminosities at longer wavelengths, 4.6$\mu$m -- 20cm 
(Figs. \ref{fig4}m -- \ref{fig4}p), are again good tracers of the youngest
stellar population independently on the EW(H$\beta$) value. We noted that
the slope of the relations in the 12$\mu$m and 22$\mu$m {\sl WISE} bands 
becomes positive and it is higher for galaxies with high EW(H$\beta$) (red
dots in Figs. \ref{fig4}m -- \ref{fig4}n). Because this emission is 
produced by dust, we interpret the positive slopes as the consequence of
increasingly hotter dust in galaxies with higher luminosities.
We also noted that the dispersion of points in Fig. \ref{fig4} is the
lowest at 22$\mu$m and 60$\mu$m, allowing for an accurate $SFR$ estimation,
in addition to that derived from the H$\beta$ luminosity.

Summarising, we concluded that the luminosities of compact galaxies with the
youngest bursts and, respectively, the highest EW(H$\beta$) are good indicators
of the most recent star-formation episode. On the other hand, 
only H$\beta$, UV, mid-infrared and radio
emission are tracers of the youngest stellar population in the
galaxies with low EW(H$\beta$).

\subsection{Luminosity-metallicity and mass-metallicity relations}

The luminosity-metallicity relation for emission-line galaxies from our
SDSS sample is shown in Fig. \ref{fig5}. We selected only 567 galaxies 
with [O {\sc iii}] $\lambda$4363 emission line fluxes, which were measured
with an accuracy better than 25\%. Oxygen abundances 
in these galaxies were derived by the direct method. 
Furthermore, we selected only compact galaxies with angular diameters 
$\la$6\arcsec.
Therefore, the oxygen abundance in these
galaxies is a characteristic of the entire galaxy, not of its individual 
H {\sc ii} regions. 
The maximum likelihood regression,
\begin{equation}
12+\log \frac{\rm O}{\rm H} = -0.032 M_g + 7.41, \label{Mg_o}
\end{equation}
is shown by a solid line.
Galaxies with high $SFR$(H$\alpha$) $\geq$ 10 $M_\odot$ yr$^{-1}$ are
encircled. They are among the most luminous objects in the sample. There is
no evident offset from the linear regression of the encircled galaxies.  

The relation in Eq. \ref{Mg_o} is flatter than that obtained by \citet{G09}
and \citet{I11a}. However, we noted that the SDSS sample is lacking
very metal-poor galaxies. On the other hand, authors of mentioned papers
considered not only SDSS galaxies but also galaxies from other samples
consisting of the most metal-deficient galaxies or of the galaxies where the
oxygen abundances were derived by the strong-line method.
Our sample, which is a variation to this, is more uniform and includes 
only objects where the oxygen abundance
was derived using a direct $T_{\rm e}$ method.

Our luminosity-metallicity relation is also much flatter compared to 
that obtained by
\citet{T04} for a sample of $\sim$ 53000 star-forming galaxies selected from
the SDSS. We noted that selection criteria are different for these two SDSS 
samples. \citet{T04} selected mainly galaxies with low-excitation
H {\sc ii} regions, where the [O {\sc iii}] $\lambda$4363 emission is not
detected and the strong lines were used for the oxygen abundance 
determination. 
Most of the galaxies in the \citet{T04} sample have 12+logO/H $>$8.6. 

In Fig. \ref{fig6}a, we showed the mass-metallicity relation for our sample
galaxies. We selected the same galaxies as in the luminosity-metallicity
relation (Fig. \ref{fig5}).
It is seen that galaxies with lower mass have systematically lower oxygen
abundances. These lower-mass galaxies are primarily 
those with high EW(H$\beta$).
The data can be fitted 
by a linear relation, as shown in Fig. \ref{fig6}a by the solid line,
\begin{equation}
12 + \log \frac{\rm O}{\rm H} = 0.102 \log \frac{M_*}{M_\odot}  + 7.13.
\label{mtot_o}
\end{equation}
The relation in Eq. \ref{mtot_o} is steeper and is better defined compared to 
the relation by \citet{I11a} for LCGs, which are the most massive galaxies of 
the SDSS sample of compact star-forming galaxies. This
is because the galaxies from our sample have a larger range of stellar masses
as compared to LCGs and include faint dwarf emission-line galaxies.

On the other hand, the mass-metallicity relation in Fig \ref{fig6}a 
(solid line)
is much flatter and is shifted to lower metallicities 
compared to that obtained by \citet{T04} (dashed line).
As for the luminosity-metallicity relation, this difference
is likely caused by a different selection criteria and different methods 
used for the oxygen abundance determination.
Furthermore, no constraints on the morphology were given
by \citet{T04}, while we used only compact objects and considered
aperture corrections. Therefore, aperture
effects in the mass determination played a minor role for our sample of
SDSS galaxies.

It was argued in some recent papers \citep[e.g. ][]{M10,L10,H12,AM13} 
that star-forming galaxies with higher $SFR$ are systematically more
metal-poor on the mass-metallicity diagrams. We checked whether this
tendency is present for galaxies from our sample. For this, we encircled 
galaxies with high $SFR$s $\geq$ 10 $M_\odot$ yr$^{-1}$. 
We concluded from Fig. \ref{fig6}a that the galaxies with high $SFR$s 
and the best-derived oxygen abundances are not 
systematically more metal-poor, which is contrary to conclusions in papers
mentioned above.

Figure \ref{fig6}b shows the dependence of the oxygen abundance 12+logO/H on 
the mass of the young stellar population formed in the most recent burst of
star formation; the ages of the papulation are between 0 and 10 Myr. 
To our knowledge, this
relation has never been discussed before. As for the mass-metallicity relation,
the mass of young stellar population is gradually decreasing with decreasing
metallicity. However, the slope of the relation in Fig. \ref{fig6}b is slightly
flatter than that for the mass-metallicity relation, which addresses the entire
stellar mass in Fig. \ref{fig6}a.

\setcounter{figure}{4}

\begin{figure}
\begin{center}
\includegraphics[angle=-90,width=1.0\linewidth]{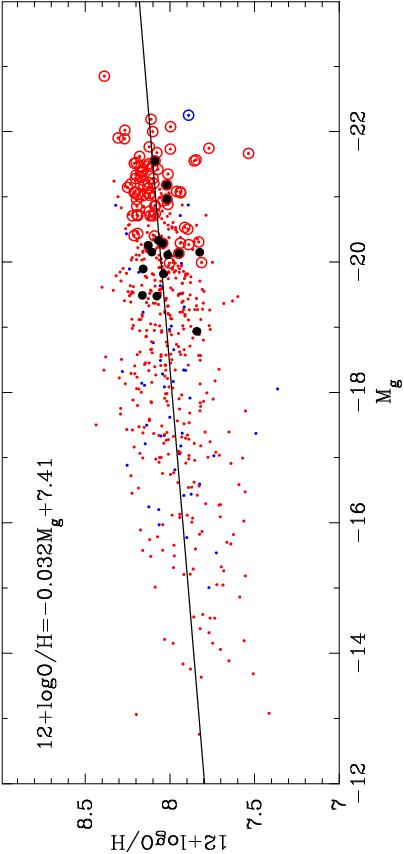}
\end{center} 
\caption{Luminosity-metallicity relation. Symbols are the same as in Fig. \ref{fig4}.
Linear likelihood regression is shown by solid line.
Only 567 galaxies where
the errors in [O {\sc iii}] 4363 emission-line flux do not exceed 25\%
are also shown.  
The galaxies with red {\sl WISE} colours $W1-W2$ $>$ 2 mag
are indicated by large black filled circles and the galaxies with 
$SFR$(H$\alpha$) $\geq$ 10 $M_\odot$ yr$^{-1}$ are encircled.
}
\label{fig5}
\end{figure}

\setcounter{figure}{5}

\begin{figure}
\begin{center}
\includegraphics[angle=-90,width=1.0\linewidth]{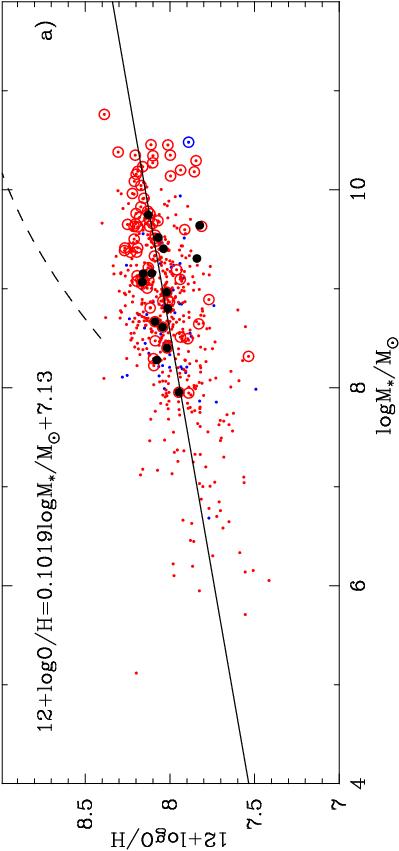}
\includegraphics[angle=-90,width=1.0\linewidth]{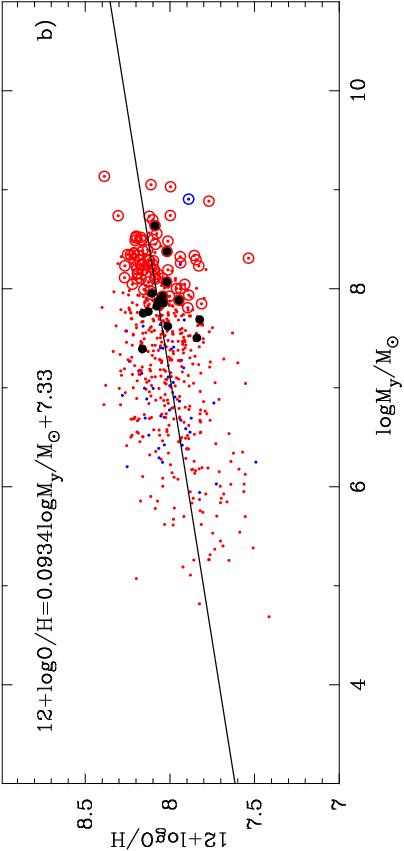}
\end{center} 
\caption{{\bf (a)} Mass-metallicity relation. Symbols are the same as in Fig. \ref{fig4}.
The linear likelihood regression is shown by solid line.
The sample of galaxies is the same as in Fig. \ref{fig5}.
The galaxies with red {\sl WISE} colours $W1-W2$ $\geq$ 2 mag
are indicated by large black filled circles and the galaxies with 
$SFR$(H$\alpha$) $\geq$ 10 $M_\odot$ yr$^{-1}$ are encircled.
For comparison, the mass-metallicity relation for 53000 SDSS 
star-forming galaxies \citep{T04} is shown by a dashed line.
{\bf (b)} Relation between the mass of the young stellar population
and the oxygen abundance. The linear likelihood regression is shown by solid 
line. The galaxies with red {\sl WISE} colours $W1-W2$ $\geq$ 2 mag
are resembled by large black filled circles and the galaxies with 
$SFR$(H$\alpha$) $\geq$ 10 $M_\odot$ yr$^{-1}$ are encircled.
}
\label{fig6}
\end{figure}

\subsection{Colour-colour diagrams and galaxies with hot dust emission}

The completion of the {\sl WISE} All-Sky Source Catalogue (ASSC) in 2011 
offered an 
opportunity in the search for star-forming galaxies with hot dust emission.
First results in this direction were demonstrated by \citet{G11} and 
\citet{I11b}, who found three and four galaxies respectively with very red 
3.4$\mu$m -- 4.6 $\mu$m
colours, indicating the rapid increase of the dust emission in the 
3.4 -- 4.6 $\mu$m range, the signature of hot dust.
These findings were based on the data from the Preliminary Release Source 
Catalogue (PRSC), which covered only half of the sky.

Several galaxy components can contribute to
the emission in the 3.4 -- 4.6 $\mu$m range: stars, ionised gas,
polycyclic aromatic hydrocarbon (PAH) emission, and hot
dust. However, the PAH emission is weak in a low-metallicity environment 
\citep{E08,H10}. Using results from the SED modelling of Sect. \ref{sec:SED}, 
we found that stellar
and interstellar ionised gas emission are characterised by $W1-W2$ colours 
of $\sim$ 0 and $\sim$ 0.7 mag, respectively 
\citep[see also Table 1 in][]{W10}. This 
indicates a steeper decline of stellar emission. 
Therefore, a colour excess above $W1-W2$ $\sim$ 0.7 mag
could be an indication of hot dust with a temperature of several 
hundred Kelvin. However, the $W1-W2$ colour alone is not 
sufficient for a precise determination of the dust temperature. 
The spectral energy distribution is needed in a wide wavelength range to fit 
and subtract stellar and gaseous emission.

In Fig. \ref{fig7}, we showed the $W1-W2$ vs $W2-W3$ colour-colour diagram
for $\sim$10000 galaxies from the SDSS sample that excludes galaxies, 
which were not detected in each of
the $W1$, $W2$, and $W3$ bands. 
The sample in Fig. \ref{fig7}a is split into two subsamples of objects with 
EW(H$\beta$) $\geq$ 50\AA\ (magenta dots) and EW(H$\beta$) $<$ 50\AA\ (cyan
dots). Twenty galaxies with $W1-W2$ $\geq$ 2 mag are shown by large 
black filled circles.
Four of these galaxies were discovered by \citet{I11b}. With the ASSC,
we added sixteen more galaxies with very red $W1-W2$ colour.
Similarly, we split the sample into two 
subsamples of objects in Fig. \ref{fig7}b with the H$\beta$ luminosity 
$L$(H$\beta$) $\geq$ 3$\times$10$^{40}$ erg s$^{-1}$ (orange dots) that
corresponds to a star-formation rate 
$SFR$(H$\alpha$) $\geq$ 0.7 $M_\odot$ yr$^{-1}$ 
\citep[according to prescriptions of ][]{K99}
and $L$(H$\beta$) $<$ 3$\times$10$^{40}$ erg s$^{-1}$ 
(green dots). 
Finally, we split the sample into two 
subsamples of objects with  EW(H$\beta$) $\geq$ 50\AA\ and
$L$(H$\beta$) $\geq$ 3$\times$10$^{40}$ erg s$^{-1}$ (red dots)
and EW(H$\beta$) $<$ 50\AA\ and 
$L$(H$\beta$) $<$ 3$\times$10$^{40}$ erg s$^{-1}$ (blue dots) 
in Fig. \ref{fig7}c. The galaxies
shown in red correspond to the LCGs discussed by \citet{I11a}. For comparison,
the cloud of grey dots is the sample of SDSS QSOs \citep{S10}. It is seen
that QSOs are well distinct from star-forming galaxies.

Most of the galaxies ($\sim$90\%) from our sample have blue $W1-W2$ colours 
of $\la$ 0.5 mag,
which are consistent with the stellar and ionised gas emission, 
when only as a main source of radiation. Similarly 
to \citet{I11b}, we found that galaxies with red $W1-W2$ colours greater than 1
mag are mainly luminous galaxies with high-excitation H {\sc ii} regions.
This is most clearly seen in Fig. \ref{fig7}c where the red points
diverge most notably from the blue ones. This finding suggests that
the emission seen in the {\sl WISE} bands and produced by hot dust is caused
by the radiation from young star-forming regions.

In Fig. \ref{fig8}, we show the {\sl GALEX} $m$(FUV) -- SDSS $r$ vs. 
{\sl WISE} $W1-W4$ colour-colour diagram. There is a clear separation
between the faint galaxies with low-excitation H {\sc ii} regions (blue dots)
from the luminous galaxies with high-excitation H {\sc ii} regions, which again
implies that star-forming regions are the main source of hot dust heating.
The galaxies with the reddest $W1-W2$ colours (large black filled circles) are
also among the galaxies with the reddest $W1-W4$ colours. 

Figure \ref{fig8} also implies that SDSS star-forming galaxies are relatively
transparent systems and only a small fraction of UV radiation is 
transformed into the emission in the mid-infrared range as seen by {\sl WISE}. 
The same conclusion comes
from the distribution of extinction coefficient $C$(H$\beta$) 
(Fig. \ref{fig3}e) and the comparison of the {\sl GALEX} FUV and NUV 
luminosities (Figs. \ref{fig4}a and \ref{fig4}b) with {\sl WISE}
luminosities (Figs. \ref{fig4}k -- \ref{fig4}n). It is seen that 
luminosities in {\sl WISE} bands are only $\sim$10\% -- 30\% of those
in {\sl GALEX} bands.

In Fig. \ref{fig9}, we showed  the relation between {\sl WISE} $W1-W2$ colour 
and oxygen abundance 12 + log O/H. The lowest-metallicity galaxies with 
12 + log O/H $<$ 7.6 were found mainly among the SDSS galaxies with low 
H$\beta$ luminosity and blue $W1-W2$ colours (blue symbols); the latter are
consistent with stellar and ionised gas emission. The contribution
of hot dust in these galaxies is small. Only few galaxies with
$L$(H$\beta$) $\geq$ 3$\times$10$^{40}$ erg s$^{-1}$  have the oxygen 
abundances of $\sim$ 7.5 -- 7.6. Most of the luminous SDSS galaxies have oxygen
abundances above 8.0 (red symbols). There is no clear dependence of the
$W1-W2$ colour on the oxygen abundance, despite the already mentioned fact that
lowest-metallicity galaxies are mainly blue. 

However, there is one exception.
This is the BCD SBS 0335--052E with 12 + log O/H =7.3
(Fig. \ref{fig9}). First, \citet{T99} with {\sl ISO} 
spectroscopic observations and later \citet{H04} with {\sl Spitzer} 
spectroscopic observations showed that mid-infrared emission in this galaxy is 
dominated by warm dust. The {\sl WISE} photometric observations \citep{G11}
confirmed these findings. All galaxies with the red $W1-W2$ colour found
in this paper (large black filled circles) are high-metallicity analogues of 
the three lower-metallicity galaxies (asterisks) discussed by \citet{G11}.
Similar to other galaxies shown in Fig. \ref{fig9}, there is no evident 
dependence of the $W1-W2$ colour with oxygen abundance among the galaxies
with the reddest {\sl WISE} colours.

\setcounter{figure}{6}

\begin{figure}
\includegraphics[angle=-90,width=0.8\linewidth]{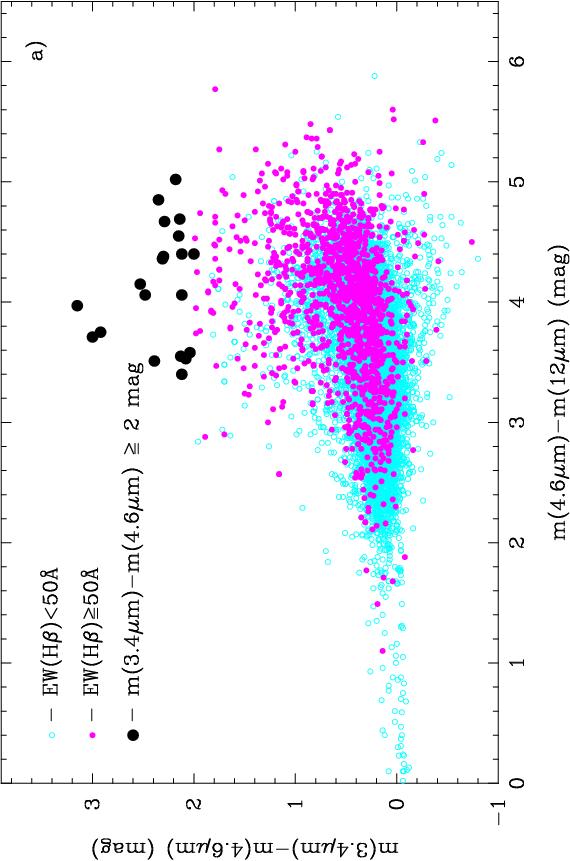} 
\vspace{0.2cm}
\includegraphics[angle=-90,width=0.8\linewidth]{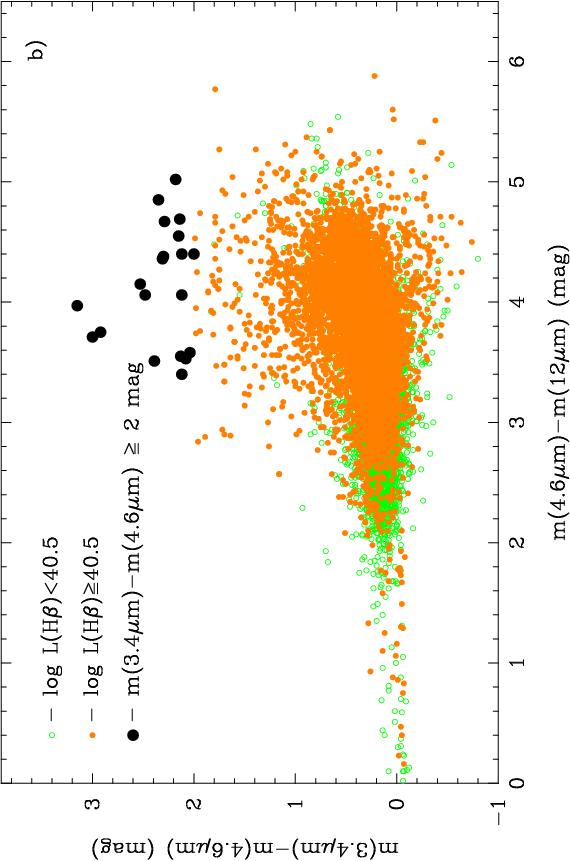}
\vspace{0.2cm}
\includegraphics[angle=-90,width=0.8\linewidth]{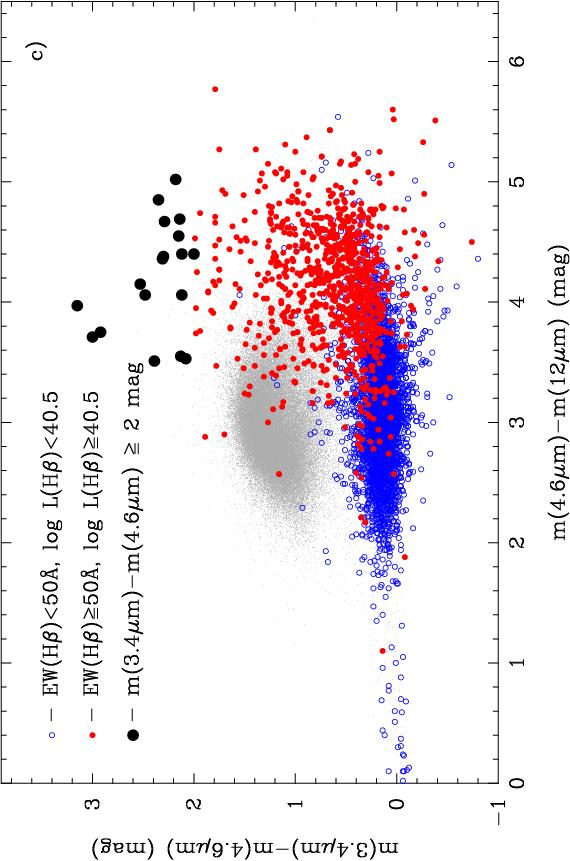}
\caption{ {\bf (a)} $m$(3.4$\mu$m) -- $m$(4.6$\mu$m) vs 
$m$(4.6$\mu$m) -- $m$(12$\mu$m) colour-colour diagram
for a sample of emission-line galaxies detected in the three {\sl WISE} 
bands 3.4$\mu$m ($W1$), 4.6$\mu$m ($W2$), and 12$\mu$m ($W3$). 
Galaxies with 
H$\beta$ equivalent width EW(H$\beta$) $\geq$ 50 \AA\ are shown
by magenta filled circles while galaxies with 
EW(H$\beta$) $<$ 50 \AA\ are shown by
cyan open circles. 
Twenty newly identified galaxies with $m$(3.4$\mu$m) -- $m$(4.6$\mu$m) $\geq$ 
2 mag are 
shown by large black filled circles. 
{\bf (b)} Same as in {\bf (a)} but galaxies with H$\beta$ luminosity 
$L$(H$\beta$) $\geq$ 3$\times$10$^{40}$ erg s$^{-1}$ are shown
by orange filled circles while galaxies with
$L$(H$\beta$) $<$ 3$\times$10$^{40}$ erg s$^{-1}$ are shown by
green open circles.
{\bf (c)} Same as in {\bf (a)} but galaxies with H$\beta$ luminosity 
$L$(H$\beta$) $\geq$ 3$\times$10$^{40}$ erg s$^{-1}$ and with 
H$\beta$ equivalent width EW(H$\beta$) $\geq$ 50 \AA\ are shown
by red filled circles while galaxies with
$L$(H$\beta$) $<$ 3$\times$10$^{40}$ erg s$^{-1}$ and with 
EW(H$\beta$) $<$ 50 \AA\ are shown by blue open circles. For comparison,
gray dots in (c) show the location of QSOs selected in the SDSS. 
}
\label{fig7}
\end{figure}

\setcounter{figure}{7}

\begin{figure}
\includegraphics[angle=-90,width=0.8\linewidth]{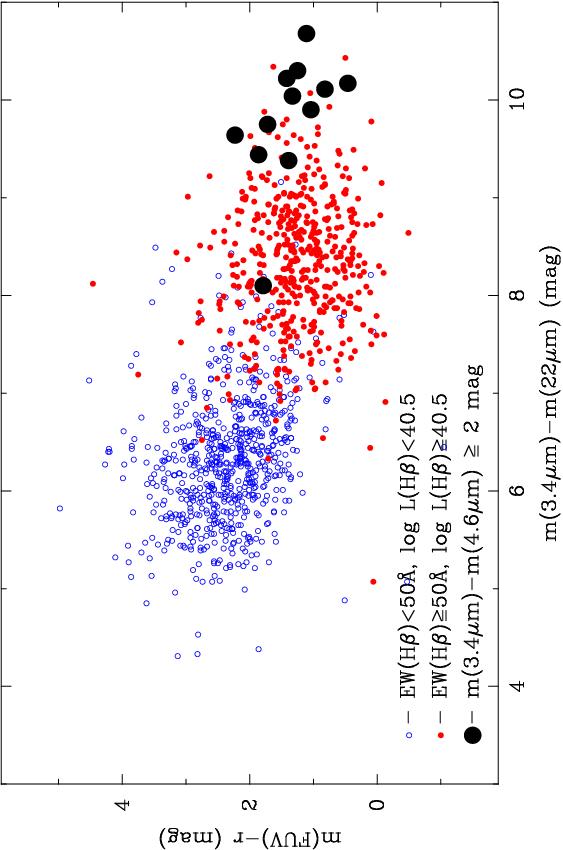} 
\caption{$m$(3.4$\mu$m) -- $m$(22$\mu$m) vs $m$(FUV) -- $r$ colour-colour 
diagram for a sample of emission-line galaxies detected in the {\sl WISE} 
3.4$\mu$m ($W1$) and 22$\mu$m ($W4$) bands and in the {\sl GALEX} FUV 
band. Galaxies with H$\beta$ equivalent width EW(H$\beta$) $\geq$ 50 \AA\ 
and with $L$(H$\beta$) $\geq$ 3$\times$10$^{40}$ erg s$^{-1}$ are shown
by red filled circles, while galaxies with 
EW(H$\beta$) $<$ 50 \AA\ 
and with $L$(H$\beta$) $<$ 3$\times$10$^{40}$ erg s$^{-1}$ are depicted by
blue open circles. 
Galaxies with $m$(3.4$\mu$m) -- $m$(4.6$\mu$m)$\geq$ 2 mag
are shown by large black filled circles. 
}
\label{fig8}
\end{figure}

\setcounter{figure}{8}

\begin{figure}
\includegraphics[angle=-90,width=0.8\linewidth]{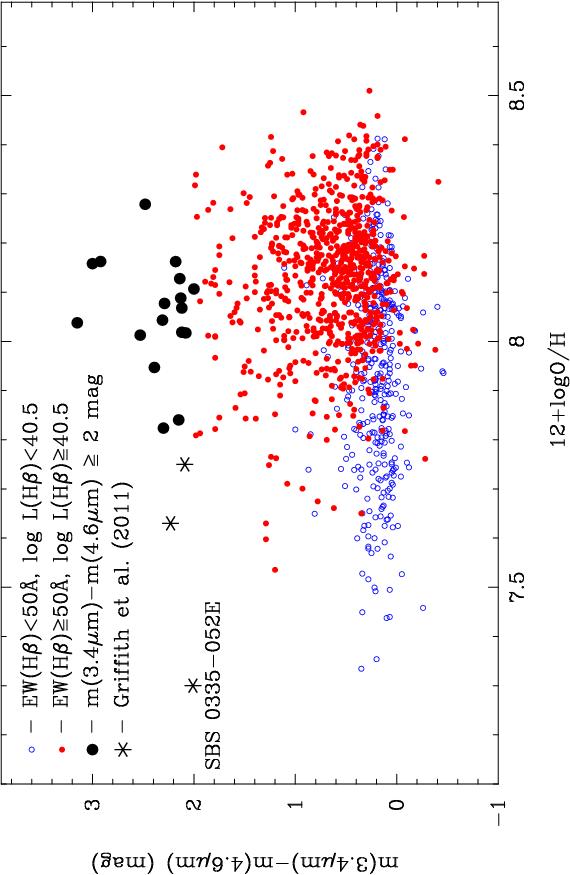} 
\caption{$m$(3.4$\mu$m) -- $m$(4.6$\mu$m) vs 12+logO/H 
diagram for a sample of emission-line galaxies detected 
in {\sl WISE} 3.4$\mu$m ($W1$) and 4.6$\mu$m ($W2$) bands. Galaxies with 
H$\beta$ equivalent width EW(H$\beta$) $\geq$ 50 \AA\ 
and with $L$(H$\beta$) $\geq$ 3$\times$10$^{40}$ erg s$^{-1}$ are shown
by red filled circles while galaxies with 
EW(H$\beta$) $<$ 50 \AA\ 
and with $L$(H$\beta$) $<$ 3$\times$10$^{40}$ erg s$^{-1}$ are shown by
blue open circles. 
Galaxies with $m$(3.4$\mu$m) -- $m$(4.6$\mu$m)$\geq$ 2 mag from this paper
are shown by large black filled circles and from
\citet{G11} by asterisks. Only the galaxies where 
the errors in [O {\sc iii}] $\lambda$4363 emission-line flux do not exceed
50\% are shown.
}
\label{fig9}
\end{figure}

\subsection{Morphology and SEDs of the galaxies with hot dust emission}
\label{sec:SED}

The galaxies with very red colours $W1-W2$ $\geq$ 2 mag are rare. 
\citet{G11} found three galaxies that are not present 
in our sample, because no SDSS spectra are available for them.
We found only twenty more galaxies like these
out of $\sim$ 14000 galaxies from our SDSS sample 
with the available {\sl WISE} data. 
The observed characteristics of these galaxies are shown
in Table \ref{tab2}; the derived parameters are present in Table \ref{tab3}.
For the sake of comparison, we also show
the parameters of all 64 galaxies with 2 mag $>$ $W1-W2$ $\geq$ 1.5 mag
and a representative sample of 10 galaxies with blue colours 
$W1-W2$ $<$ 0.5 mag in Tables \ref{tab2} and \ref{tab3}. In general, 
galaxies with the reddest $W1-W2$ colours
are apparently faint but intrinsically bright with $L$(H$\beta$),
which corresponds
to $SFR$ $\sim$ 3 -- 47 $M_\odot$ yr$^{-1}$ and is similar to that in
high-redshift LBGs. Despite the high absolute brightness, their stellar
masses of 10$^8$ -- 10$^9$ $M_\odot$ are typical of dwarf galaxies.

The SDSS composite $g,r,i$ images of galaxies with $W1-W2$ $\geq$ 2 mag, 
2 mag $>$ $W1-W2$ $\geq$ 1.5 mag, and $W1-W2$ $<$ 0.5 mag are shown
in Figs. \ref{fig10}, \ref{fig11}, and \ref{fig12}, respectively.
All galaxies with the reddest $W1-W2$ colours are compact (Fig. \ref{fig10})
with angular diameters $\leq$ 1\arcsec\ -- 2\arcsec, corresponding to linear 
diameters $\sim$ 3 -- 6 kpc at the redshifts of 0.15 - 0.3. 
They can be classified
as ``green pea'' \citep{C09} and/or LCGs \citep{I11a}.
For comparison, the sample of galaxies with bluer colours 
(Figs. \ref{fig11} and \ref{fig12}) include not only compact but also extended
objects and objects with multiple knots.

Modelled and aperture-corrected observed spectral energy 
distributions (SEDs) for twenty galaxies with the reddest $W1-W2$ are shown in 
Fig. \ref{fig13}. We produced modelled SEDs (blue lines) based on the observed
SDSS spectra (black lines), which were extrapolated to the UV and IR ranges. 
They include the modelled stellar SEDs 
(green lines) and modelled 
ionised gas SEDs (magenta lines). For comparison, we show photometric
data from {\sl GALEX}, SDSS, 2MASS and {\sl WISE} by red
symbols and red lines. These data were transformed to units 
erg s$^{-1}$cm$^{-2}$\AA$^{-1}$ using equations in Sect. \ref{phot}. 
However, we note that photometric data are not used for modelling of the SEDs. 
Nevertheless, photometric data agree
with the modelled SEDs, such as in the {\sl GALEX} FUV and NUV ranges.
This is because all objects with the reddest $W1-W2$ colours are compact with
angular diameters that are comparable to that of the SDSS spectroscopic 
aperture. Therefore, the uncertainties introduced by aperture corrections are 
small.

Some photometric data in the visual range (SDSS $g$, $r$, $i$ and $z$) show an 
excess above the continuum, which can exceed $\sim$1.5 mag in some cases. 
This excess is due to a significant contribution of strong emission lines.
We also noted that the contribution of the ionised gas emission is increased 
with increasing wavelength, and in some galaxies with high
EW(H$\beta$) (e.g. J1327+6151 in Fig. \ref{fig13}), it is dominant 
longward of the $K$ band.
Therefore, it is important to consider this emission
as to not to misinterpret it as the emission of cool stars and hence not to
overestimate the galaxy stellar mass \citep[see, e.g. ][]{I11a}.
On the other hand, strong excess emission is seen in the {\sl WISE} bands
(red filled circles connected by solid lines) suggesting a dominating
contribution of dust emission in this wavelength range. While the emission
at 3.4$\mu$m is still close to the value predicted for the combined 
stellar+ionised gas emission, it is higher by 1-2 orders of magnitude at 
4.6$\mu$m and continues to rise longward.

The SEDs of all galaxies with 2 mag $>$ $W1-W2$ $\geq$ 1.5 mag and
of ten representative galaxies with $W1-W2$ $<$ 0.5 mag are shown in 
Figs. \ref{fig14} and \ref{fig15}, respectively. The continuum in the 
mid-infrared range for objects in Fig. \ref{fig14} does not rise so steeply,
as in the case of objects in Fig. \ref{fig13}. As for galaxies with 
blue $W1-W2$ colours (Fig. \ref{fig15}), the emission decreases in the
wavelength range 3.4$\mu$m -- 4.6$\mu$m and agrees with the modelled combined
stellar and ionised gas emission but increases at longer wavelengths,
suggesting that the contribution of the hot dust is not as large as in the case
of the galaxies with redder $W1-W2$ colours. As noted, in the case
of extended galaxies (e.g., J0849+1114 and J0949+1750 in Fig. \ref{fig14}),
photometric data strongly deviate from the spectroscopic fits.

\setcounter{figure}{15}

\begin{figure*}
\centering{
\hbox{
\includegraphics[angle=-90,width=0.48\linewidth]{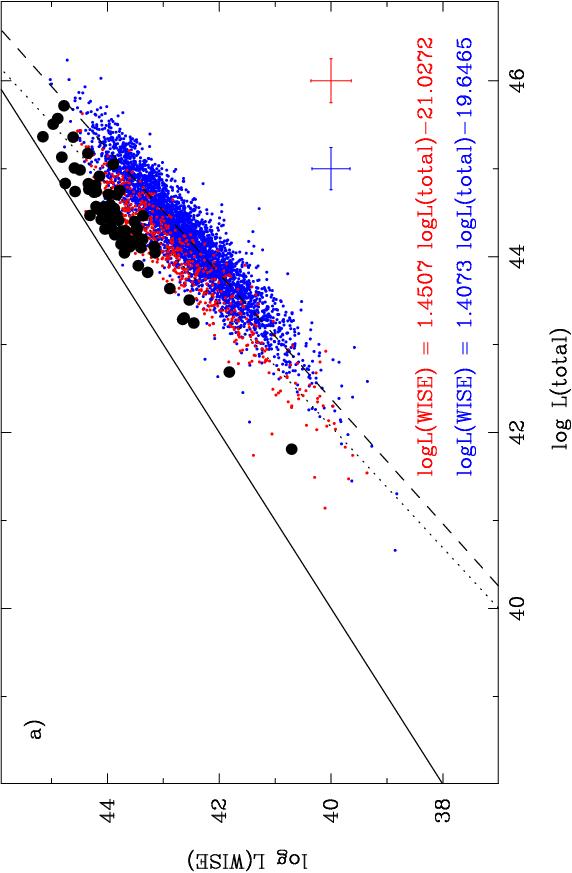}
\hspace{0.2cm}\includegraphics[angle=-90,width=0.48\linewidth]{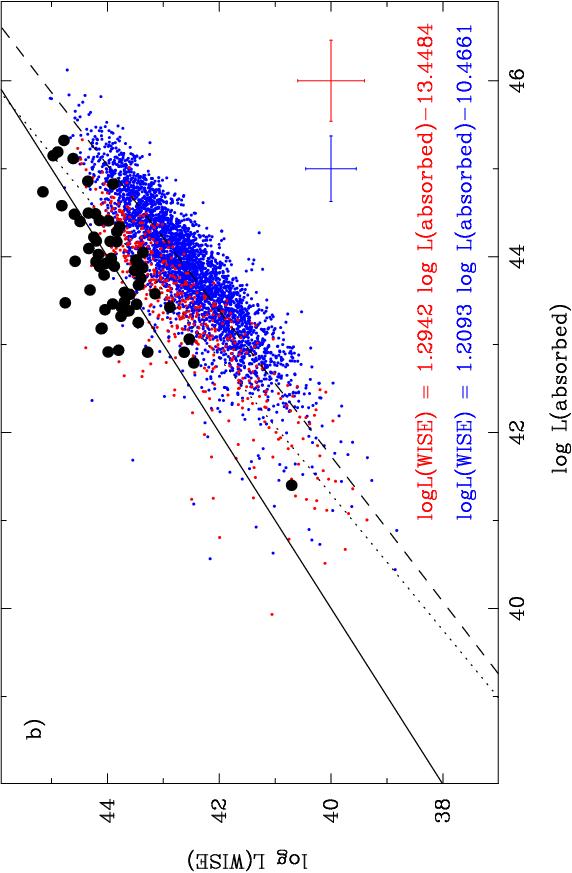}}
} 
\caption{{\bf (a)} Dependence of the luminosity $L$({\sl WISE}) integrated
over all {\sl WISE} bands in the wavelength range 3.4$\mu$m -- 22$\mu$m 
on the total extinction-corrected luminosity $L$(total) in the wavelength 
range 0.1$\mu$m -- 22$\mu$m. Galaxies with 
EW(H$\beta$) $\geq$ 50\AA\ and EW(H$\beta$) $<$ 50\AA\ are shown by red and 
blue symbols, respectively. Galaxies with $m$(3.4$\mu$m) -- $m$(4.6$\mu$m) 
$\geq$ 1.5 mag are shown by large black filled circles.
Solid line is the line of equal values; dotted and dashed lines are
maximum likelihood fits to the sample galaxies with EW(H$\beta$) $\geq$ 50\AA\
and EW(H$\beta$) $<$ 50\AA, respectively. Error bars are average 
dispersions of the data around the fits.
{\bf (b)} Dependence of the luminosity $L$({\sl WISE}) on the 
luminosity $L$(absorbed) of the absorbed emission in the 
wavelength range 0.0912$\mu$m -- 3$\mu$m. The meaning of lines and 
symbols is the same as in (a).}
\label{fig16}
\end{figure*}

\subsection{The energy balance for the SDSS dwarf emission-line galaxies}

The large wavelength coverage from 0.1$\mu$m to 22$\mu$m of our galaxy fits and
 the large sample allowed us to statistically study the energy balance
for all SDSS spectra of emission-line galaxies. 
Some part of the galaxy emission is absorbed by dust at short wavelengths.
It heats dust and is re-emitted in the infrared
range. In particular, it is interesting to study what fraction of emission
in the infrared is due to the hot dust. 

We used the modelled
SEDs (some examples are present in Figs. \ref{fig13} -- \ref{fig15}) 
and {\sl WISE} photometric data to 
calculate the total extinction-corrected luminosity of the galaxy $L$(total) 
in the wavelength range 0.1$\mu$m - 22$\mu$m:
\begin{equation}
L({\rm total})=\int_{0.1\mu{\rm m}}^{22\mu{\rm m}}L_\lambda \times 
2.5^{A_{\lambda(1+z)}({\rm MW})+A_\lambda({\rm int})} d\lambda, \label{Ltot}
\end{equation}
where $\lambda$ and $\lambda(1+z)$ are rest-frame and observed wavelengths,
$L_\lambda$ is the rest-frame monochromatic luminosity non-corrected for
extinction, $A_{\lambda(1+z)}$(MW) is the Milky Way extinction at the observed
wavelength, and $A_\lambda$(int) is the internal galaxy extinction at the 
rest-frame wavelength. 
We also
calculated the luminosity of the absorbed emission $L$(absorbed) 
in the wavelength range 0.1$\mu$m - 3$\mu$m by subtracting the
galaxy luminosity corrected for the Milky Way extinction 
(but not for the internal extinction) from the luminosity corrected
for both the Milky Way and galaxy internal extinction:
\begin{eqnarray}
L({\rm absorbed})&=&\int_{0.1\mu{\rm m}}^{3\mu{\rm m}}L_\lambda \times 
2.5^{A_{\lambda(1+z)}({\rm MW})} \\ \nonumber
&&\times\left(2.5^{A_\lambda({\rm int})}-1\right) d\lambda. \label{Labs}
\end{eqnarray}

Finally, we calculated the galaxy luminosity $L$({\sl WISE}) in the wavelength
range 3.4$\mu$m -- 22$\mu$m by using {\sl WISE} photometric data.
To reduce the uncertainties of aperture corrections, however,
we restricted ourselves to the
consideration of compact galaxies, which only comprises roughly half of 
the entire SDSS sample.

In Fig. \ref{fig16}a, we showed the dependence of $L$({\sl WISE}) on the
total galaxy luminosity $L$(total). The total luminosity was varied in
a wide range from $\sim$10$^7$$L_\odot$ to $\sim$10$^{12}$$L_\odot$.
The solid line in the Figure is a line of equal luminosities.
It is seen that the fraction of the 
luminosity in the {\sl WISE} bands at the low-luminosity end is $\la$ 1\%
of the total luminosity (red and blue dots), but it is gradually
increased to $\sim$10\% at the high-luminosity end. This fraction is also
higher for galaxies with high EW(H$\beta$) $\geq$50\AA, which contain the
dominating part of the hot dust emission. The largest fraction of the
{\sl WISE} luminosity is emitted in galaxies with red $W1-W2$ colours.
For galaxies with $W1-W2$ $\geq$ 1.5 mag (large black filled circles),
it may be as high as $\sim$50\% of $L$(total).

The dependence of the luminosity $L$({\sl WISE}) in the {\sl WISE} bands on
the luminosity $L$(absorbed) 
is shown in Fig. \ref{fig16}b. The solid line is the line
of equal luminosities. The fraction of the radiation emitted by dust in the 
{\sl WISE} bands is $\sim$10\% in galaxies with low EW(H$\beta$) $<$ 50\AA\ 
(blue dots), and it is higher by a factor of $\sim$3 in galaxies with high
EW(H$\beta$) $\geq$ 50\AA\ (red dots), which again indicates the dominant role
of young starbursts in heating the dust to high temperatures. As for galaxies
with red $W1-W2$ colours ($\geq$ 1.5 mag), $L$({\sl WISE}) is comparable to
$L$(absorbed) (large black filled circles). 
We concluded that 
the contribution of hot dust in the infrared emission is high 
and possibly is dominated in galaxies with the red $W1-W2$ colours. 

We have only scarce data from the {\sl IRAS} 60$\mu$m band because of the low
{\sl IRAS} sensitivity. For compact galaxies, which were detected by both
{\sl IRAS} and {\sl WISE}, we show 
the dependence of the luminosity ratio $L$({\sl WISE})/$\nu L_\nu$(60 $\mu$m) 
vs $W1-W2$ colour in Fig. \ref{fig17}. Because 
$L$({\sl WISE}) and $\nu L_\nu$(60 $\mu$m) are tracers of hot
(several hundred K) and cooler (several ten K) dust, respectively, 
the luminosity ratio tells us what
the fraction of the infrared emission from the hot dust is. This is seen in 
Fig. \ref{fig17} when $L$({\sl WISE})/$\nu L_\nu$(60 $\mu$m) is higher
in galaxies with redder $W1-W2$ colours and approaches $\sim$1 in the reddest
galaxies. It is also higher in galaxies with high $L$(H$\beta$) (red and black
symbols), which again implies that dust is heated to high temperatures by young
star-forming regions with an age of a few Myr.

\setcounter{figure}{16}

\begin{figure}
\includegraphics[angle=-90,width=0.98\linewidth]{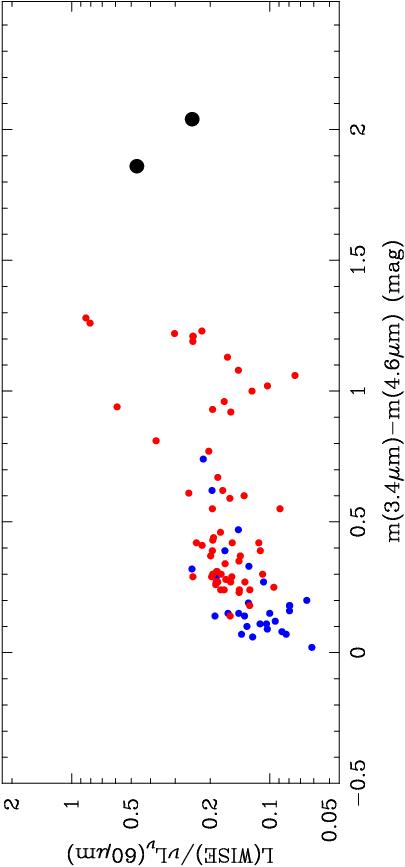}
\caption{The ratio $L$({\sl WISE})/$\nu$$L_\nu$(60$\mu$m) vs. 
{\sl WISE} $W1-W2$ colour for the compact galaxies from the SDSS sample. 
Galaxies with $L$(H$\beta$) $\geq$ 3$\times$10$^{40}$ erg s$^{-1}$ and
$L$(H$\beta$) $<$ 3$\times$10$^{40}$ erg s$^{-1}$ are shown by red and blue
dots, respectively. Galaxies with $W1-W2$ $\geq$ 1.5 mag are shown
by large black filled circles.}
\label{fig17}
\end{figure}

\section{Summary \label{sum}}

We studied a large sample of
$\sim$ 14000 Sloan Digital Sky Survey (SDSS) emission-line star-forming 
galaxies. These data are supplemented by data from 
the {\sl GALEX} survey in the UV range, the
2MASS survey in the near-infrared range, the {\sl Wide-field Infrared Survey 
Explorer} ({\sl WISE}) survey in the mid-infrared
range, the {\sl IRAS} survey in the far-infrared range, and the NVSS survey
in the radio continuum at 20 cm. Using the SDSS spectra 
in the visible range, we fitted the spectral
energy distributions (SEDs) by considering both the stellar and the
ionised gas emission. These fits were extrapolated to the UV and mid-infrared
ranges. 

We also searched for star-forming SDSS galaxies with an aim to find galaxies 
with hot dust emission
at wavelengths of $\lambda$3.4 - 4.6 $\mu$m ({\sl WISE} $W1$ and $W2$ bands).

Our main results are as follows.

1. We found that $\sim$ 12500 and $\sim$ 13500 galaxies out of the total sample
of SDSS galaxies were detected by {\sl GALEX} and {\sl WISE}, respectively. 
Roughly half of the selected SDSS galaxies are compact systems. This allowed us
to compare the global galaxy characteristics derived from the spectroscopic and
photometric data.

2. The luminosities obtained from the photometric observations in a wide 
wavelength range from the UV to the radio range are correlated with 
the luminosity $L$(H$\beta$) of the H$\beta$ emission line, which implies 
that young star-forming
regions strongly contribute to the emission of the galaxies. This contribution
becomes more notable with a rising $L$(H$\beta$) and the equivalent width
EW(H$\beta$) of the H$\beta$ emission line. The luminosities of star-forming
galaxies vary over a large range. At highest luminosities of 
$\sim$10$^{12}$ $L_\odot$, they approach the luminosities of high-redshift
Lyman-break galaxies (LBGs), implying that these galaxies can be 
considered as the local counterparts of distant forming galaxies
in the early Universe. On the other hand, their stellar masses are low
but similar to that of LBGs with
an average value of $\sim$10$^9$ $M_\odot$. This suggests that selected SDSS
galaxies are dwarf systems with luminosities elevated by the strong ongoing
star formation with $SFR$s up to $\sim$100 $M_\odot$ yr$^{-1}$.

3. A major fraction of galaxies has {\sl WISE} $W1-W2$ colours 
of $\sim$ 0.0 -- 0.4 mag, which are
consistent with the colour for the emission from stars and the ionised 
interstellar medium. The
contribution of hot dust with temperatures of several hundred degrees
to the mid-infrared emission is small in these galaxies. On the other hand, 
we found 20 galaxies with redder colours, $W1-W2$ $\geq$ 2 mag. Most of these 
galaxies are luminous compact galaxies (LCGs) with
star-formation rates $SFR$(H$\alpha$) $>$ 0.7 $M_\odot$ yr$^{-1}$ and high
H$\beta$ equivalent widths 
EW(H$\beta$) $>$ 50\AA, which
implies a very recent starburst, that can efficiently heat interstellar dust
to high temperatures.

4. We analysed the energy balance between the emission that are absorbed by 
dust at
short wavelengths from the UV to near-infrared ranges and those that
are emitted by dust
at longer wavelengths in the mid-infrared range. We found that the fraction of 
the emission by hot dust in the {\sl WISE} bands at 3.4$\mu$m -- 22$\mu$m 
relative to the absorbed emission varies from a few percent in the 
lowest luminosity galaxies to several ten percent in the highest luminosity
galaxies. This fraction also increases with increasing EW(H$\beta$).
In galaxies with the
reddest $W1-W2$ colours the luminosity in the {\sl WISE} bands compares
well to the luminosity of the emission absorbed at shorter wavelengths
and to the luminosity in the {\sl IRAS} 60$\mu$m band, which is emitted by 
cooler dust.

\acknowledgements
Y.I.I., N.G.G. and K.J.F. are grateful to the staff of the Max Planck 
Institute for Radioastronomy (MPIfR) for their warm hospitality. 
Y.I.I. and N.G.G. acknowledge financial support by the MPIfR.
{\sl GALEX} is a NASA mission managed by the Jet Propulsion Laboratory. 
This publication makes use of data products from the Two Micron All Sky Survey,
which is a joint project of the University of Massachusetts and the Infrared 
Processing and Analysis Center/California Institute of Technology, funded by 
the National Aeronautics and Space Administration and the National Science 
Foundation.
This publication
makes use of data products from the {\sl Wide-field Infrared
Survey Explorer}, which is a joint project of the University of
California, Los Angeles, and the Jet Propulsion Laboratory,
California Institute of Technology, funded by theNational Aeronautics
and Space Administration. Funding for the Sloan
Digital Sky Survey (SDSS) and SDSS-II has been provided by
the Alfred P. Sloan Foundation, the Participating Institutions,
the National Science Foundation, the U.S. Department of
Energy, the National Aeronautics and Space Administration,
the Japanese Monbukagakusho, and the Max Planck Society,
and the Higher Education Funding Council for England.
This research has made use of the NASA/IPAC Extragalactic Database (NED) 
which is operated by the Jet Propulsion Laboratory, California Institute 
of Technology, under contract with the National Aeronautics and Space 
Administration. 


\Online

\setcounter{table}{1}

 \begin{longtable}{lccccccccrrl}
 \caption{Observed parameters of the galaxies.}
 \label{tab2} \\ \hline \hline
 \endfirsthead
 \caption{---{\sl Continued.}} \\
 \hline\hline
 Name      &R.A.(J2000)$^a$&Dec.(J2000)$^a$&   $z$$^b$ & \multicolumn{7}{c}{Apparent magnitude}&Other names  \\ \cline{5-11}
          &         &            &       &$g$$^c$&FUV$^d$&NUV$^d$&3.4$\mu$m$^e$&4.6$\mu$m$^e$&12$\mu$m$^e$&22$\mu$m$^e$& \\ \hline
\endhead
 \hline 
 \endfoot
 Name      &R.A.(J2000)$^a$&Dec.(J2000)$^a$&   $z$$^b$ & \multicolumn{7}{c}{Apparent magnitude}&Other names  \\ \cline{5-11}
          &         &            &       &$g$$^c$&FUV$^d$&NUV$^d$&3.4$\mu$m$^e$&4.6$\mu$m$^e$&12$\mu$m$^e$&22$\mu$m$^e$& \\ \hline
\multicolumn{12}{c}{a) Galaxies with $m$(3.4$\mu$m)--$m$(4.6$\mu$m) $\geq$ 2 mag} \\
J0135$+$1455&01:35:37.57& $+$14:55:10.91& 0.2178& 18.96&  ... &  ... & 14.13& 12.09&  8.51&  5.78\\
J0239$+$0018&02:39:00.79& $+$00:18:35.88& 0.2166& 20.72&  ... &  ... & 17.76& 15.47& 10.80&  7.42\\
J0333$+$0017&03:33:19.22& $+$00:17:31.33& 0.1938& 20.22& 20.83& 21.22& 17.83& 15.71& 11.31&  7.79\\
J0936$+$0900&09:36:23.29& $+$09:00:01.03& 0.2237& 20.32& 20.82& 21.45& 18.24& 16.10& 11.41&  7.94\\
J1018$+$4106&10:18:03.24& $+$41:06:21.09& 0.2371& 20.43& 21.07& 20.41& 17.92& 15.62& 11.24&  7.70\\
J1046$+$3023&10:46:45.76& $+$30:23:30.88& 0.1271& 19.17&  ... &  ... & 14.78& 11.86&  8.11&  5.05\\
J1101$+$4022&11:01:20.35& $+$40:22:42.32& 0.2798& 20.86&  ... & 21.12& 18.04& 15.51& 11.36&  7.79\\
J1141$+$4515&11:41:06.18& $+$45:15:39.51& 0.1262& 18.89& 20.31& 20.62& 14.68& 12.56&  9.16&  6.35\\
J1143$+$3242&11:43:48.30& $+$32:42:57.94& 0.0740& 18.62& 20.21& 19.63& 16.79& 14.61&  9.59&  6.11\\
J1228$+$3219&12:28:50.47& $+$32:19:09.39& 0.1744& 21.28& 22.46& 22.54& 17.48& 15.33& 10.78&  7.84\\
J1238$+$4618&12:38:03.77& $+$46:18:20.13& 0.0988& 18.95& 20.01& 19.75& 15.52& 13.52&  9.12&  5.62\\
J1311$-$0027&13:11:15.04& $-$00:27:57.91& 0.2309& 20.21& 20.90& 21.19& 17.44& 14.96& 10.90&  8.06\\
J1315$+$2618&13:15:14.05& $+$26:18:41.32& 0.3054& 21.30&  ... & 22.04& 16.45& 13.45&  9.74&  6.70\\
J1327$+$6151&13:27:34.45& $+$61:51:02.99& 0.3166& 20.79& 21.28& 21.52& 18.21& 16.13& 12.60&$>$9.00\\
J1439$+$2453&14:39:05.24& $+$24:53:53.39& 0.2119& 19.54& 20.45& 20.33& 14.27& 12.14&  8.59&  6.17\\
J1457$+$2232&14:57:35.14& $+$22:32:01.79& 0.1488& 19.43& 20.23& 19.86& 16.26& 13.95&  9.59&  6.51\\
J1514$+$3852&15:14:08.63& $+$38:52:07.31& 0.3329& 20.45& 20.69& 20.39& 17.90& 15.55& 10.70&  7.79\\
J1537$+$5847&15:37:37.27& $+$58:47:40.51& 0.2143& 20.43& 20.25& 20.62& 17.55& 14.40& 10.43&  7.38\\
J1541$+$1753&15:41:01.32& $+$17:53:34.03& 0.2599& 21.31&  ... & 21.85& 18.66& 16.27& 12.76&  9.11\\
J1604$+$0819&16:04:36.66& $+$08:19:59.11& 0.3123& 20.80& 21.60& 21.64& 17.93& 15.81& 11.75&  8.49\\
\multicolumn{12}{c}{b) Galaxies with 2 mag $>$ $m$(3.4$\mu$m)--$m$(4.6$\mu$m) $\geq$ 1.5 mag} \\
J0042$+$1602&00:42:36.94& $+$16:02:02.63& 0.2473& 20.38& 21.55& 21.21& 16.65& 14.90& 10.63&  7.02\\
J0122$+$0100&01:22:18.12& $+$01:00:26.02& 0.0554& 16.84& 17.41& 17.00& 12.24& 10.64&  6.97&  4.04&HS 0119+0044\\
J0132$-$0853&01:32:58.56& $-$08:53:37.70& 0.0952& 18.82&  ... &  ... & 14.61& 12.63&  8.89&  5.80\\
J0303$-$0759&03:03:21.42& $-$07:59:23.20& 0.1649& 19.38& 19.63& 19.61& 15.98& 14.39& 10.24&  7.33\\
J0327$+$0101&03:27:50.15& $+$01:01:34.88& 0.1088& 18.49& 19.66& 19.49& 14.92& 13.22&  9.74&  6.65\\
J0729$+$3949&07:29:30.30& $+$39:49:41.62& 0.0788& 16.85& 17.83& 17.59& 13.30& 11.44&  7.36&  3.65\\
J0740$+$3209&07:40:41.05& $+$32:09:41.79& 0.1103& 19.60&  ... &  ... & 15.89& 14.34& 10.23&  6.70\\
J0808$+$2814&08:08:16.91& $+$28:14:31.13& 0.3262& 20.57& 20.76& 20.61& 17.43& 15.65& 12.18&$>$8.31\\
J0817$+$4014&08:17:47.65& $+$40:14:49.41& 0.1456& 18.71&  ... &  ... & 13.53& 11.83&  8.93&  6.02\\
J0820$+$5050&08:20:01.72& $+$50:50:39.20& 0.2173& 19.12& 20.27& 20.13& 15.95& 14.33& 10.37&  7.11\\
J0822$+$2241&08:22:47.66& $+$22:41:44.10& 0.2163& 19.45& 20.34& 20.17& 14.37& 12.80&  9.27&  6.27\\
J0836$+$3130&08:36:54.91& $+$31:30:02.02& 0.2897& 20.52& 21.82& 21.56& 17.13& 15.49& 12.60&$>$8.53\\
J0840$+$1344&08:40:34.11& $+$13:44:51.34& 0.2270& 20.05& 20.47& 20.62& 17.28& 15.47& 11.30&  7.86\\
J0848$+$5106&08:48:44.50& $+$51:06:26.76& 0.0719& 19.49& 21.11& 20.27& 17.30& 15.76& 11.34&  8.42\\
J0849$+$1114&08:49:05.40& $+$11:14:45.87& 0.0773& 18.53& 18.95& 18.34& 12.40& 10.76&  7.22&  4.12\\
J0852$+$1216&08:52:21.72& $+$12:16:51.76& 0.0760& 17.35& 18.09& 17.94& 14.92& 13.27&  8.83&  5.43\\
J0927$+$1740&09:27:28.68& $+$17:40:18.61& 0.2883& 20.29& 21.14& 20.72& 17.52& 15.66& 11.57&  8.71\\
J0928$+$1502&09:28:10.52& $+$15:02:27.97& 0.0784& 18.16& 19.44& 19.15& 14.17& 12.34&  8.41&  5.30\\
J0949$+$1750&09:49:17.95& $+$17:50:45.07& 0.0498& 15.85& 18.36& 17.81& 12.50& 10.78&  7.84&  5.04&CGCG 092-053\\
J1011$+$1308&10:11:57.08& $+$13:08:22.12& 0.1439& 19.85&  ... &  ... & 18.09& 16.30& 10.53&  7.23\\
J1018$+$5155&10:18:55.44& $+$51:55:27.81& 0.1294& 18.77& 19.43& 19.30& 15.44& 13.84&  9.85&  6.84\\
J1023$+$2421&10:23:59.21& $+$24:21:06.39& 0.2093& 18.55& 19.90& 19.42& 15.08& 13.26&  9.04&  5.86\\
J1034$+$1155&10:34:49.22& $+$11:55:50.33& 0.1830& 18.31& 19.19& 18.92& 14.59& 12.99&  9.32&  6.79\\
J1047$+$0739&10:47:55.93& $+$07:39:51.19& 0.1683& 19.90& 20.84& 21.11& 15.10& 13.35&  9.17&  6.12\\
J1050$+$1538&10:50:32.51& $+$15:38:06.31& 0.0845& 18.22& 18.74& 18.55& 16.44& 14.93& 10.04&  6.66\\
J1111$+$0713&11:11:41.32& $+$07:13:08.52& 0.2280& 19.55& 20.96& 20.18& 16.86& 15.36& 12.22&  9.18\\
J1132$+$3128&11:32:57.32& $+$31:28:31.86& 0.0332& 18.50& 19.94& 19.39& 16.20& 14.41&  9.75&  6.60\\
J1135$+$6025&11:35:27.96& $+$60:25:32.99& 0.4299& 19.14& 19.81& 19.33& 15.55& 13.98& 10.43&  7.52\\
J1148$+$1756&11:48:40.87& $+$17:56:33.02& 0.0792& 18.51& 19.93& 19.39& 16.16& 14.37&  9.66&  6.44\\
J1151$+$3756&11:51:35.33& $+$37:56:03.63& 0.2969& 20.56& 21.49& 21.47& 16.70& 14.89& 10.98&  7.77\\
J1151$+$0106&11:51:52.09& $+$01:06:06.02& 0.0888& 19.15& 20.64& 20.21& 16.14& 14.39&  9.12&  5.80\\
J1155$+$5739&11:55:28.34& $+$57:39:51.97& 0.0171& 16.41& 17.66& 17.49& 14.70& 13.20&  8.52&  5.32&Mrk 193, SBS 1152+579\\
J1158$-$0203&11:58:16.36& $-$02:03:15.37& 0.1451& 19.07& 21.03& 20.62& 14.99& 13.18&  9.45&  6.25\\
J1159$+$1344&11:59:22.12& $+$13:44:14.23& 0.1164& 18.67& 20.74& 20.20& 13.87& 11.91&  9.07&  6.32\\
J1201$+$0211&12:01:22.30& $+$02:11:08.35& 0.0034& 17.72& 18.64& 18.58& 15.38& 13.55&  9.63&  6.73\\
J1205$+$2856&12:05:22.38& $+$28:56:48.53& 0.1076& 19.76& 20.09& 19.77& 16.62& 15.08& 10.57&  7.32\\
J1219$+$1526&12:19:03.99& $+$15:26:08.52& 0.1957& 19.54& 19.31& 19.73& 15.88& 14.38& 11.14&  8.01\\
J1245$+$1043&12:45:09.05& $+$10:43:40.15& 0.1658& 18.40& 19.14& 18.84& 15.45& 13.48&  9.23&  5.80\\
J1245$+$5254&12:45:34.60& $+$52:54:41.84& 0.1810& 18.87& 20.22& 19.86& 16.10& 14.41& 10.43&  7.37\\
J1248$+$1234&12:48:34.64& $+$12:34:02.94& 0.2634& 19.90& 19.65& 20.20& 16.73& 14.94& 10.80&  7.89\\
J1253$-$0312&12:53:05.96& $-$03:12:58.94& 0.0228& 15.19& 16.41& 16.32& 12.36& 10.74&  6.11&  3.35\\
J1258$+$1405&12:58:02.05& $+$14:05:37.83& 0.3077& 20.89& 21.86& 21.21& 16.68& 14.79& 11.91&  8.57\\
J1308$+$2533&13:08:05.98& $+$25:33:38.17& 0.2024& 20.93& 21.33& 20.93& 15.70& 13.72&  9.77&  6.65\\
J1318$+$0336&13:18:47.09& $+$03:36:56.53& 0.1762& 19.69& 20.36& 20.65& 15.66& 13.72&  9.96&  6.74\\
J1323$-$0132&13:23:47.47& $-$01:32:51.95& 0.0225& 18.12& 19.22& 19.18& 17.18& 15.61& 10.89&  7.76&UM 570\\
J1334$+$5349&13:34:46.27& $+$53:49:27.49& 0.1679& 19.90& 20.58& 19.70& 17.81& 16.17& 12.39&  8.90\\
J1341$+$3502&13:41:41.62& $+$35:02:14.05& 0.0229& 17.88& 21.16& 20.39& 14.95& 13.33&  8.29&  5.01\\
J1347$+$3456&13:47:06.91& $+$34:56:24.21& 0.0539& 16.61& 18.02& 17.74& 13.05& 11.30&  6.78&  3.38&CG 1189, HS 1344+3511\\
J1418$+$2809&14:18:43.40& $+$28:09:57.46& 0.0413& 17.51&  ... & 18.72& 14.86& 12.87&  8.34&  5.00\\
J1430$+$4802&14:30:55.90& $+$48:02:01.58& 0.4775& 20.70&  ... & 21.44& 15.66& 13.94&  9.35&  7.07\\
J1437$+$1719&14:37:57.17& $+$17:19:20.57& 0.2021& 20.59& 21.39& 21.09& 18.43& 16.49& 11.75&  8.55\\
J1455$+$0036&14:55:33.67& $+$00:36:57.31& 0.0753& 18.22& 19.57& 19.17& 15.67& 14.17& 10.35&  7.26\\
J1503$+$1843&15:03:10.43& $+$18:43:08.81& 0.0464& 18.75& 21.83& 20.82& 14.09& 12.54&  8.62&  5.65\\
J1519$+$3945&15:19:47.14& $+$39:45:37.82& 0.0467& 17.82& 18.49& 17.95& 12.59& 10.83&  6.72&  2.98&CG 684\\
J1541$+$4536&15:41:20.04& $+$45:36:19.17& 0.2029& 19.38& 19.43& 19.62& 16.71& 14.99& 10.06&  6.28\\
J1555$+$3543&15:55:16.39& $+$35:43:24.65& 0.4519& 19.79& 20.44& 20.45& 15.43& 13.79& 10.00&  7.12\\
J1559$+$0047&15:59:57.37& $+$00:47:41.10& 0.2469& 20.74&  ... & 21.86& 17.63& 15.84& 11.37&  7.89\\
J1616$+$2138&16:16:06.67& $+$21:38:17.56& 0.2882& 20.74& 21.53& 22.14& 15.60& 14.09& 10.64&  7.54\\
J1639$+$2707&16:39:37.27& $+$27:07:50.16& 0.0406& 19.90& 22.00& 21.61& 16.75& 14.96& 10.60&  7.66\\
J1719$+$3037&17:19:58.31& $+$30:37:35.71& 0.1419& 18.94& 21.40& 20.99& 15.46& 13.89&  9.87&  6.91\\
J2049$+$0009&20:49:29.06& $+$00:09:23.78& 0.0715& 19.26& 21.51& 20.97& 17.00& 15.49& 11.88&$>$8.74\\
J2212$+$0033&22:12:20.20& $+$00:33:40.90& 0.0581& 21.09& 18.53& 18.19& 13.15& 11.60&  7.54&  4.34\\
J2215$+$0002&22:15:23.04& $+$00:02:46.75& 0.0774& 18.98& 20.31& 20.05& 16.14& 14.56& 10.02&  6.83\\
J2238$+$1400&22:38:31.12& $+$14:00:29.78& 0.0206& 18.89& 18.46& 18.50& 15.83& 14.03&  9.55&  6.73&HS 2236+1344\\
J2308$+$2121&23:08:38.24& $+$21:21:39.14& 0.2091& 20.96&  ... &  ... & 18.43& 16.74& 11.84&  8.47\\
J2318$-$0041&23:18:13.00& $-$00:41:26.15& 0.2516& 18.43& 19.38& 19.11& 14.88& 13.18&  9.84&  7.05\\
\multicolumn{12}{c}{c) Representative galaxies with $m$(3.4$\mu$m)--$m$(4.6$\mu$m) $<$ 0.5 mag} \\
J0135$-$0014&01:35:02.72& $-$00:14:31.04& 0.1540& 19.53& 20.18& 20.10& 16.40& 16.48& 11.90&  8.03\\
J0202$-$0027&02:02:23.53& $-$00:27:27.99& 0.0860& 19.36& 19.44& 19.45& 16.35& 16.63& 12.52&$>$8.85\\
J0934$+$2225&09:34:24.09& $+$22:25:22.64& 0.0844& 18.48& 19.48& 19.26& 16.11& 15.91& 11.51&  7.81\\
J1126$+$3803&11:26:37.77& $+$38:03:02.89& 0.2469& 19.23& 20.45& 19.85& 17.06& 16.75& 12.47&  8.74\\
J1201$+$2806&12:01:49.90& $+$28:06:10.67& 0.0559& 18.08& 19.17& 19.10& 16.93& 16.58& 12.48&  8.58\\
J1321$+$4708&13:21:51.95& $+$47:08:35.92& 0.1163& 18.09& 19.04& 18.73& 15.46& 15.18& 10.48&  7.02\\
J1347$+$3112&13:47:23.56& $+$31:12:54.36& 0.1191& 19.35& 20.08& 20.09& 15.34& 14.93& 10.79&  7.93\\
J1354$+$2149&13:54:34.25& $+$21:49:53.86& 0.1107& 18.18&  ... &  ... & 16.21& 15.82& 11.58&  8.20\\
J1534$+$1454&15:34:33.36& $+$14:54:47.56& 0.0733& 18.16& 19.18& 18.80& 15.16& 14.74& 10.27&  6.84\\
J1628$+$3054&16:28:27.54& $+$30:54:53.69& 0.1143& 19.40& 20.45& 20.03& 16.17& 16.01& 11.86&  8.11\\ \hline

\multicolumn{12}{l}{$^a$Equatorial coordinates.} \\
\multicolumn{12}{l}{$^b$Redshift.} \\
\multicolumn{12}{l}{$^c$SDSS $g$ magnitude.} \\
\multicolumn{12}{l}{$^d${\sl GALEX} far-UV and near-UV magnitudes.} \\
\multicolumn{12}{l}{$^e$Magnitudes in {\sl WISE} bands.} \\
 \end{longtable}

\setcounter{table}{2}

 \begin{longtable}{lcccccccccrccr}
 \caption{Derived parameters of the galaxies.}
 \label{tab3} \\ \hline \hline
 \endfirsthead
 \caption{---{\sl Continued.}} \\
 \hline\hline
Name     &12+logO/H$^a$&EW(H$\beta$)$^b$&$M_g$$^c$&\multicolumn{7}{c}{Log of luminosity}&log$M_*$$^f$&log$M_{\rm y}$$^g$&$SFR$$^h$ \\ \cline{5-11}
         &         & (\AA)     &      &H$\beta$$^d$&FUV$^e$&NUV$^e$&3.4$\mu$m$^e$&4.6$\mu$m$^e$&12$\mu$m$^e$&22$\mu$m$^e$&($M_\odot$)&($M_\odot$) \\ \hline
 \endhead
 \hline
 \endfoot
Name     &12+logO/H$^a$&EW(H$\beta$)$^b$&$M_g$$^c$&\multicolumn{7}{c}{Log of luminosity}&log$M_*$$^f$&log$M_{\rm y}$$^g$&$SFR$$^h$ \\ \cline{5-11}
         &         & (\AA)     &      &H$\beta$$^d$&FUV$^e$&NUV$^e$&3.4$\mu$m$^e$&4.6$\mu$m$^e$&12$\mu$m$^e$&22$\mu$m$^e$&($M_\odot$)&($M_\odot$) \\ \hline
\multicolumn{13}{c}{a) Galaxies with $m$(3.4$\mu$m)--$m$(4.6$\mu$m) $\geq$ 2 mag} \\
J0135$+$1455&  8.02&~\,\,46&$-$21.19& 42.58&  ... &  ... & 39.39& 39.68& 39.54& 39.52&  9.45&  9.35&46.8\\
J0239$+$0018&  8.08& 127&$-$19.42& 41.67&  ... &  ... & 37.93& 38.32& 38.62& 38.86&  8.28&  7.82   & 5.8\\
J0333$+$0017&  8.07& 138&$-$19.66& 41.69& 41.01& 40.39& 37.81& 38.13& 38.32& 38.62&  9.53&  7.88   & 6.0\\
J0936$+$0900&  8.13& 148&$-$19.90& 41.74& 40.86& 40.18& 37.78& 38.10& 38.41& 38.69&  9.75&  7.78   & 6.8\\
J1018$+$4106&  7.82& 166&$-$19.92& 41.68& 40.69& 40.55& 37.96& 38.35& 38.53& 38.84&  9.63&  7.68   & 5.9\\
J1046$+$3023&  8.16&~\,\,75&$-$19.73& 41.69&  ... &  ... & 38.63& 39.27& 39.20& 39.32&  9.02&  7.69& 6.0\\
J1101$+$4022&  8.01& 154&$-$19.88& 41.63&  ... & 40.43& 38.07& 38.55& 38.65& 38.97&  8.81&  7.63   & 5.3\\
J1141$+$4515&  7.63&~\,\,37&$-$20.00& 41.56& 40.90& 40.42& 38.66& 38.98& 38.77& 38.79&  9.89&  7.99& 4.5\\
J1143$+$3242&  8.16& 185&$-$19.06& 41.47& 40.08& 40.02& 37.34& 37.68& 38.12& 38.40&  9.07&  7.39   & 3.6\\
J1228$+$3219&  7.84& 327&$-$18.35& 41.39& 40.16& 39.71& 37.85& 38.18& 38.43& 38.50&  9.31&  7.51   & 3.0\\
J1238$+$4618&  8.11& 127&$-$19.38& 41.65& 40.73& 40.53& 38.11& 38.38& 38.57& 38.86&  9.16&  7.96   & 5.5\\
J1311$-$0027&  8.28& 112&$-$20.08& 41.99& 41.20& 40.56& 38.13& 38.59& 38.64& 38.67&  9.28&  7.97   &12.0\\
J1315$+$2618&  8.16& 288&$-$19.65& 41.76&  ... & 40.14& 38.78& 39.45& 39.37& 39.48&  9.15&  7.75   & 7.1\\
J1327$+$6151&  8.02& 151&$-$20.25& 42.11& 41.49& 40.75& 38.12& 38.42& 38.27&$<$38.60&  8.40&  8.38 &15.9\\
J1439$+$2453&  8.09& 144&$-$20.55& 42.40& 41.58& 41.07& 39.31& 39.63& 39.48& 39.34&  8.67&  8.64   &30.9\\
J1457$+$2232&  8.04& 178&$-$19.83& 42.00& 40.77& 40.54& 38.19& 38.58& 38.77& 38.88&  8.61&  7.93   &12.3\\
J1514$+$3852&  8.40& 100&$-$20.72& 41.77& 41.07& 40.79& 38.29& 38.70& 39.07& 39.13&  9.97&  8.20   & 7.3\\
J1537$+$5847&  8.04& 166&$-$19.69& 41.69& 40.81& 40.29& 38.01& 38.74& 38.76& 38.87&  9.41&  7.86   & 6.0\\
J1541$+$1753&  7.95& 229&$-$19.26& 41.90& 41.08& 40.45& 37.75& 38.18& 38.01& 38.36&  7.95&  7.88   & 9.8\\
J1604$+$0819&  8.02& 146&$-$20.21& 42.04& 41.64& 40.88& 38.22& 38.54& 38.60& 38.79&  8.97&  8.07   &13.5\\
\multicolumn{13}{c}{b) Galaxies with 2 mag $>$ $m$(3.4$\mu$m)--$m$(4.6$\mu$m) $\geq$ 1.5 mag} \\
J0042$+$1602&  8.10& 140&$-$20.08& 42.08& 40.95& 40.57& 38.50& 38.67& 38.81& 39.15&  8.28&  8.27   &14.8\\
J0122$+$0100&  8.12&~\,\,32&$-$20.19& 42.10& 41.70& 41.71& 38.90& 39.01& 38.91& 38.97& 10.31~\,\,&  8.82&15.5\\
J0132$-$0853&  8.34&~\,\,55&$-$19.42& 41.58&  ... &  ... & 38.44& 38.70& 38.63& 38.76&  9.77&  7.89& 4.7\\
J0303$-$0759&  7.86& 103&$-$20.12& 41.73& 40.76& 40.42& 38.40& 38.50& 38.59& 38.65&  8.51&  7.79   & 6.6\\
J0327$+$0101&  8.11&~\,\,49&$-$20.05& 41.63& 40.34& 40.07& 38.43& 38.58& 38.40& 38.53&  9.17&  7.58& 5.3\\
J0729$+$3949&  8.27&~\,\,59&$-$20.96& 42.11& 41.51& 41.36& 38.79& 39.00& 39.06& 39.44&  9.38&  8.11&15.9\\
J0740$+$3209&  8.04&~\,\,80&$-$18.97& 41.48&  ... &  ... & 38.06& 38.15& 38.23& 38.53&  9.19&  7.56& 3.7\\
J0808$+$2814&  8.17&~\,\,79&$-$20.55& 41.99& 42.23& 41.45& 38.46& 38.64& 38.46&$<$38.90&  8.61&  8.54 &12.0\\
J0817$+$4014&  8.54&~\,\,52&$-$20.51& 42.26&  ... &  ... & 39.26& 39.41& 39.00& 39.06& 10.41~\,\,&  8.50&22.4\\
J0820$+$5050&  8.21&~\,\,53&$-$21.03& 42.10& 41.34& 40.91& 38.66& 38.78& 38.79& 38.99&  9.32&  8.08&15.5\\
J0822$+$2241&  8.08& 140&$-$20.69& 42.35& 41.63& 41.14& 39.29& 39.38& 39.23& 39.32&  9.47&  8.55   &27.6\\
J0836$+$3130&  8.15&~\,\,19&$-$20.31& 41.39& 41.42& 40.83& 38.47& 38.59& 38.18&$<$38.70&  9.83&  8.07& 3.0\\
J0840$+$1344&  8.13&~\,\,91&$-$20.20& 42.10& 41.59& 40.96& 38.17& 38.37& 38.47& 38.74&  9.78&  8.39   &15.5\\
J0848$+$5106&  8.04& 118&$-$18.12& 40.81& 39.38& 39.37& 37.11& 37.19& 37.39& 37.45&  8.83&  6.83   & 0.8\\
J0849$+$1114&  8.36&~~~\,\,3&$-$19.24& 40.95& 41.93& 42.03& 39.13& 39.26& 39.10& 39.24&  9.67&  9.00& 1.1\\
J0852$+$1216&  8.12& 146&$-$20.38& 42.20& 41.14& 40.92& 38.11& 38.24& 38.45& 38.70&  9.63&  8.28   &19.5\\    
J0927$+$1740&  8.01&~\,\,93&$-$20.53& 41.92& 41.04& 40.73& 38.31& 38.52& 38.59& 38.62&  9.72&  8.02   &10.2\\
J0928$+$1502&  8.12&~\,\,86&$-$19.64& 41.95& 41.11& 41.00& 38.44& 38.64& 38.64& 38.78& 10.23~\,\,&  8.38&11.0\\
J0949$+$1750&  7.42&~\,\,12&$-$20.94& 42.18& 42.16& 42.39& 38.70& 38.86& 38.46& 38.48& 11.02~\,\,&  9.53&18.6\\
J1011$+$1308&  8.01& 299&$-$19.33& 41.81&  ... &  ... & 37.42& 37.61& 38.35& 38.56&  8.37&  7.74   & 8.0\\
J1018$+$5155&  8.06& 103&$-$20.17& 41.80& 41.26& 40.95& 38.38& 38.49& 38.52& 38.62&  9.29&  7.89   & 7.8\\
J1023$+$2421&  8.36&~\,\,45&$-$21.50& 42.46& 42.12& 41.69& 38.97& 39.17& 39.29& 39.46&  9.84&  8.58&35.5\\
J1034$+$1155&  8.35&~\,\,32&$-$21.43& 42.19& 41.98& 41.59& 39.05& 39.15& 39.05& 38.96& 10.59~\,\,&  8.86&19.1\\
J1047$+$0739&  7.85& 123&$-$19.65& 42.33& 42.01& 41.32& 38.77& 38.94& 39.04& 39.15& 10.30~\,\,&  8.30&26.3\\
J1050$+$1538&  7.94& 183&$-$19.76& 42.01& 40.89& 40.67& 37.60& 37.68& 38.06& 38.31&  8.52&  8.01   &12.6\\
J1111$+$0713&  8.12&~\,\,28&$-$20.71& 41.54& 41.01& 40.85& 38.35& 38.42& 38.10& 38.21&  9.35&  7.53& 4.3\\
J1132$+$3128&  7.82&~\,\,75&$-$17.40& 40.57& 39.71& 39.61& 36.86& 37.04& 37.34& 37.49&  7.80&  6.73& 0.5\\
J1135$+$6025&  7.58&~\,\,25&$-$22.66& 42.65& 42.17& 41.72& 39.48& 39.58& 39.43& 39.49& 10.66~\,\,&  9.75&55.0\\
J1148$+$1756&  7.97& 165&$-$19.31& 41.78& 40.37& 40.29& 37.65& 37.83& 38.15& 38.33&  8.92&  7.72   & 7.4\\
J1151$+$3756&  8.28& 117&$-$20.33& 41.91& 41.05& 40.53& 38.66& 38.85& 38.85& 39.03&  9.78&  7.86   &10.0\\
J1151$+$0106&  8.43&~\,\,58&$-$18.93& 41.14& 40.00& 39.85& 37.78& 37.95& 38.48& 38.71&  9.64&  7.65& 1.7\\
J1155$+$5739&  7.95& 176&$-$18.03& 41.01& 39.99& 39.80& 36.88& 36.95& 37.26& 37.43&  7.86&  7.09   & 1.3\\
J1158$-$0203&  8.19&~\,\,37&$-$29.14& 41.90& 41.11& 40.85& 38.67& 38.87& 38.80& 38.96&  9.37&  7.94& 9.8\\
J1159$+$1344&  7.79&~\,\,37&$-$20.03& 41.84& 41.20& 41.09& 38.92& 39.17& 38.74& 38.73& 10.30~\,\,&  8.65& 8.5\\
J1201$+$0211&  7.51& 199&$-$13.17& 39.40& 38.05& 37.78& 35.19& 35.39& 35.39& 35.44&  6.14&  5.37   & 0.03\\
J1205$+$2856&  7.89& 214&$-$18.76& 41.74& 40.69& 40.50& 37.74& 37.83& 38.06& 38.25&  7.79&  7.78   & 6.8\\
J1219$+$1526&  7.89& 196&$-$20.36& 41.95& 41.12& 40.58& 38.60& 38.67& 38.39& 38.54&  8.50&  7.81   &11.0\\
J1245$+$1043&  8.25& 113&$-$21.11& 42.44& 41.50& 41.20& 38.61& 38.87& 39.00& 39.27& 10.20~\,\,&  8.63&33.9\\
J1245$+$5254&  8.48&~\,\,24&$-$20.85& 41.76& 41.13& 40.84& 38.43& 38.58& 38.60& 38.72&  9.79&  7.86& 7.1\\
J1248$+$1234&  8.11& 114&$-$20.70& 41.85& 41.30& 40.69& 38.53& 38.72& 38.81& 38.86&  8.87&  7.95   & 8.7\\
J1253$-$0312&  8.04& 226&$-$19.87& 41.78& 40.48& 40.22& 38.07& 38.18& 38.47& 38.46&  8.53&  7.68   & 7.4\\
J1258$+$1405&  8.12& 232&$-$20.09& 42.01& 41.11& 40.82& 38.70& 38.93& 38.51& 38.74&  9.15&  7.84   &12.6\\
J1308$+$2533&  7.81& 409&$-$19.05& 41.77& 40.72& 40.43& 38.70& 38.96& 38.97& 39.11&  8.80&  7.65   & 7.3\\
J1318$+$0336&  8.08&~\,\,70&$-$19.96& 41.79& 40.94& 40.41& 38.58& 38.83& 38.76& 38.94&  9.32&  7.84& 7.6\\
J1323$-$0132&  7.78& 245&$-$16.92& 40.36& 39.14& 38.82& 36.13& 36.23& 36.55& 36.69&  7.55&  6.19   & 0.3\\
J1334$+$5349&  8.05&~\,\,98&$-$19.65& 41.43& 40.70& 40.66& 37.68& 37.81& 37.75& 38.04&  9.32&  7.65   & 3.3\\
J1341$+$3502&  8.10&~\,\,49&$-$17.19& 40.89& 39.34& 39.47& 37.03& 37.15& 37.60& 37.80&  9.17&  7.34& 1.0\\
J1347$+$3456&  8.19&~\,\,65&$-$20.36& 42.31& 41.78& 41.78& 38.59& 38.76& 39.00& 39.21& 10.26~\,\,&  8.94&25.2\\
J1418$+$2809&  8.32&~\,\,91&$-$18.86& 41.45&  ... & 40.35& 37.59& 37.86& 38.10& 38.33&  9.33&  7.63& 3.5\\
J1430$+$4802&  7.78&~\,\,24&$-$21.36& 42.14&  ... & 41.40& 39.54& 39.70& 39.96& 39.77&  9.74&  8.37&17.0\\
J1437$+$1719&  7.81& 203&$-$19.39& 41.86& 40.76& 40.42& 37.60& 37.85& 38.18& 38.35&  9.63&  7.85   & 8.9\\
J1455$+$0036&  8.20&~\,\,40&$-$19.50& 41.20& 40.02& 39.84& 37.80& 37.87& 37.83& 37.96&  9.68&  7.85& 2.0\\
J1503$+$1843&  8.25&~\,\,38&$-$17.88& 41.26& 40.11& 40.42& 38.00& 38.09& 38.09& 38.17&  9.41&  7.81& 2.2\\
J1519$+$3945&  8.32&~\,\,37&$-$18.83& 41.12& 41.05& 41.10& 38.61& 38.78& 38.86& 39.25&  9.70&  7.85& 1.6\\
J1541$+$4536&  8.39&~\,\,85&$-$20.61& 41.99& 41.41& 40.90& 38.30& 38.45& 38.86& 39.26&  9.14&  8.19   &12.0\\
J1555$+$3543&  8.27&~\,\,65&$-$22.13& 42.77& 43.11& 41.90& 39.58& 39.71& 39.66& 39.70& 10.38~\,\,&  8.92&72.5\\
J1559$+$0047&  8.15& 136&$-$19.71& 41.64&  ... & 39.88& 38.11& 38.29& 38.51& 38.80&  9.70&  7.73   & 5.4\\
J1616$+$2138&  8.30&~\,\,90&$-$20.08& 42.09& 41.54& 40.60& 39.07& 39.15& 38.96& 39.09&  9.81&  8.11   &15.2\\
J1639$+$2707&  8.01&~\,\,58&$-$16.43& 40.56& 39.56& 39.53& 36.82& 37.01& 37.18& 37.25&  8.16&  6.90& 0.4\\
J1719$+$3037&  8.21&~\,\,15&$-$20.21& 41.86& 41.65& 41.35& 38.46& 38.56& 38.60& 38.68&  9.81&  8.83& 8.9\\
J2049$+$0009&  8.37&~\,\,23&$-$18.33& 40.60& 39.90& 39.88& 37.22& 37.30& 37.17&$<$37.32&  9.41&  7.48& 0.5\\
J2212$+$0033&  7.77&~\,\,20&$-$16.04& 40.55& 41.23& 41.19& 38.57& 38.66& 38.72& 38.89&  9.40&  7.43& 0.4\\
J2215$+$0002&  8.07& 157&$-$18.79& 41.32& 40.19& 40.01& 37.64& 37.74& 37.99& 38.15&  8.62&  7.38   & 2.6\\
J2238$+$1400&  7.61& 177&$-$15.95& 40.02& 39.41& 39.07& 36.59& 36.78& 37.00& 37.03&  7.55&  6.08   & 0.1\\
J2308$+$2121&  8.30&~\,\,74&$-$19.09& 41.75&  ... &  ... & 37.63& 37.78& 38.17& 38.41&  9.52&  7.86& 6.9\\
J2318$-$0041&  8.18&~\,\,34&$-$22.06& 42.35& 42.27& 41.74& 39.23& 39.38& 39.15& 39.16&  9.76&  8.40&27.6\\
\multicolumn{13}{c}{c) Galaxies with $m$(3.4$\mu$m)--$m$(4.6$\mu$m) $<$ 0.5 mag} \\
J0135$-$0014&  8.15&~\,\,88&$-$19.81& 41.37& 40.65& 40.32& 38.16& 37.60& 37.86& 38.30&  9.29&  7.66   & 2.9\\
J0202$-$0027&  7.76&~\,\,97&$-$18.65& 41.14& 40.44& 40.12& 37.65& 37.01& 37.09&$<$37.45&  9.23&  7.26 & 1.7\\
J0934$+$2225&  8.05& 102&$-$19.49& 41.64& 40.55& 40.34& 37.72& 37.27& 37.46& 37.84&  9.26&  8.03   & 5.4\\
J1126$+$3803&  8.04&~\,\,83&$-$21.22& 42.06& 41.08& 40.88& 38.34& 37.93& 38.07& 38.46&  9.82&  8.17   &14.1\\
J1201$+$2806&  8.01& 162&$-$18.97& 41.42& 40.18& 39.91& 37.03& 36.64& 36.71& 37.16&  9.45&  7.47   & 3.2\\
J1321$+$4708&  8.11&~\,\,90&$-$20.61& 41.93& 41.01& 40.79& 38.28& 37.86& 38.18& 38.45&  9.04&  8.00   &10.5\\
J1347$+$3112&  8.01& 111&$-$19.40& 41.56& 40.69& 40.34& 38.35& 37.98& 38.07& 38.11&  9.48&  7.74   & 4.5\\
J1354$+$2149&  8.17& 102&$-$20.40& 42.02&  ... &  ... & 37.94& 37.57& 37.69& 37.94&  9.48&  8.34   &12.9\\
J1534$+$1454&  8.15& 132&$-$19.50& 41.74& 40.60& 40.47& 37.98& 37.62& 37.84& 38.10&  9.84&  7.86   & 6.8\\
J1628$+$3054&  8.03& 105&$-$19.26& 41.45& 40.53& 40.36& 37.98& 37.52& 37.61& 38.00&  9.38&  7.67   & 3.5\\ \hline
\multicolumn{14}{l}{$^a$Oxygen abundance.}\\
\multicolumn{14}{l}{$^b${Rest-frame equivalent width of the H$\beta$ emission line.}}\\
\multicolumn{14}{l}{$^c$Extinction-corrected absolute SDSS $g$ magnitude.}\\
\multicolumn{14}{l}{$^d${Log of the aperture- and extinction-corrected luminosity in units erg s$^{-1}$cm$^{-2}$.}}\\
\multicolumn{14}{l}{$^e$Log of the extinction-corrected luminosity in units erg s$^{-1}$cm$^{-2}$\AA$^{-1}$.}\\
\multicolumn{14}{l}{$^f${Log of the aperture-corrected stellar mass.}}\\
\multicolumn{14}{l}{$^g${Log of the aperture-corrected mass of the young stellar population.}}\\
\multicolumn{14}{l}{$^h${Star formation rate in $M_\odot$yr$^{-1}$ derived from $L$(H$\alpha$)=2.8$L$(H$\beta$) and using relation 
$SFR$($M_\odot$yr$^{-1}$)=4.4$\times$10$^{-42}$$L$(H$\alpha$)(erg s$^{-1}$) \citep{K99}.}}\\
 \end{longtable}

\setcounter{figure}{9}

\begin{figure*}
\hbox{
\includegraphics[angle=0,width=1.0\linewidth]{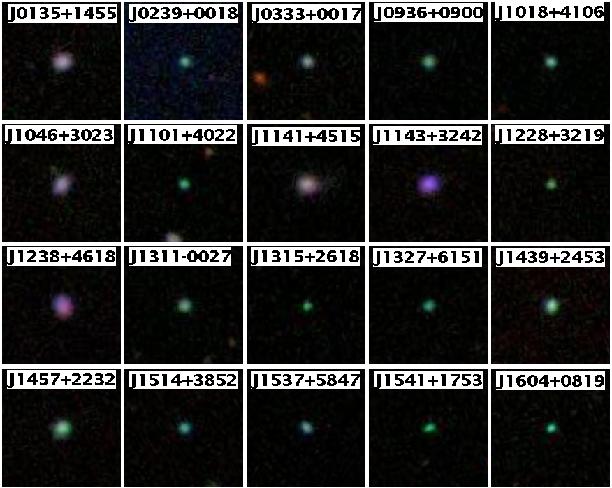} 
}
\caption{20\arcsec$\times$20\arcsec\ SDSS composite $g,r,i$ images of 
galaxies with
$m$(3.4$\mu$m) -- $m$(4.6$\mu$m) $\geq$ 2 mag.} 
\label{fig10}
\end{figure*}

\setcounter{figure}{10}

\begin{figure*}
\includegraphics[angle=0,width=1.0\linewidth]{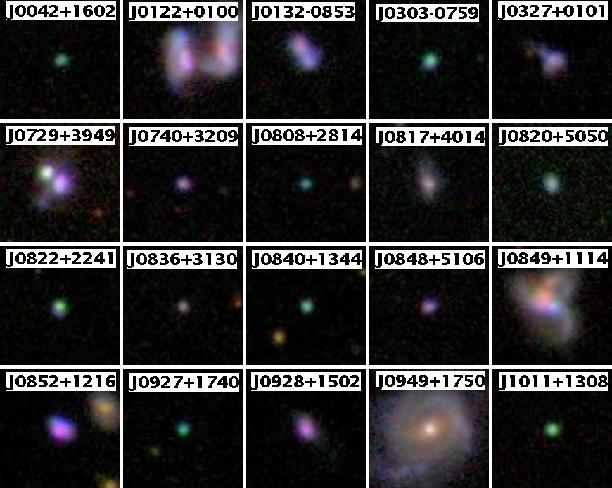} 
\caption{20\arcsec$\times$20\arcsec\ SDSS composite 
$g,r,i$ images of galaxies with
2 mag $>$ $m$(3.4$\mu$m) -- $m$(4.6$\mu$m) $\geq$ 1.5 mag.} 
\label{fig11}
\end{figure*}

\setcounter{figure}{10}

\begin{figure*}
\includegraphics[angle=0,width=1.0\linewidth]{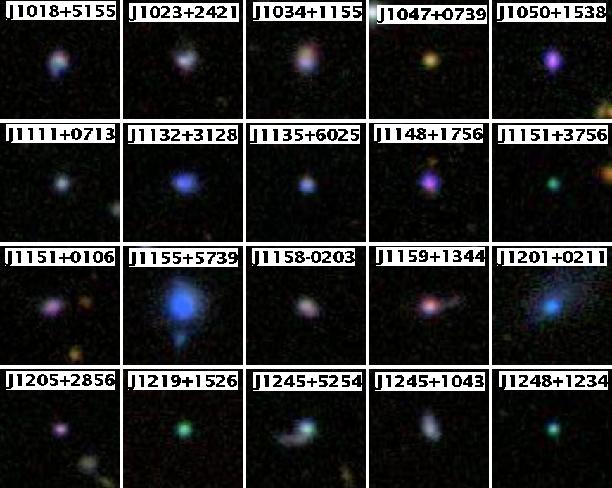} 
\caption{{\sl ---Continued.}} 
\end{figure*}

\setcounter{figure}{10}

\begin{figure*}
\includegraphics[angle=0,width=1.0\linewidth]{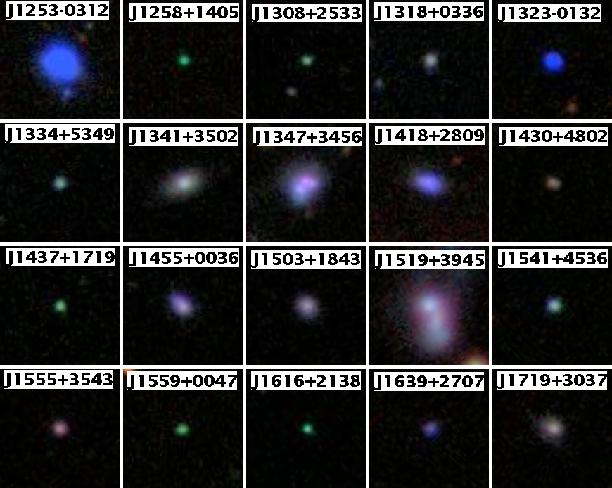} 
\caption{{\sl ---Continued.}} 
\end{figure*}

\setcounter{figure}{10}

\begin{figure*}
\includegraphics[angle=0,width=1.0\linewidth]{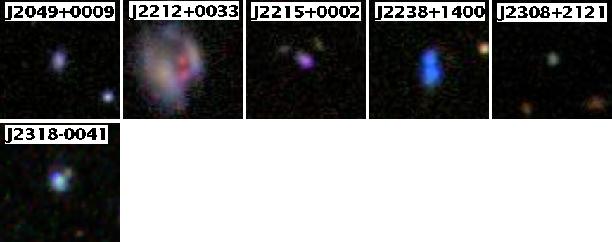} 
\caption{{\sl ---Continued.}} 
\end{figure*}

\setcounter{figure}{11}

\begin{figure*}
\includegraphics[angle=0,width=1.0\linewidth]{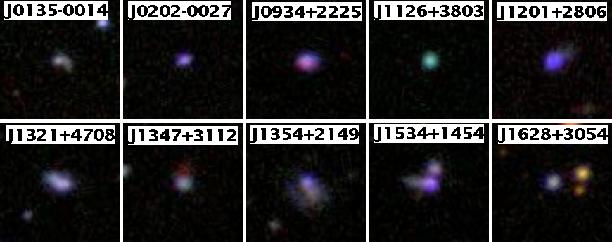} 
\caption{20\arcsec$\times$20\arcsec\ SDSS composite 
$g,r,i$ images of ten representative
galaxies with $m$(3.4$\mu$m) -- $m$(4.6$\mu$m) $<$ 0.5 mag.} 
\label{fig12}
\end{figure*}

\setcounter{figure}{12}

\begin{figure*}
\includegraphics[angle=-90,width=1.0\linewidth]{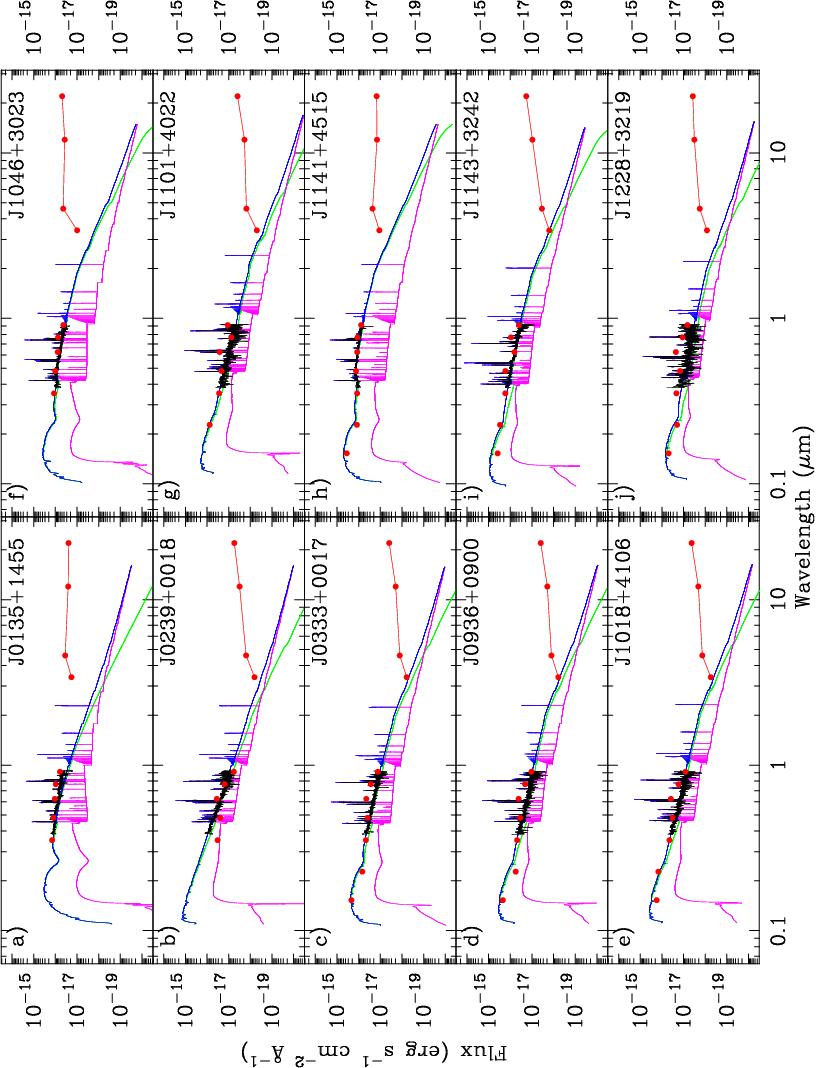} 
\caption{
Spectral energy distributions (SED) of galaxies with 
$m$(3.4$\mu$m) -- $m$(4.6$\mu$m) $\geq$ 2 mag. 
Observed SDSS optical spectra are shown by black solid 
lines; observed {\sl GALEX}, SDSS and {\sl WISE} monochromatic fluxes are shown
by filled red circles. The {\sl WISE} data are connected by a red solid line.
The modelled SED shown by the blue line is a sum of stellar emission (green
line) and ionised gas emission (magenta line).
}
\label{fig13}
\end{figure*}

\setcounter{figure}{12}

\begin{figure*}
\includegraphics[angle=-90,width=1.0\linewidth]{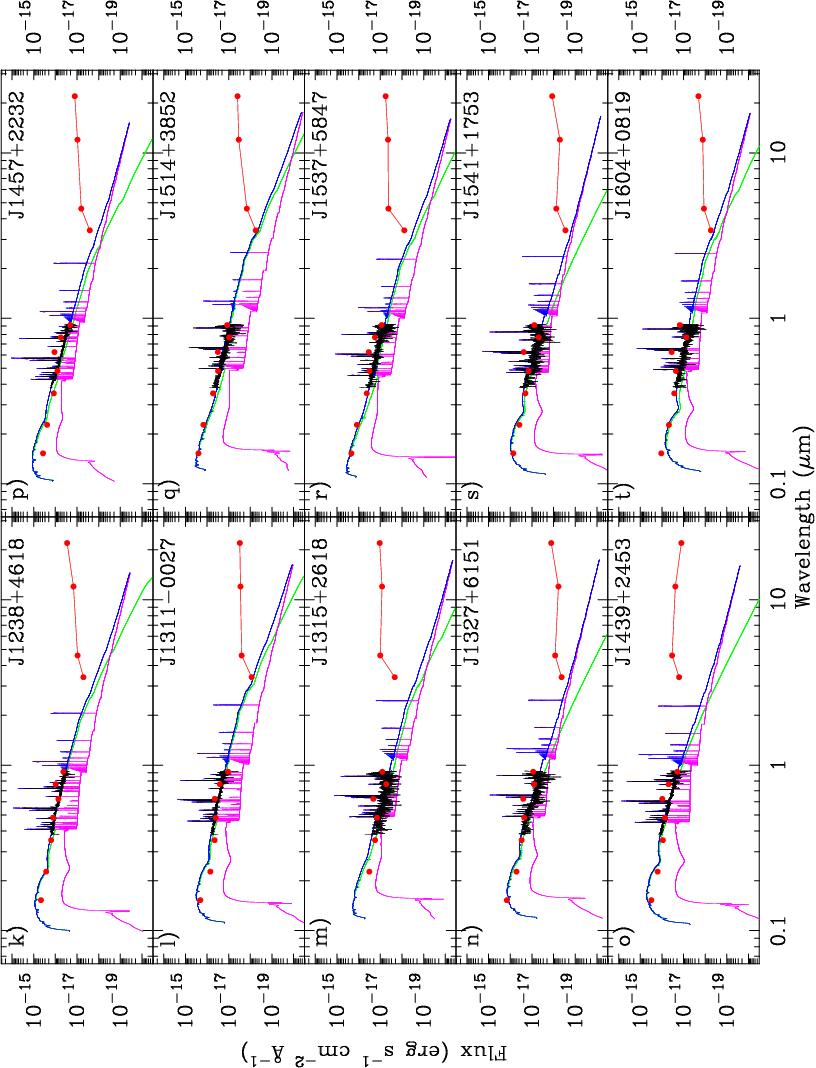} 
\caption{{\sl ---Continued.}} 
\end{figure*}

\setcounter{figure}{13}

\begin{figure*}
\includegraphics[angle=-90,width=1.0\linewidth]{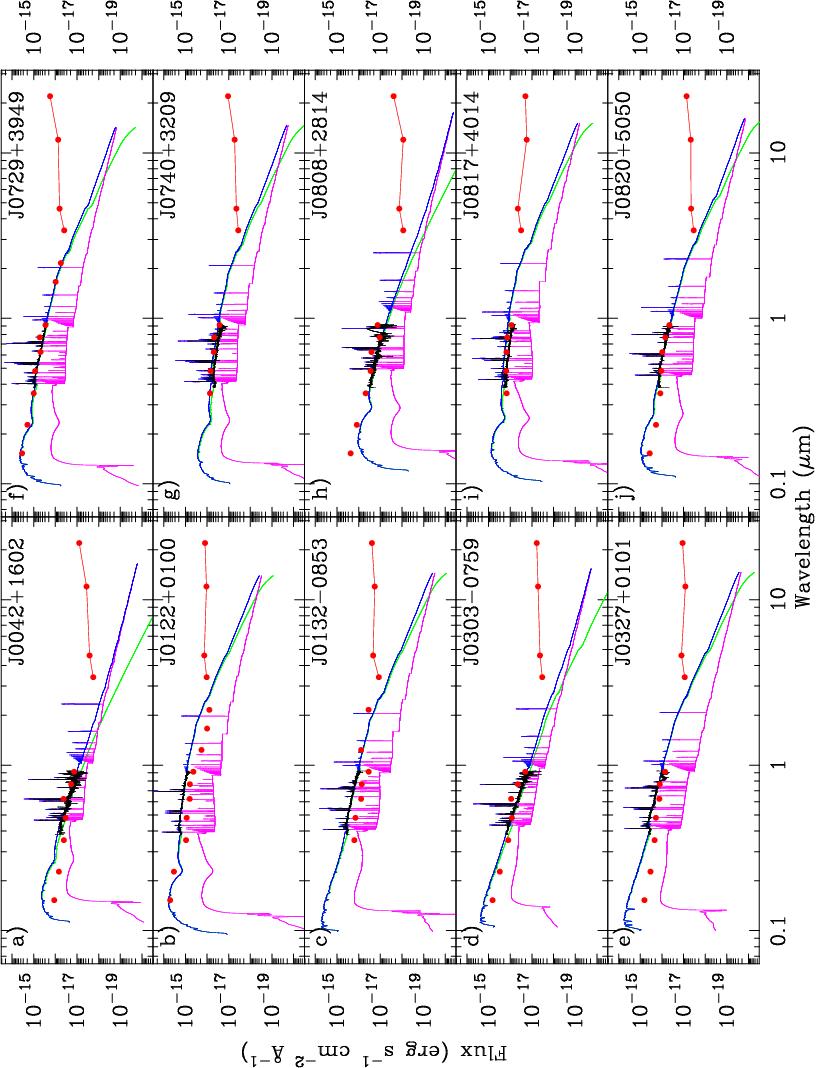} 
\caption{Same as in Fig. \ref{fig13} but for galaxies with
2 mag $>$ $m$(3.4$\mu$m) -- $m$(4.6$\mu$m) $\geq$ 1.5 mag.}
\label{fig14}
\end{figure*}

\setcounter{figure}{13}

\begin{figure*}
\includegraphics[angle=-90,width=1.0\linewidth]{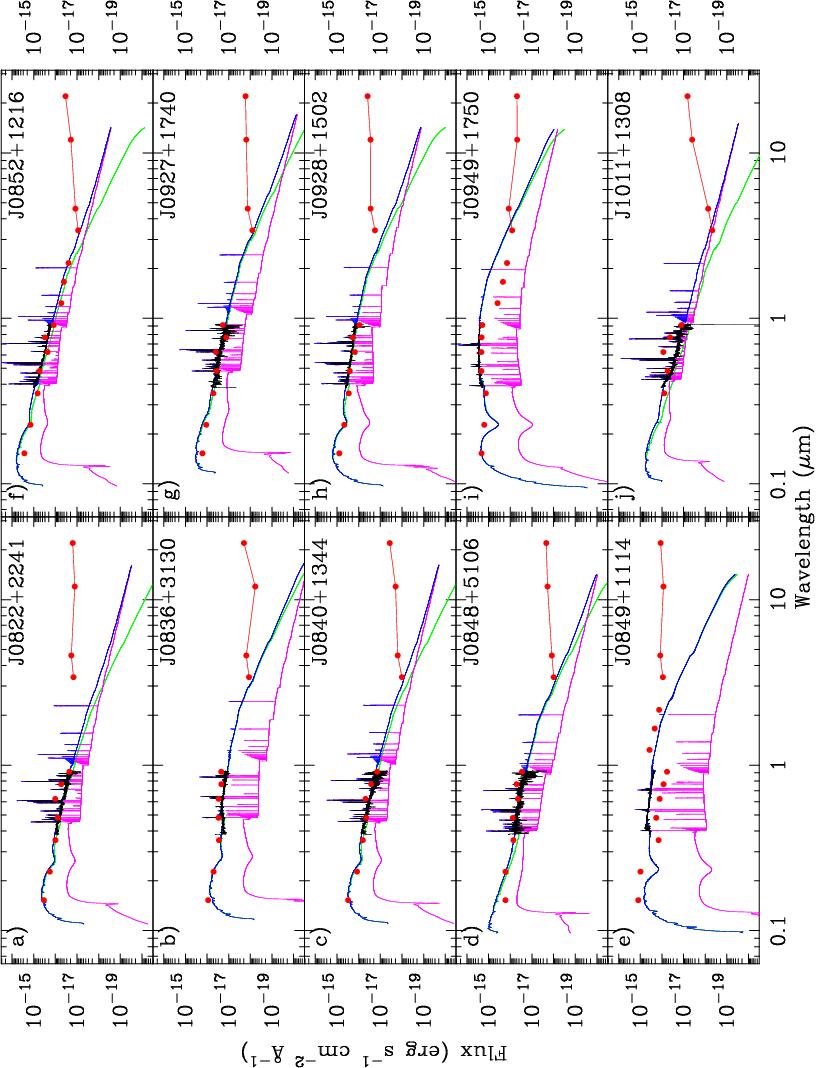} 
\caption{{\sl ---Continued.} }
\end{figure*}

\setcounter{figure}{13}

\begin{figure*}
\includegraphics[angle=-90,width=1.0\linewidth]{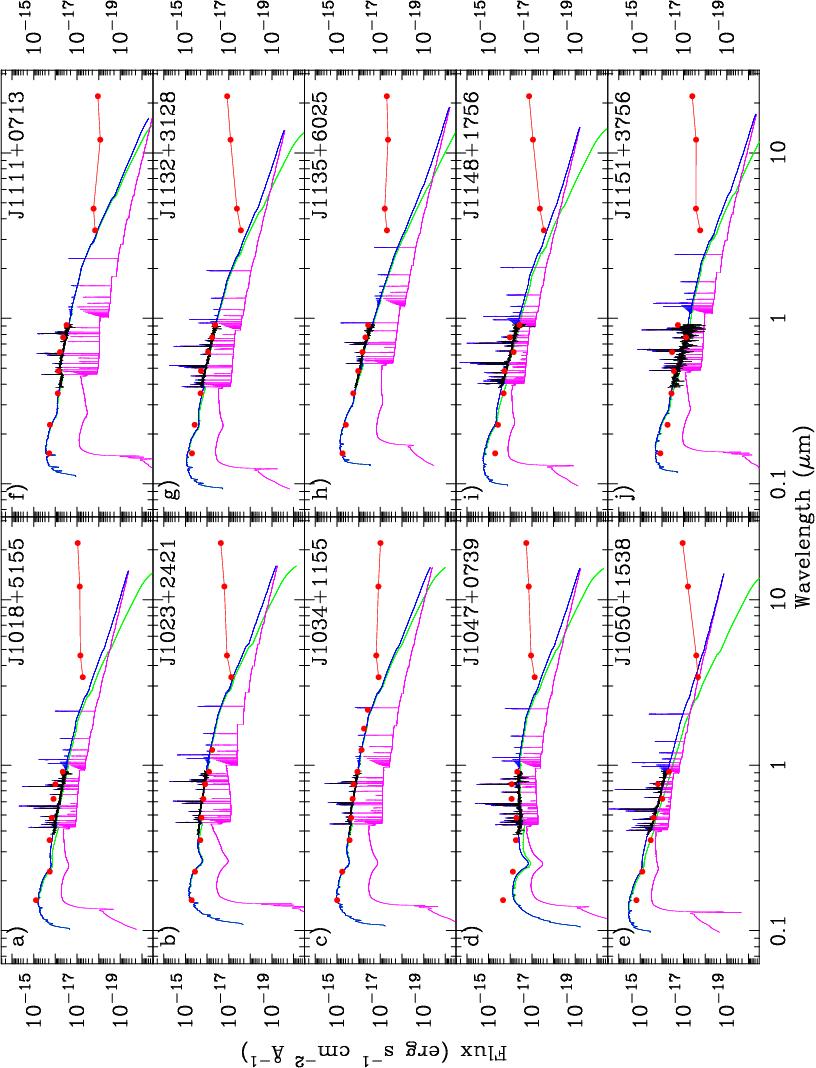} 
\caption{{\sl ---Continued.} }
\end{figure*}

\setcounter{figure}{13}

\begin{figure*}
\includegraphics[angle=-90,width=1.0\linewidth]{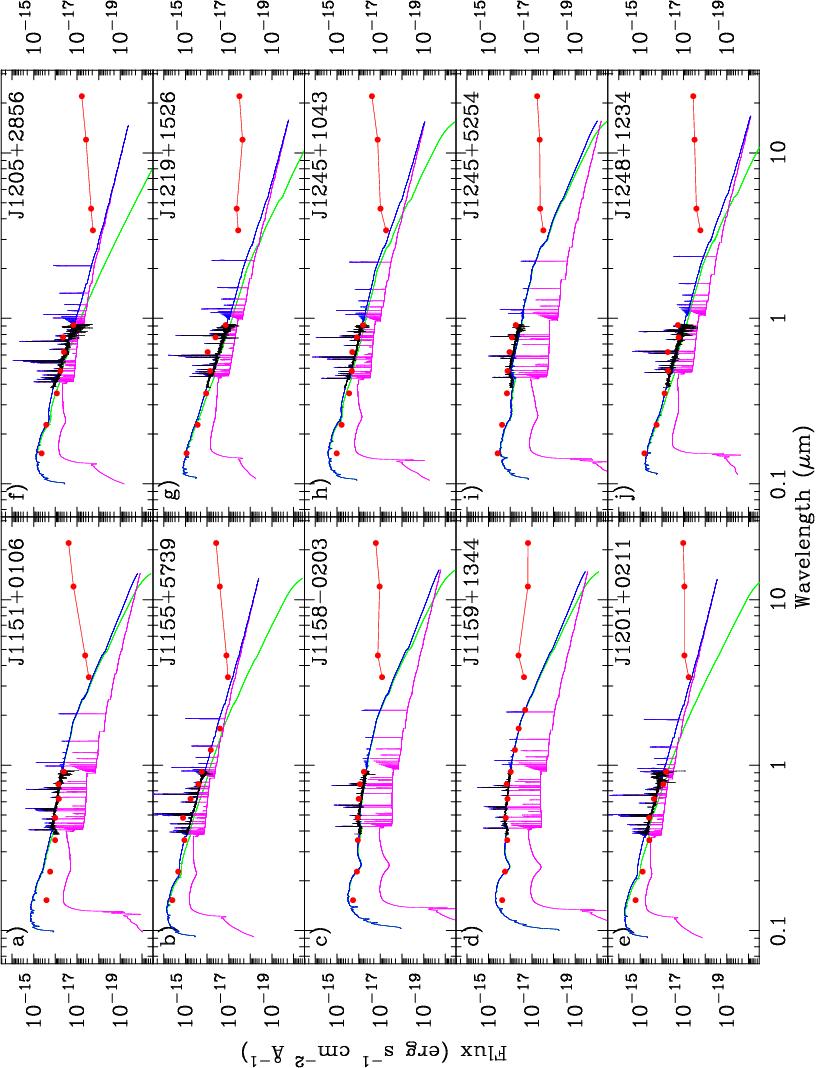} 
\caption{{\sl ---Continued.} }
\end{figure*}

\setcounter{figure}{13}

\begin{figure*}
\includegraphics[angle=-90,width=1.0\linewidth]{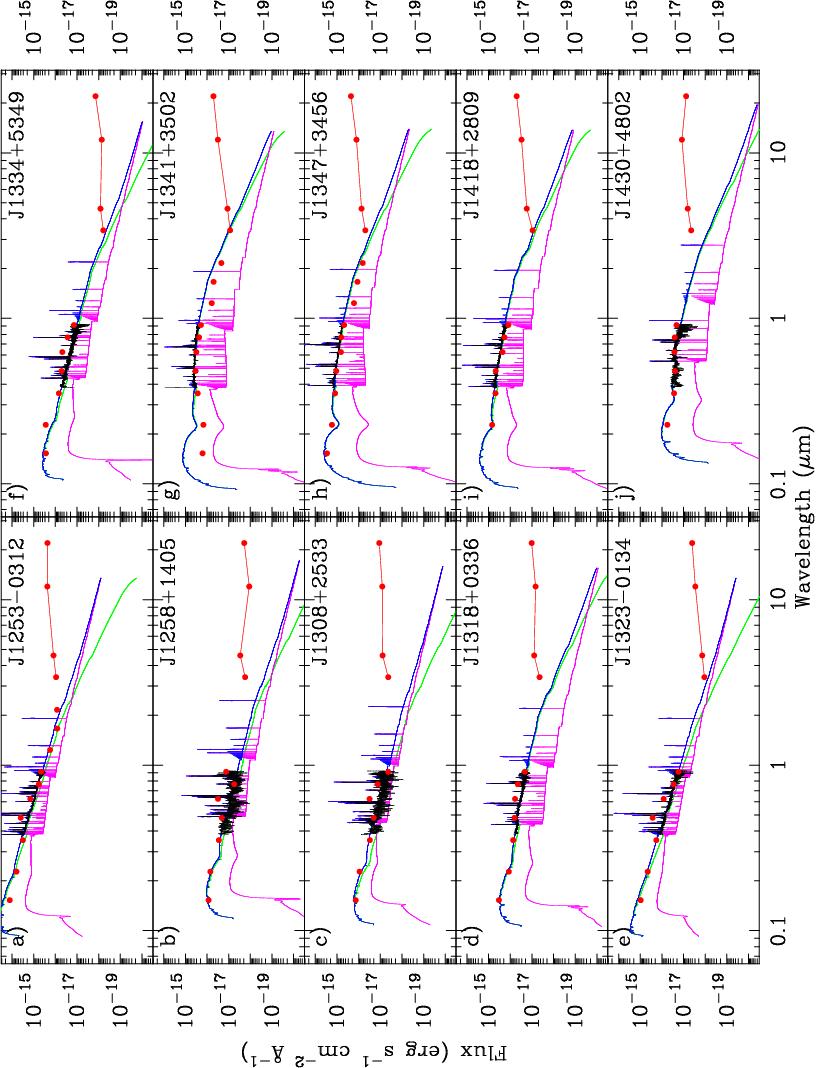} 
\caption{{\sl ---Continued.} }
\end{figure*}

\setcounter{figure}{13}

\begin{figure*}
\includegraphics[angle=-90,width=1.0\linewidth]{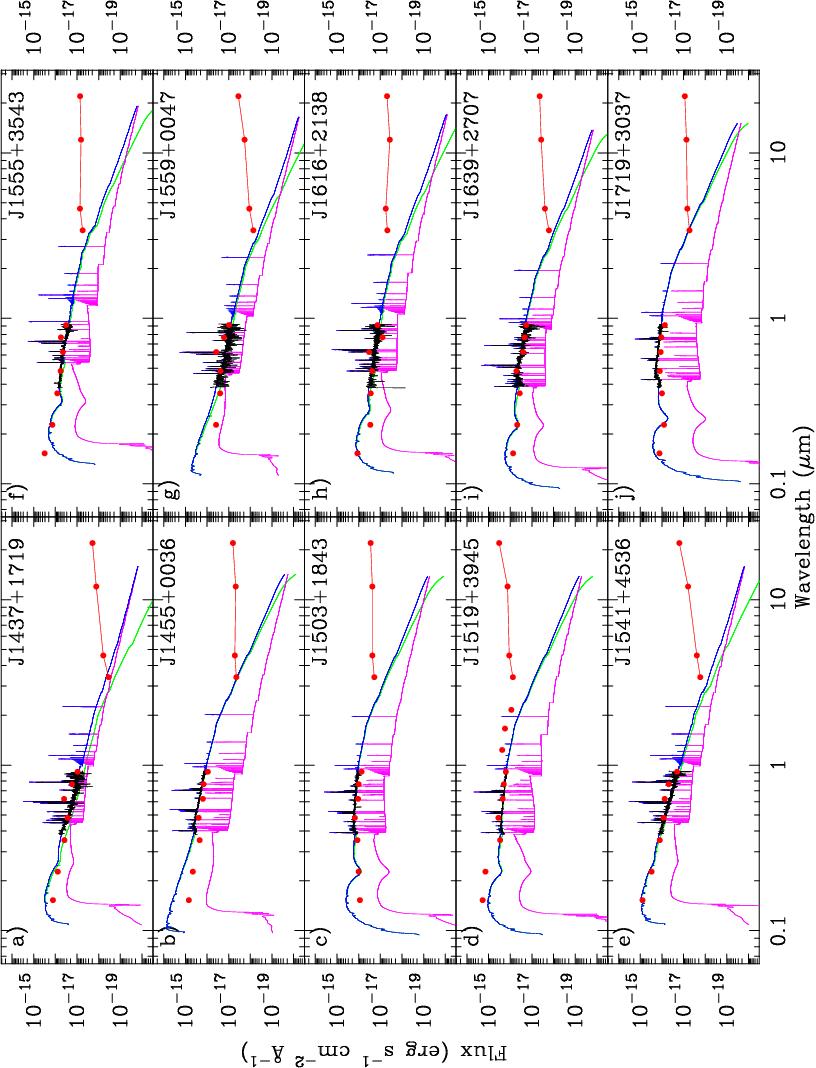} 
\caption{{\sl ---Continued.} }
\end{figure*}

\setcounter{figure}{13}

\begin{figure*}
\includegraphics[angle=-90,width=1.0\linewidth]{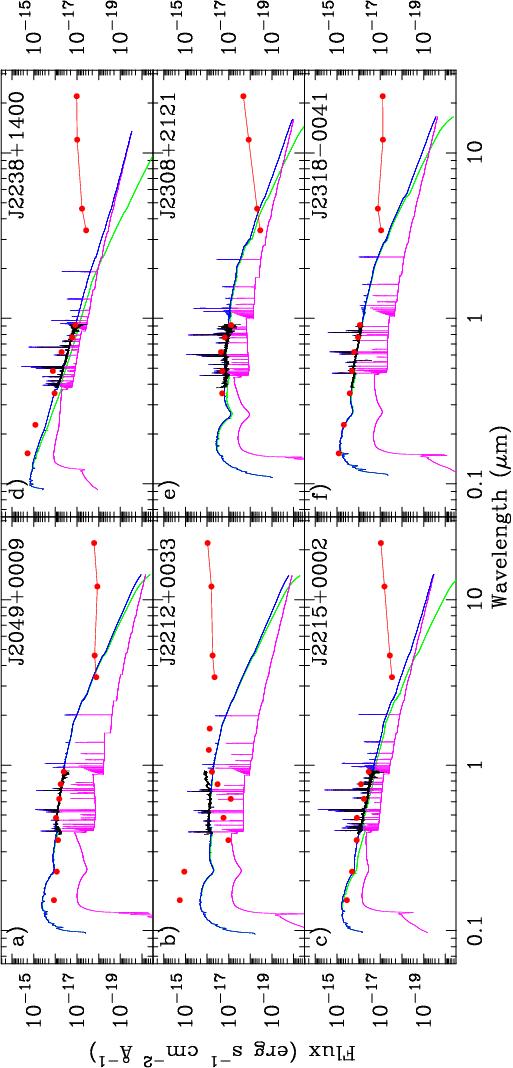} 
\caption{{\sl ---Continued.} }
\end{figure*}

\setcounter{figure}{14}

\begin{figure*}
\centering{
\includegraphics[angle=-90,width=1.0\linewidth]{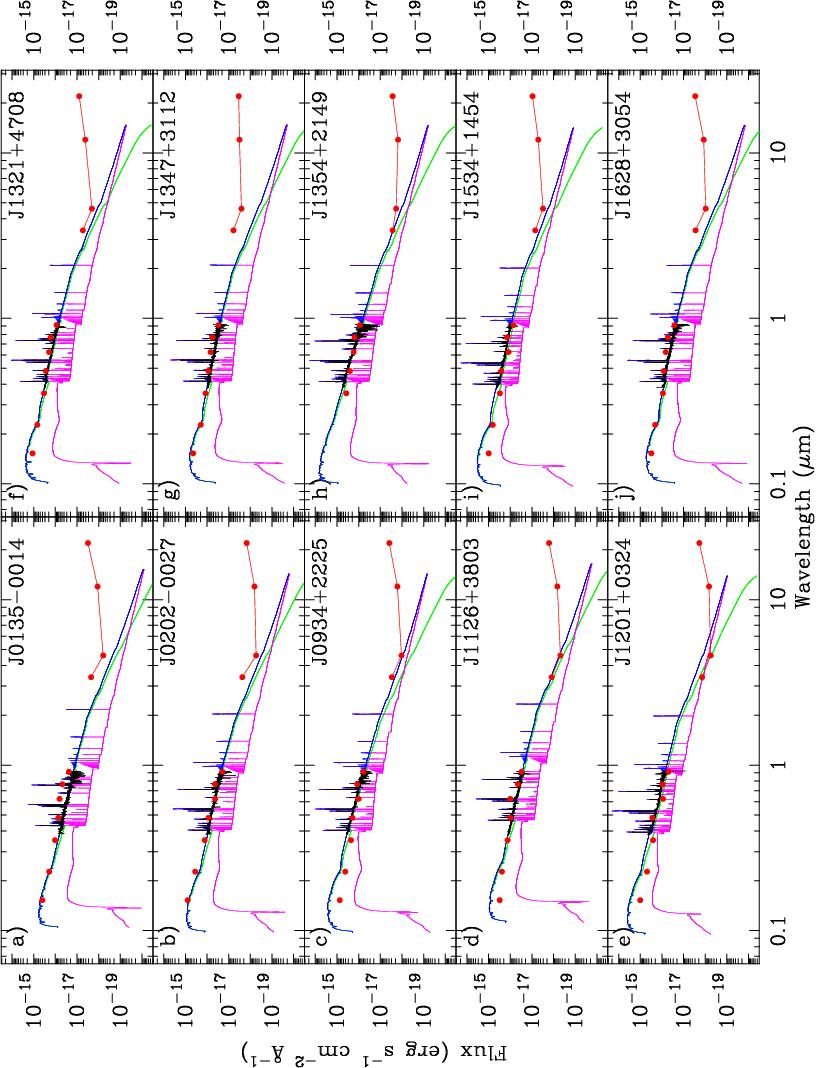}} 
\caption{
Same as in Fig. \ref{fig13} but for galaxies with
$m$(3.4$\mu$m) -- $m$(4.6$\mu$m) $<$ 0.5 mag.}
\label{fig15}
\end{figure*}

\end{document}